\documentclass{article}

\usepackage{fullpage}
\usepackage{enumerate}
\usepackage{graphicx}
\usepackage{setspace}
\usepackage{listings}
\usepackage[margin=1in]{geometry}
\usepackage{xcolor}
\usepackage{forest}
\usepackage{adjustbox}
\usepackage{float}
\usetikzlibrary{arrows.meta,backgrounds,calc,positioning}

\usepackage{amsfonts}
\usepackage{amssymb}
\usepackage{amsthm}

\newcommand{\CCfont}[1]{\ensuremath{\mathsf{#1}}}
\newcommand{\AC}{\CCfont{AC}}
\newcommand{\NC}{\CCfont{NC}}

\newtheorem{theorem}{Theorem}[subsection]

\theoremstyle{definition}

\newtheorem*{definition}{Definition}

\usepackage{hyperref}

\usepackage[most]{tcolorbox}

\title{\bf Formalizing PARITY Circuit Lower Bounds in Lean}

\author{
  Saint Wesonga\thanks{
  Department of Electrical Engineering and Computer Science,
  University of Wyoming,
  \texttt{swesonga@uwyo.edu}}
}

\forestset{
  decision tree/.style={
    for tree={
      grow=south,
      parent anchor=south,
      child anchor=north,
      edge={semithick},
      l sep=8mm,
      s sep=3mm,
      font=\small
    }
  },
  variable/.style={
    circle,
    draw,
    semithick,
    minimum size=7.5mm,
    inner sep=1pt,
    fill=blue!6
  },
  leaf/.style={
    rectangle,
    draw,
    semithick,
    rounded corners=1pt,
    minimum width=7mm,
    minimum height=5.5mm,
    inner sep=1pt,
    fill=gray!8
  },
  zero edge/.style={
    edge path'={(!u.south west) -- (.north)},
    edge label={node[midway,left,font=\scriptsize] {$0$}}
  },
  one edge/.style={
    edge path'={(!u.south east) -- (.north)},
    edge label={node[midway,right,font=\scriptsize] {$1$}}
  }
}

\definecolor{ink}{HTML}{172033}
\definecolor{muted}{HTML}{667085}
\definecolor{blue}{HTML}{2563EB}
\definecolor{bluepale}{HTML}{EFF6FF}
\definecolor{teal}{HTML}{0F766E}
\definecolor{tealpale}{HTML}{ECFDF5}
\definecolor{amber}{HTML}{B45309}
\definecolor{amberpale}{HTML}{FFFBEB}
\definecolor{line}{HTML}{CBD5E1}
\definecolor{panelblue}{HTML}{F8FAFF}

\tikzset{
  wire/.style={draw=ink, line width=1.05pt},
  signal/.style={wire,-{Latex[length=2.1mm,width=1.5mm]}},
  input/.style={rounded corners=2pt, draw=line, fill=white, line width=.8pt,
                minimum width=14mm, minimum height=7mm, font=\small\bfseries, text=ink},
  output/.style={rounded corners=3pt, draw=teal, fill=tealpale, very thick,
                 minimum width=23mm, minimum height=8mm, font=\small\bfseries, text=teal},
  not/.style={circle, draw=amber, fill=amberpale, thick, minimum size=7.5mm,
              inner sep=0pt, font=\scriptsize\bfseries, text=amber},
  andgate/.style={circle, draw=ink, fill=white, thick, minimum size=9.5mm,
                  inner sep=0pt, font=\small\bfseries, text=ink},
  orgate/.style={circle, draw=blue, fill=bluepale, very thick, minimum size=10mm,
                 inner sep=0pt, font=\small\bfseries, text=blue},
  macro/.style={rounded corners=5pt, draw=blue!35, line width=1pt, fill=panelblue},
  callout/.style={rounded corners=4pt, draw=line, fill=white, line width=.8pt},
  junction/.style={circle, fill=ink, inner sep=0pt, minimum size=2.4pt},
}

\tikzset{
  wire/.style={draw=ink, line width=1.05pt},
  signal/.style={wire,-{Latex[length=2.1mm,width=1.5mm]}},
  input/.style={rounded corners=2pt, draw=line, fill=white, line width=.8pt,
                minimum width=14mm, minimum height=7mm, font=\small\bfseries, text=ink},
  result/.style={rounded corners=3pt, draw=teal, fill=tealpale, very thick,
                 minimum width=31mm, minimum height=8mm, font=\small\bfseries, text=teal},
  not/.style={circle, draw=amber, fill=amberpale, thick, minimum size=7.5mm,
              inner sep=0pt, font=\scriptsize\bfseries, text=amber},
  andgate/.style={circle, draw=ink, fill=white, thick, minimum size=9.5mm,
                  inner sep=0pt, font=\small\bfseries, text=ink},
  orgate/.style={circle, draw=blue, fill=bluepale, very thick, minimum size=10mm,
                 inner sep=0pt, font=\small\bfseries, text=blue},
  macro/.style={rounded corners=5pt, draw=blue!35, line width=1pt, fill=panelblue},
  callout/.style={rounded corners=4pt, draw=line, fill=white, line width=.8pt},
  junction/.style={circle, fill=ink, inner sep=0pt, minimum size=2.4pt},
}

\definecolor{blue}{HTML}{2B6CB0}
\definecolor{paleblue}{HTML}{EAF3FB}
\definecolor{teal}{HTML}{1F7A78}
\definecolor{paleteal}{HTML}{E8F5F3}
\definecolor{orange}{HTML}{D97706}
\definecolor{paleorange}{HTML}{FFF4E5}
\definecolor{purple}{HTML}{6B46C1}
\definecolor{palepurple}{HTML}{F1ECFA}
\definecolor{ink}{HTML}{243447}
\definecolor{muted}{HTML}{607080}
\definecolor{rulegray}{HTML}{CCD6E0}
\definecolor{lightgray}{HTML}{F7F9FB}
\definecolor{codegray}{rgb}{0.95, 0.95, 0.95}

\newtcolorbox{callout}{
  enhanced,
  colback=paleblue,
  colframe=blue,
  boxrule=0.8pt,
  arc=0pt,
  left=8pt,right=8pt,top=7pt,bottom=7pt,
  before skip=5pt,after skip=7pt
}

\begin{document}

\lstset{
  basicstyle=\ttfamily,
  backgroundcolor=\color{codegray},
  frame=single,
  tabsize=2,
  breaklines=true,
}

\maketitle

\begin{abstract}
  We formalize H{\aa}stad's PARITY lower bound in Lean using the switching
  lemma.  For every fixed $d \geq 2$, formulas and DAG circuits of computation
  depth at most $d$ computing PARITY on $n$ inputs require size
  $\exp(\Omega_d(n^{1/(d-1)}))$ for all sufficiently large $n$.  This matches
  the classical upper bound up to constants in the exponent and implies that
  PARITY is not in nonuniform $\AC^0$.  We also construct a polynomial-size,
  logarithmic-depth bounded-fan-in formula family for PARITY, providing a
  witness to $\NC^1 \not\subseteq \AC^0$ for the formalized models.  The Lean
  source code is available at
  \url{https://github.com/formalcs/circuit-complexity} and is checked with
  Lean~4.33.1 and mathlib~4.33.1.
\end{abstract}

\section{Introduction}

We formalize in Lean H{\aa}stad's lower bound for constant-depth circuits
computing PARITY, the Boolean function that tests whether an odd number of input
bits are 1. We work with circuits with unbounded-fan-in AND and OR gates and
unary NOT gates.  A circuit family $(C_n)$ provides, for each input length $n$,
an $n$-input circuit, and no computational restriction is imposed on the map $n
\mapsto C_n$. A language belongs to nonuniform $\mathsf{AC}^0$ when it is
decided by such a family with a common polynomial bound on size and a common
constant bound on depth.

For each fixed $d \ge 2$ and all sufficiently large $n$, we establish
H{\aa}stad’s quantitative lower bound: every depth-$d$ circuit computing PARITY
has size at least $\exp(\Omega_d(n^{1/(d-1)}))$, matching the classical upper
bound up to constants in the exponent. Our formalization proves this exponent
both for formulas and for circuits with shared gates.  In particular, we
formally prove that PARITY is not in nonuniform $\mathsf{AC}^0$.

Furst, Saxe, and Sipser \cite{FuSaSi81} introduced the random-restriction
approach to show that PARITY is not in $\AC^0$: restrictions simplify small
constant-depth circuits while turning PARITY into either PARITY or its negation
on the remaining variables. Ajtai independently established the PARITY lower
bound in 1983, in work on the expressive power of logical formulas over finite
structures \cite{Ajtai1983}.  Yao \cite{Yao1985} strengthened the
superpolynomial size lower bound to $\Omega(2^{n^{1/4^d}})$, as recorded in
H{\aa}stad's comparison with the earlier result \cite{hastad1986almost}.
H{\aa}stad \cite{hastad1986almost} sharpened the size lower bound to
$\exp(\Omega_d(n^{1/(d-1)}))$, with negations confined to inputs and $d$ layers
of AND and OR gates.  The multiplicative constant hidden by $\Omega_d$ is a
separate quantity: it multiplies $n^{1/(d-1)}$ and may depend on $d$. Our
formalization establishes the same power $1/(d-1)$, with its own explicit
multiplicative constants.  H{\aa}stad's switching lemma provided exact DNF/CNF
conversions under suitable random restrictions with sharper bounds on the
probability of failure whereas Yao's argument used approximate conversions. As
H{\aa}stad noted, the upper bound $\exp(O(n^{1/(d-1)}))$ was already known as
folklore \cite{Hast86}. His lower bound matched it up to constants in the
exponent, establishing essentially optimal size bounds for every fixed depth
\cite{hastad1986almost}.

Our switching-lemma proof follows the combinatorial presentation in Beame's
\emph{A Switching Lemma Primer} \cite{Beame1994Primer} and O'Donnell's
Lecture~14, \emph{The Switching Lemma} \cite{online:SwitchingLemma}.  The
formalized lower-bound proof uses the switching lemma to reduce formula
depth. Once the bottom fan-in is bounded, the implementation extracts the bottom
subformulas and records their DNF or CNF form. The lemma is applied to DNF views
of these subformulas: a DNF is used directly, while a CNF is represented by a
DNF computing its negation.  A single restriction makes all these DNF views have
shallow canonical decision trees. Accounting for the negation in the CNF case,
these trees yield narrow CNFs or DNFs computing the restricted bottom
subformulas, with the form chosen to match the surrounding gate. Substitution
then merges adjacent gates of the same type and reduces the depth of the
formula. Each restriction of PARITY computes either PARITY or its negation on
the remaining variables. Iteration therefore gives a contradiction once the
resulting DNF has width smaller than its number of live variables: every
satisfiable term of a DNF computing PARITY or its negation must mention every
live variable.

Formalizing this argument makes its assumptions and intermediate transformations
explicit. The circuit and formula definitions specify evaluation, size, depth,
and the well-formedness conditions on which the lower bound depends. The proofs
check that normalization and unfolding circuits into formulas preserve
evaluation with controlled size and depth, and that each restriction and
depth-reduction step computes the restricted function while maintaining the
bounds needed for the next step. These definitions and lemmas provide reusable
components for further circuit lower bounds and restriction arguments.  Lean
checks the proof of every stated theorem relative to its definitions,
hypotheses, and axioms. Verifying that these express the intended mathematics
remains necessary.

The main theorem gives an explicit integer form of the circuit lower bound for
every fixed computation depth $d \geq 2$.
Listing~\ref{lst:MainQuantitativeParityCircuitLowerBound} gives its Lean
statement. Circuit size counts non-input gates, including the separate output
vertex. The stored circuit depth counts edges on paths ending at that vertex,
including the final output wire. Thus $d$ computation layers correspond to the
hypothesis \lstinline|circuit.depth| $\leq d+1$.  Signed inputs have depth
zero. An explicit NOT gate counts as a computation layer. The constants may
depend on $d$. We make no claim of optimal constants.

\noindent\begin{minipage}{\linewidth}
\begin{lstlisting}[caption={Main Theorem: Quantitative PARITY Circuit Lower Bound},
                   mathescape=true,
                   label={lst:MainQuantitativeParityCircuitLowerBound}]{}
theorem circuit_parity_size_lower_bound_root_sharp
    (d  : Nat)
    (hd : 2 $\leq$ d) :
  $\exists$ N, $\forall$ n, N $\leq$ n $\rightarrow$
    $\forall$ (circuit : Circuit),
      circuit.inputWidth = n $\rightarrow$
      circuit.depth $\leq$ d + 1 $\rightarrow$
      CircuitComputesParity n circuit $\rightarrow$
      2 ^ (Nat.nthRoot (d - 1) (n / (360 * 40 ^ (d - 2))) /
          (8 * (d + 3))) $\leq$ circuit.circuitSize
\end{lstlisting}
\end{minipage}

Since this bound is superpolynomial, it yields the following corollary: for all
$d$ and every polynomial size bound $p$, there is an $N$ such that, for all $n >
N$, no well-formed $n$-input circuit in the formalized model with depth at most
$d$ and size at most $p(n)$ computes $\operatorname{PARITY}_n$.
Listing~\ref{lst:MainParityCircuitLowerBound} gives the Lean statement, where
$c$ and $k$ specify the polynomial size bound $c n^k$.

\noindent\begin{minipage}{\linewidth}
\begin{lstlisting}[caption={Corollary: No Polynomial-Size Constant-Depth PARITY Circuits},
                   mathescape=true,
                   label={lst:MainParityCircuitLowerBound}]{}
theorem hastad_parity_lower_bound_general (c k d : Nat) :
  $\exists$ N, $\forall$ n, N < n $\rightarrow$
    $\forall$ (circuit : UFICircuitOfSizeAtMostPolyNAndDepthAtMostD n c k d),
      $\lnot$ CircuitComputesParity n circuit.val
\end{lstlisting}
\end{minipage}

This work makes the following contributions:

\begin{enumerate}
  \item It formalizes the switching lemma for unbounded-fan-in DNFs, including
    the canonical decision tree and the injective encoding and decoding of bad
    restrictions. The development also covers $\AC^0$ formula normalization,
    initial (round-zero) fan-in reduction, parity-independent iterated depth
    collapse, properties of the PARITY language, and a quantitative size lower
    bound for formulas computing PARITY. For every fixed $d \geq 2$,
    Listing~\ref{lst:FormulaParitySizeLowerBoundRoot} gives size at least
    $\exp(\Omega_d(n^{1/(d-1)}))$ for formulas of depth at most $d$.  Literals
    and constants have depth zero, and each AND, OR, or explicit NOT gate adds
    one. The exponent $1/(d-1)$ follows by performing exactly $d-2$ switching
    rounds and using a terminal live-variable reserve linear in the bottom-width
    cutoff.
  \item It defines well-formed directed-acyclic-graph circuits and transfers the
    formula lower bound to circuits with shared gates.  Unfolding preserves
    evaluation, and representing nullary AND/OR gates as depth-zero constants
    removes the artificial depth increase they would otherwise
    introduce. Removing the output wire then preserves computation
    depth. Controlling the size increase from unfolding yields the circuit bound
    $\exp(\Omega_d(n^{1/(d-1)}))$ in
    Listing~\ref{lst:DAGParityCircuitLowerBound}.
  \item It constructs a polynomial-size, logarithmic-depth bounded-fan-in
    formula family for PARITY, which, together with the lower bound, witnesses
    $\NC^1 \not\subseteq \AC^0$ for the formalized models.
\end{enumerate}

\paragraph{Development workflow.}
The author chose the proof strategy, representations, theorem interfaces, and
module structure, and directed AI coding agents in implementing and revising the
Lean proofs.  The models used include Claude Sonnet 4.5, Opus 4.6, Opus 4.7,
Opus 4.8, GPT-5.5, and GPT-5.6 Sol.

\paragraph{Related formalizations.}
Dong~\cite{Dong2026Switching} formalizes a fixed-cardinality switching lemma for
DNFs in Lean and derives the corresponding CNF bound by duality.
Fredriksen~\cite{Fredriksen2026Switching} formalizes a counting switching lemma
for simple DNFs, its CNF dual, and multistage collapse results under explicit
structural and scheduling hypotheses. The cited release does not claim a lower
bound for arbitrary $\AC^0$ circuits. These developments already formalize
versions of the switching lemma in Lean.

Complexitylib~\cite{Schlesinger2026Complexitylib} includes circuit-to-formula
normalization with explicit size and depth bounds, width-sensitive DNF/CNF
switching, iterated restrictions through arbitrary finite formula depth, and a
finite PARITY counting theorem. The normalization preserves the selected circuit
output's semantics without increasing its depth. Complexitylib's circuit depth
counts AND/OR gates, including output gates, with free negation on gate inputs.
Our DAG model has explicit NOT gates and separate output vertices, so the
numerical size and depth conventions differ. At the cited revision,
Complexitylib's roadmap identifies the remaining step as the arithmetic and
family-level specialization needed to conclude that PARITY is not in nonuniform
$\AC^0$.

Thiemann~\cite{CliqueAndMonotoneCircuitsAFP} gives an Isabelle/HOL formalization
of a superpolynomial monotone-circuit lower bound for CLIQUE.  Other Lean
formalizations include certificates for arithmetic-circuit and
arithmetic-formula lower bounds for the permanent~\cite{OpenAI2026Permanent}.

The focus of this paper is completing the restriction argument to obtain the
optimal exponent $1/(d-1)$ in the PARITY lower bounds for both formulas and DAG
circuits, together with the resulting separation from nonuniform $\AC^0$.  The
comparisons above are based on documentation and selected theorem statements at
the cited revisions.

\paragraph{Organization.}
Section~2 introduces formulas, restrictions, decision trees, the finite space of
restrictions with a fixed number of live variables, and the formal statement of
the switching lemma. Section~3 proves properties of PARITY, Section~4 defines
formula families and establishes the $\NC^1$ upper bound, and Section~5 presents
normalization, depth reduction, and the formula lower bound. Section~6 unfolds
DAG circuits into formulas and transfers the lower bound to circuits. The
appendix gives supporting definitions and the detailed canonical-tree and
encoder--decoder proof of the switching lemma.

The repository for this paper is available at
\url{https://github.com/formalcs/circuit-complexity/}. The implementation is in
the \texttt{release/0.1.0} tag and uses Lean~4.33.1 and mathlib~4.33.1.
Listings retain theorem statements and representative definitions while omitting
proof bodies.

\paragraph{Acknowledgements.}
I would like to thank Setareh Harfi, Hadi Shafei, and Morgan Sinclaire for
helpful discussions on formalization. I am also grateful for many helpful
discussions, research guidance, and feedback on the manuscript from my advisor,
John Hitchcock.

\section{Preliminaries}
Lean is an open-source programming language and proof assistant that enables
correct, maintainable, and formally verified code \cite{lean}. We selected Lean
because it is being widely adopted for the formalization of a broad swath of
mathematics (in mathlib \cite{mathlib2020}, for example), thereby ensuring that
a broad set of libraries will be available for further formalization efforts.

\subsection{DNFs and CNFs}

As mentioned earlier, the lower bound for PARITY is shown using the switching
lemma, which is a theorem about DNFs and CNFs.  The key concepts and definitions
of \cite{online:SwitchingLemma} required for the formalization are included
below.

\begin{definition}[DNF]
  A DNF is an OR (disjunction) of terms, where each term is an AND (conjunction)
  of literals.  An example is $f = x_0 x_1 x_2 \lor \overline{x_1} x_3 \lor
  \overline{x_2} x_4$.
\end{definition}

\begin{definition}[CNF]
  A CNF is an AND (conjunction) of clauses, where each clause is an OR
  (disjunction) of literals.  An example is $g = (x_0 \lor x_2) \land
  (\overline{x_1} \lor x_4 \lor x_5 \lor x_9) \land (\overline{x_3} \lor x_7)$.
\end{definition}

\begin{definition}[Size]
  The size of a DNF is its number of terms.  The size of a CNF is its number of
  clauses.
\end{definition}

\begin{definition}[Width]
  The width of a DNF is the maximum number of literals in any term of the DNF.
  Similarly, the width of a CNF is the maximum number of literals in any clause.
\end{definition}

To formalize these concepts in Lean, we first define the basic structure of a
propositional formula in listing \ref{lst:ufiFormula}.  The inputs to the
formulas and circuits will be represented using the Lean \lstinline|Bool| type.
The inductive definition of \lstinline|UnboundedFanInFormula| uses the
\lstinline|deriving| keyword to construct an instance of Lean's \lstinline|Repr|
class automatically \cite{online:FuncPInLeanDerivingStdClasses}.  This instance
provides a printable representation of formula values.

\begin{lstlisting}[caption={Defining an Unbounded Fan-in Formula.}, label={lst:ufiFormula}]{}
inductive UnboundedFanInFormula where
  | inputGate :                Nat -> Bool -> UnboundedFanInFormula
  | constant  :               Bool -> Nat  -> UnboundedFanInFormula
  | notGate   :      UnboundedFanInFormula -> UnboundedFanInFormula
  | andGate   : List UnboundedFanInFormula -> UnboundedFanInFormula
  | orGate    : List UnboundedFanInFormula -> UnboundedFanInFormula
  deriving Repr
\end{lstlisting}

The first argument to the \lstinline|inputGate| constructor of an
\lstinline|UnboundedFanInFormula| is a natural number identifying which variable
this input refers to. The second argument is a \lstinline|Bool| indicating
whether or not the reference is to the negated variable. For example, $x_2$
would be represented as \lstinline|inputGate 2 false| and $\overline{x_2}$ as
\lstinline|inputGate 2 true|.  The arguments to the \lstinline|andGate| and
\lstinline|orGate| constructors are lists of circuits. The absence of length
bounds on these lists makes the inductive definition suitable for representing
unbounded fan-in circuits.

CNFs and DNFs are special cases of propositional formulas, i.e. Boolean circuits
with a single output node in which every internal node has a fan-out of 1
\cite{online:Formula}.  Therefore, CNFs and DNFs are formalized as
\lstinline|UnboundedFanInFormula|s. To differentiate them from other circuits,
we define several helper functions (Listing \ref{lst:isDNF}). The first,
\lstinline|isInput|, is a predicate (defined as a Lean match expression) that
only holds for input gates. Underscores in the match expressions are
placeholders that prevent unused arguments and constructors from having to be
named. The \lstinline|isAndOfInputsOnly| predicate uses Lean's
\href{https://lean-lang.org/doc/api/Init/Data/List/Basic.html#List.all}{List.all}
function to check that every input to an AND gate is a literal. Finally, the
\lstinline|isDNF| function differentiates DNFs from other
\lstinline|UnboundedFanInFormula|s by matching the formula with an OR gate whose
inputs all satisfy the \lstinline|isAndOfInputsOnly| predicate.  CNFs are
identified by matching the \lstinline|UnboundedFanInFormula| with an AND gate
whose inputs all satisfy \lstinline|isOrOfInputsOnly|. The definitions of
\lstinline|isOrOfInputsOnly| and \lstinline|isCNF| are similar and therefore
omitted.

\noindent
\begin{minipage}{\linewidth}
\begin{lstlisting}[caption={Detecting DNFs}, label={lst:isDNF}]{}
def isInput (gate : UnboundedFanInFormula) : Bool :=
  match gate with
  | inputGate _ _ => true
  | _             => false

def isAndOfInputsOnly (gate : UnboundedFanInFormula) : Bool :=
  match gate with
  | andGate gates => (gates.all isInput)
  | _             => false

def isDNF (gate : UnboundedFanInFormula) : Bool :=
  match gate with
  | orGate gates => (gates.all isAndOfInputsOnly)
  | _            => false
\end{lstlisting}
\end{minipage}

While the \lstinline|isDNF| function indicates whether an
\lstinline|UnboundedFanInFormula| has the shape of a DNF, two additional
properties are required to make proofs about DNFs straightforward. Every clause
(term) of the DNF should be non-empty and no term should have any variable
repeated. Requiring these properties of DNFs does not result in any loss of
generality.  We define the \lstinline|IsProperDNF| proposition to indicate when
a formula has the shape required of DNFs and satisfies the no empty clauses and
no duplicate variables in any clause requirements. \lstinline|IsProperDNF|
depends on the \lstinline|dnfClauses| function, which extracts the list of
clauses from a DNF, each returned as a lists of (\textit{literal, negation
  flag}) pairs.  See listing \ref{lst:dnfClauses} in the appendix for the
implementation of the \lstinline|dnfClauses| function.

\noindent
\begin{minipage}{\linewidth}
\begin{lstlisting}[caption={Defining a Proper DNF},
                   mathescape=true,
                   label={lst:IsProperDNF}]{}
def IsProperDNF (f : UnboundedFanInFormula) : Prop :=
  isDNF f = true
    $\land$ ($\forall ~ c ~ \in$ dnfClauses f, c $\neq$ [])
    $\land$ ($\forall ~ c ~ \in$ dnfClauses f, (c.map Prod.fst).Nodup)
\end{lstlisting}
\end{minipage}

Every \lstinline|UnboundedFanInFormula| depends on finitely many inputs, so we
track an upper bound on its input indices. Since the switching lemma concerns
DNFs, stating it directly for unrestricted formulas would repeatedly expose
shape hypotheses. Lean subtypes bundle a value with a proof of a predicate
\cite{lean-lang-ref}. We therefore use \lstinline|UnboundedFanInProperDNF| to
bundle a formula with proofs that its input indices lie below $n$ and that it is
a proper DNF (Listing~\ref{lst:UnboundedFanInProperDNF}).

\noindent
\begin{minipage}{\linewidth}
\begin{lstlisting}[caption={Defining a Proper DNF},
                   mathescape=true,
                   label={lst:UnboundedFanInProperDNF}]{}

def UnboundedFanInProperDNF (n : Nat) :=
  {
    circuit : UnboundedFanInFormula //
    ufiLargestInput circuit < n
      $\land$ IsProperDNF circuit
  }
\end{lstlisting}
\end{minipage}

\subsection{Restrictions}

The switching lemma is a theorem about what happens when some variables of a DNF
are assigned (i.e. restricted), while others are left free. Following
\cite{online:SwitchingLemma}, we present the concepts related to restricting
variables of a function $f$.

\begin{definition}[Restriction]
A restriction of variables is a partial assignment of bits to any number of the
inputs. Any input which is not assigned a bit is said to remain free or to be
set to star (*).
\end{definition}

\begin{definition}[Stars]
Let $\alpha$ be a restriction of some inputs to a circuit.  $stars(\alpha)$ is
the set of inputs/coordinates which $\alpha$ leaves free.  $f|_\alpha$ is the
restricted function $\{0,1\}^{stars(\alpha)} \rightarrow \{0,1\}$.
\end{definition}

\begin{definition}[Restriction Set]
$\mathcal{R}_s$ is the set of all restrictions with exactly $s$ stars.
\end{definition}

\begin{definition}[Random Restriction]
A random restriction with $s$ stars is a uniformly random restriction $\alpha$
from $R_s$.
\end{definition}

A precise formal definition of a restriction needs a finite input width, a
finite set of variables that remain free, and an assignment of \lstinline|Bool|
values to the variables that the restriction fixes.  We use the uniform
distribution on restrictions with exactly $s=\lceil\sigma n\rceil$ free
variables, where $\sigma\in(0,1)$ and $n$ is the input width.  We define a
bounded finite set (\lstinline|BoundedFinset| in Lean) to represent the set of
free/live variables (those set to $*$) in a restriction. The value in this
subtype is a finite set $S$ of natural numbers (each identifying a variable) and
the property of the subtype is a proof that $S$ is a subset of the first $n$
natural numbers.  The subtype (\lstinline|OpenUnitIntervalQ|) of the rationals
in the open unit interval represents the fractional upper bound on the number of
stars the random restrictions of interest should have. Next, the
\lstinline|RandomRestrictionVars| subtype pairs the bounded set with the
requirement that its cardinality must be $\lceil \sigma n \rceil$.  Displayed
theorem and lemma excerpts omit their proof bodies.  Where \lstinline|sorry|
appears in a displayed function definition, it abbreviates an implementation
rather than a proof statement.  The corresponding declarations in the repository
contain checked implementations and proofs.  We also define the Lean function
\lstinline|totalRestrictionCount| (listing \ref{lst:totalRestrictionCount}) to
compute $|\mathcal{R}_s| = \binom{n}{s} 2^{n - s}$.

\begin{lstlisting}[caption={Restrictions in Lean},mathescape=true,
                   label={lst:RandomRestrictionVars}]{}
def BoundedFinset (n : Nat) :=
  {
    s : Finset Nat //
    s $\subseteq$ Finset.range n
  }

def OpenUnitIntervalQ :=
  {
    x : $\mathbb{Q}$ //
    0 < x $\land$ x < 1
  }

def RandomRestrictionVars ($\sigma$ : OpenUnitIntervalQ) (n : Nat) :=
  {
    liveVars : BoundedFinset n
    liveVars.val.card = Nat.ceil ($\sigma$.val * (n : $\mathbb{Q}$))
  }
\end{lstlisting}

\begin{lstlisting}[caption={Computing the Total Number of Restrictions},
                   mathescape=true,
                   label={lst:totalRestrictionCount}]{}
def totalRestrictionCount (n : Nat)
                          ($\sigma$ : OpenUnitIntervalQ) : $\mathbb{Q}$ :=
  let s := Nat.ceil ($\sigma$.val * (n : $\mathbb{Q}$))
  (Nat.choose n s * 2 ^ (n - s))
\end{lstlisting}

To represent a random restriction, we create a structure called
\lstinline|AssignedRandomRestriction|, which specifies the set of live
variables, a list of values of the dead variables, and the requirement that the
cardinality of the set of live variables and the length of values of the dead
variables must equal $n$.

\noindent\begin{minipage}{\linewidth}
\begin{lstlisting}[caption={Restrictions in Lean},
                   mathescape=true,
                   label={lst:AssignedRandomRestriction}]{}
structure AssignedRandomRestriction ($\sigma$ : OpenUnitIntervalQ) (n : Nat) where
  starAssignment : RandomRestrictionVars $\sigma$ n
  varAssignments : List Bool
  non_starred_vars_fully_assigned : starAssignment.val.val.card + varAssignments.length = n
\end{lstlisting}
\end{minipage}

\subsection{Decision Trees}\label{sec:DecisionTrees}

\begin{definition}[Decision Tree]
  A decision tree is a binary tree in which each internal node represents a test
  on an input variable, each branch represents the outcome of the test, and each
  leaf node is labeled with a 0 or a 1.
\end{definition}

We define a decision tree inductively as shown in listing
\ref{lst:DecisionTree}.  The first argument to the \lstinline|dtNode|
constructor is a natural number that identifies the variable that determines
which branch of the decision tree should be taken when evaluating an input.

\begin{lstlisting}[caption={Defining Decision Trees},
                   label={lst:DecisionTree}]{}
inductive DecisionTree where
  | dtLeaf : Bool -> DecisionTree
  | dtNode : Nat  -> DecisionTree -> DecisionTree -> DecisionTree
  deriving Repr
\end{lstlisting}

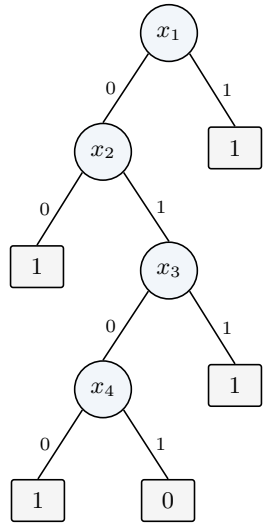
\begin{figure}[H]
  \centering
  \begin{adjustbox}{max width=0.96\textwidth,max totalheight=0.65\textheight,center}
  \begin{forest}
    decision tree,
      for tree={s sep=10mm}
    [$x_1$, variable
      [$x_2$, variable, zero edge
        [$1$, leaf, zero edge]
        [$x_3$, variable, one edge
          [$x_4$, variable, zero edge
            [$1$, leaf, zero edge]
            [$0$, leaf, one edge]
          ]
          [$1$, leaf, one edge]
        ]
      ]
      [$1$, leaf, one edge]
    ]
  \end{forest}
  \end{adjustbox}
  \caption{Example of a Decision Tree.}
  \label{fig:DecisionTreeExample}
\end{figure}

Any function from $\{0,1\}^* \rightarrow \{0,1\}$ can be represented using a
decision tree.  Consider the function $f : \{0, 1\}^6 \rightarrow \{0, 1\}$
given by the DNF $f = (x_1 \land x_2 \land \neg x_3) \lor (x_4 \land \neg x_2)
\lor x_3 \lor \neg x_4$.  $f$ is computed by the decision tree in
figure~\ref{fig:DecisionTreeExample}.

Evaluation follows a path to a leaf, whose label is the value of the function on
the given input. See Listing~\ref{lst:evalDecisionTree}.

\noindent\begin{minipage}{\linewidth}
\begin{lstlisting}[caption={Evaluating a Decision Tree},
                   mathescape=true,
                   label={lst:evalDecisionTree}]{}
def evalDecisionTree (tree : DecisionTree) (inputs : List Bool) : Bool :=
  match tree with
  | dtLeaf bit          => bit
  | dtNode i left right => match inputs[i]? with
                           | none       => false
                           | some input =>
                               match input with
                               | false => evalDecisionTree left inputs
                               | true  => evalDecisionTree right inputs
\end{lstlisting}
\end{minipage}

One measure of the complexity of a function is its decision tree depth, i.e. the
length of the longest path from the root of the tree to a leaf. This is computed
by the \lstinline|decisionTreeDepth| recursive function in listing
\ref{lst:decisionTreeDepth} (this is the function $\text{DT}_{\text{depth}}$ in
\cite{online:SwitchingLemma}).

\begin{lstlisting}[caption={Computing Decision Tree Depth},
                   mathescape=true,
                   label={lst:decisionTreeDepth}]{}
def decisionTreeDepth (tree : DecisionTree) : Nat :=
  match tree with
  | dtLeaf _ => 0
  | dtNode _ left right =>
      1 + (max (decisionTreeDepth left) (decisionTreeDepth right))
\end{lstlisting}

\subsection{The Switching Lemma}\label{subsec:SwitchingLemma}
As mentioned earlier, applying a random restriction $\alpha$ to a function $f$
yields the function $f|_\alpha : \{0,1\}^{stars(\alpha)} \rightarrow
\{0,1\}$. $f|_\alpha$ can be represented as a decision tree. Theorem
\ref{SwitchingLemmaThm} states the switching lemma from
\cite{online:SwitchingLemma}.

\begin{theorem}
\label{SwitchingLemmaThm}
Let $f$ be a DNF of width at most $w$ over $n$ variables. Let $\alpha$ be a
random restriction with $s = \sigma n$ stars, where $\sigma \leq 1/5$. Then for
each $d \geq 0$ (and $\leq s$),

$$ Pr[DT_{depth}(f|_\alpha) > d] \leq (10 \sigma w)^d $$
\end{theorem}

$n$, $d$, $f$, and $w$ are fixed in the statement of the switching lemma. To
state it in Lean, we will use the approach presented by
\cite{online:SwitchingLemma} and compute the probability as a rational fraction
using a combinatorial proof from Razborov. A restriction $\beta$ is described as
\textit{bad} if it makes $\text{DT}_{\text{depth}}(f|_\beta) > d$.  Let
$\mathcal{B}$ be the set of all bad restrictions. Recall that $\mathcal{R}_s$ is
the set of all restrictions with $s$ stars. Then the switching lemma can be
stated as the following inequality:

$$ \displaystyle \frac{|\mathcal{B}|}{|\mathcal{R}_s|} \leq (10 \sigma w)^d $$

Recall that $|\mathcal{R}_s|$ is computed by the Lean function
\lstinline|totalRestrictionCount|.  To compute $|\mathcal{B}|$, we define the
\lstinline|badRestrictionCount| function in listing
\ref{lst:badRestrictionCount} .  It operates in two stages: it first generates
the set of all possible restrictions for any given $\sigma$ and $n$ then uses
the Mathlib \lstinline|countP| function to count the number of restrictions that
are bad.  See Section~\ref{sec:GeneratingAllRestrictions} for the implementation
of \lstinline|generateAllRestrictions| and its correctness theorems.

\noindent\begin{minipage}{\linewidth}
\begin{lstlisting}[caption={Computing the Number of Bad Restrictions},
                   mathescape=true,
                   label={lst:badRestrictionCount}]{}
def badRestrictionCount (n : Nat)
                        (d : Nat)
                        (f : UnboundedFanInProperDNF n)
                        ($\sigma$ : OpenUnitIntervalQ) : $\mathbb{Q}$ :=
((generateAllRestrictions n $\sigma$).countP fun $\rho$ =>
    isBadRestriction d n $\sigma$ f $\rho$)
\end{lstlisting}
\end{minipage}

With these functions defined, the switching lemma is then stated in Lean as
shown in listing \ref{lst:switching_lemma_exact}.  The \lstinline|hexact|
hypothesis states that the $\lceil \sigma n \rceil$ live-set size has no
rounding loss.  This is implicit in the switching lemma stated in
\cite{online:SwitchingLemma} (without this hypothesis, the constant increases
from 10 to 40).

\noindent\begin{minipage}{\linewidth}
\begin{lstlisting}[caption={The Switching Lemma in Lean},
                   mathescape=true,
                   label={lst:switching_lemma_exact}]{}
theorem switching_lemma_exact (n w d : Nat)
                       (f : UnboundedFanInProperDNF n)
                       (hwidth : dnfWidth f.val $\leq$ w)
                       ($\sigma$  : OpenUnitIntervalQ)
                       (h$\sigma$ : $\sigma$.val $\leq$ 1 / 5)
                       (hs_exact : (Nat.ceil ($\sigma$.val * n) : $\mathbb{Q}$)
                         = $\sigma$.val * n) :
    (badRestrictionCount n d f $\sigma$)
      / (totalRestrictionCount n $\sigma$)
        $\leq$ (10 * $\sigma$.val * w) ^ d
\end{lstlisting}
\end{minipage}

The canonical decision-tree construction, the encoding and decoding of bad
restrictions, and the counting argument supporting this theorem are presented in
Appendix~\ref{sec:CoreSwitchingLemmaIdea}.  The remainder of the main text uses
the switching lemma only through the interface above.

\section{The PARITY Function}

The parity of a list of bits is defined to be 0 if the sum of bits modulo 2 is
even and 1 otherwise.  See the \lstinline|parityBit| function in listing
\ref{lst:FormulaComputesParity} for the Lean implementation of this idea.  A DNF
on $n$ variables computes parity when its evaluation agrees with the
\lstinline|parityBit| function on every length-$n$ input list.  This is stated
in Lean as the \texttt{DNFComputesParity} proposition.

\begin{lstlisting}[caption={What it Means to Compute PARITY},
                   mathescape=true,
                   label={lst:FormulaComputesParity}]{}
def parityBit (bits : List Bool) : Bool :=
  if List.count true bits % 2 = 0 then
    false
  else
    true

def FormulaComputesParity (n : Nat) (f : UnboundedFanInFormula) : Prop :=
  $\forall$ inputs : List Bool, inputs.length = n $\rightarrow$
    ufiFormulaEval f inputs = parityBit inputs

def DNFComputesParity (n : Nat) (f : UnboundedFanInDNF n) : Prop :=
  FormulaComputesParity n f.val
\end{lstlisting}

Trevisan's notes \cite{online:TrevisanParityLowerBounds} present two well-known
facts. First, an unbounded-fan-in DNF on $n\geq1$ variables that computes PARITY
has at least $2^{n-1}$ terms.  Next, if a DNF on $n$ variables computes their
parity, then every clause of that DNF has exactly $n$ literals (one for every
variable).  We formalize these facts in Lean as the
\texttt{proper\_dnf\_computing\_parity\_clause\_count} and
\texttt{proper\_dnf\_computing\_parity\_clause\_length} theorems respectively.

\begin{lstlisting}[caption={Lower Bound on Clause Count for DNFs Computing parityBit},
                   mathescape=true,
                   label={lst:proper_dnf_computing_parity_clause_count}]{}
theorem proper_dnf_computing_parity_clause_count
    (n : Nat)
    (hn : 1 $\leq$ n)
    (f : UnboundedFanInProperDNF n)
    (h_parity : DNFComputesParity n f.val) :
  2 ^ (n - 1) $\leq$ (dnfClauses f.val).length
\end{lstlisting}

\begin{lstlisting}[caption={Lower Bound on Number of Literals in Every Clause of a DNF Computing parityBit},
                   mathescape=true,
                   label={lst:proper_dnf_computing_parity_clause_length}]{}
theorem proper_dnf_computing_parity_clause_length
    (n : Nat)
    (f : UnboundedFanInProperDNF n)
    (h_parity : DNFComputesParity n f.val) :
  $\forall$ c $\in$ dnfClauses f.val, c.length = n
\end{lstlisting}

\subsection{PARITY Under Restrictions}\label{subsec:ParityUnderRestrictions}

An important property of the PARITY function is that any restriction of its
inputs is an instance of the PARITY function.  In other words, partitioning a
set of variables into dead variables and live variables and fixing the values of
the dead variables yields a function of the live variables. This function is the
PARITY function on a smaller input and the exclusive OR of its output with the
parity of the dead bits is equal to the PARITY of the $n$-bit assignment to all
the variables.

To state this in Lean, we define a function (\texttt{assembleInput}) to assemble
a full $n$-bit input from a partial assignment.  Its arguments are the list of
unassigned variable indices in $[0,n)$ (\texttt{live}), the values to place at
  those positions (paired positionally with \texttt{live}), and the values to
  place at the remaining positions in $[0,n)$ in increasing order of position
    (\texttt{deadBits}).

\begin{lstlisting}[caption={Assembling Live and Dead Bits},
                   mathescape=true,
                   label={lst:assembleInput}]{}
def assembleInput (n : Nat)
                  (live : List Nat)
                  (liveBits : List Bool)
                  (deadBits : List Bool) : List Bool :=
  (List.range n).map (fun i =>
    match live.findIdx? ($\cdot$ = i) with
    | some j => liveBits.getD j false
    | none =>
        deadBits.getD
          (((List.range i).filter (fun k => !live.contains k)).length) false)
\end{lstlisting}

Once an $n$-bit input has been assembled, the
\texttt{exists\_offset\_odd\_countP\_assembleInput\_iff} lemma in listing
\ref{lst:ParityRestrictionIsOffsetParity} proves the existence of a Boolean
offset, determined by the dead bits, such that the restricted function is PARITY
or its complement.

\noindent\begin{minipage}{\linewidth}
\begin{lstlisting}[caption={parityBit Restriction is Offset parityBit},
                   mathescape=true,
                   label={lst:ParityRestrictionIsOffsetParity}]{}
lemma exists_offset_odd_countP_assembleInput_iff
      (n : Nat)
      (live : List Nat)
      (deadBits : List Bool)
      (h_live_lt : $\forall$ v $\in$ live, v < n)
      (h_live_nodup : live.Nodup)
      (h_card : deadBits.length + live.length = n) :
$\exists$ (offset : Bool),
  $\forall$ (liveBits : List Bool),
    liveBits.length = live.length
    $\rightarrow$ (Odd ((assembleInput n live liveBits deadBits).countP ($\cdot$ == true))
       $\leftrightarrow$
       (Odd (liveBits.countP ($\cdot$ == true)) $\leftrightarrow$ offset = false))
\end{lstlisting}
\end{minipage}

Invoking the \texttt{exists\_offset\_odd\_countP\_assembleInput\_iff} lemma with
exactly $n$ dead bits yields an offset that makes the ``computes parity''
proposition hold.  With this Boolean offset known, we can prove a stronger
version of \texttt{proper\_dnf\_computing\_offset\_parity\_clause\_length} by
showing that the clause length must be $n$ for the complement of PARITY as well:

\begin{lstlisting}[caption={Per-Clause Literals Lower Bound in a DNF Computing PARITY or its Complement},
                   mathescape=true,
                   label={lst:proper_dnf_computing_offset_parity_clause_length}]{}
theorem proper_dnf_computing_offset_parity_clause_length
    (n : Nat)
    (f : UnboundedFanInProperDNF n)
    (offset : Bool)
    (h_parity : $\forall$ inputs, inputs.length = n $\rightarrow$
      ((ufiFormulaEval f.val inputs = true) $\leftrightarrow$
        (Odd (inputs.countP ($\cdot$ == true)) $\leftrightarrow$ offset = false))) :
    $\forall$ c $\in$ dnfClauses f.val, c.length = n
\end{lstlisting}

\texttt{proper\_dnf\_computing\_offset\_parity\_clause\_length} is a key
component of the PARITY $\AC^0$ formula lower bounds proof.

\subsection{Minimum Clause Length for DNFs Computing Parity}

The offset that makes the ``computes parity'' proposition hold is also useful
for stating an important corollary of the fact that a DNF computing PARITY of
$n$ variables must have width $n$ (listing
\ref{lst:proper_dnf_computing_parity_clause_length}): DNFs whose width is less
than the number of variables must misclassify PARITY. This lemma is also a key
component in the PARITY $\AC^0$-formula size lower bounds proof.

\begin{lstlisting}[caption={Narrow DNFs Misclassify PARITY},
                   mathescape=true,
                   label={lst:narrow_dnf_misclassifies_parity}]{}
theorem narrow_dnf_misclassifies_parity (m : Nat)
                                        (g : UnboundedFanInProperDNF n)
                                        (hgw : dnfWidth g.val < n)
                                        (offset : Bool) :
  $\exists$ (inputs : List Bool),
    inputs.length = n
      $\land$ ((ufiFormulaEval g.val inputs = true) $\neq$
              (Odd (inputs.countP ($\cdot$ == true)) $\leftrightarrow$ offset = false))
\end{lstlisting}

\section{Families of Formulas}
DNFs are a formulas with a specific restricted shape.  We are interested in the
computation capabilities of more general formulas but a formula computes a
function at exactly 1 input length. Therefore, we define the concept of a family
of formulas such that each family contains a formula that computes the function
of interest at that input length.

\subsection{\texorpdfstring{$\AC^0$}{AC0} Formulas}
An $\AC^0$ formula family is a mapping of each input length to a constant-depth,
polynomial-size formula.  It is defined in lean as the
\texttt{UFIFormulaOfSizeAtMostPolyNAndDepthAtMostD} subtype. It bundles an
unbounded fan-in formula with bounds on its input indices, size, and depth.  The
\texttt{AC0FormulaFamily} definition maps every positive natural number to such
a formula and its fixed parameters are the coefficient $c$ and degree $k$ of the
polynomial size bound and the depth bound $d$.

\noindent\begin{minipage}{\linewidth}
\begin{lstlisting}[caption={AC$^0$ Formula Family},
                   mathescape=true,
                   label={lst:AC0Formula}]{}
def UFIFormulaOfSizeAtMostPolyNAndDepthAtMostD
  (n c k d : Nat) :=
  {
    circuit : UnboundedFanInFormula //
    d > 0
    $\land$ ufiLargestInput circuit < n
    $\land$ ufiFormulaCircuitSize circuit $\leq$ c * n ^ k
    $\land$ ufiFormulaDepth circuit $\leq$ d
  }

def AC0FormulaFamily (c k d : Nat) :=
  (n : PNat) ->
    UFIFormulaOfSizeAtMostPolyNAndDepthAtMostD n.val c k d
\end{lstlisting}
\end{minipage}

\subsection{\texorpdfstring{$\NC^1$}{NC1} Formulas}
For $\NC^1$ Formulas, the
\texttt{BFIFormulaOfSizeAtMostPolyNAndDepthAtMostLogCircuitSize} subtype has a
value of type \texttt{BoundedFanInFormula} while the same circuit size and input
index upper bounds hold. However, its depth is bounded by the logarithm of its
circuit size. Like the \texttt{AC0FormulaFamily}, the \texttt{NC1FormulaFamily}
is also a function of 3 constants. $c$ and $k$ have identical meaning across
both and $c2$ is a coefficient on the circuit depth (which is logarithm of the
circuit size).

\begin{lstlisting}[caption={NC$^1$ Formula Family},
                   mathescape=true,
                   label={lst:NC1Formula}]{}
def BFIFormulaOfSizeAtMostPolyNAndDepthAtMostLogCircuitSize
  (n c k c2 : Nat) :=
  {
    circuit : BoundedFanInFormula //
    $\land$ bfiLargestInput circuit < n
    $\land$ bfiFormulaDepth circuit $\leq$ ((Nat.clog 2 (c * n ^ k)) * c2)
    $\land$ bfiFormulaCircuitSize circuit $\leq$ c * n ^ k
  }

def NC1FormulaFamily (c k c2 : Nat) :=
  (n : PNat) ->
    BFIFormulaOfSizeAtMostPolyNAndDepthAtMostLogCircuitSize
      n.val c k c2
\end{lstlisting}

The PARITY function can be computed by an $\NC^1$ formula family.  Consider the
\lstinline|parityCircuit| function in listing \ref{lst:parityCircuit}. It builds
a balanced binary tree computing PARITY on $n$ input bits (indexed from zero)
using bounded fan-in formulas.

\noindent\begin{minipage}{\linewidth}
\begin{lstlisting}[caption={NC$^1$ Formula for PARITY of $n$ Bits},
                   mathescape=true,
                   label={lst:parityCircuit}]{}
def xorBFI (left right : BoundedFanInFormula) : BoundedFanInFormula :=
  .orGate (.andGate left (.notGate right))
    (.andGate right (.notGate left))

def parityCircuitRange : Nat $\rightarrow$ Nat $\rightarrow$ BoundedFanInFormula
  | _,     0       => .constant false 0
  | start, 1       => .inputGate start false
  | start, len + 2 => let half := (len + 2) / 2
                      xorBFI
                        (parityCircuitRange start half)
                        (parityCircuitRange (start + half)
                                              (len + 2 - half))

def parityCircuit (n : Nat) : BoundedFanInFormula :=
  parityCircuitRange 0 n
\end{lstlisting}
\end{minipage}

The Lean theorem \lstinline|parityCircuit_is_correct| shows that the
construction correctly computes the $n$-bit parity.  The interval
characterization of the collected input indices shows that every input index is
less than the input size $n$.

\begin{lstlisting}[caption={Correctness of the NC$^1$ Formula for PARITY},
                   mathescape=true,
                   label={lst:parityCircuit_uses_all_inputs}]{}
theorem parityCircuit_input_indices_upper_bound (n : PNat) :
    bfiLargestInput (parityCircuit n) < n

def BFIFormulaComputesParity (n : Nat) (f : BoundedFanInFormula) : Prop :=
  $\forall$ inputs : List Bool, inputs.length = n $\rightarrow$
    bfiFormulaEval f inputs = parityBit inputs

theorem parityCircuit_is_correct (n : Nat) :
    BFIFormulaComputesParity n (parityCircuit n)
\end{lstlisting}

An $\NC^1$ formula family computing PARITY is shown in listing
\ref{lst:parityNC1FormulaFamily}. It has a circuit size upper bound of $5n^2$ at
every input length $n$.

\begin{lstlisting}[caption={NC$^1$ Formula Family for PARITY},
                   mathescape=true,
                   label={lst:parityNC1FormulaFamily}]{}
theorem parityCircuit_output_has_logarithmic_depth_upper_bound (n : Nat) :
    bfiFormulaDepth (parityCircuit n) $\leq$
      Nat.clog 2 (5 * n ^ 2) * 3

theorem parityCircuit_output_has_quadratic_upper_bound (n : Nat)
    (h_n : 1 $\leq$ n) :
    bfiFormulaCircuitSize (parityCircuit n) $\leq$ 5 * n ^ 2

def parityNC1FormulaFamily : NC1FormulaFamily 5 2 3 :=
  fun n =>
    $\langle$parityCircuit n.val, by
      constructor
      $\cdot$ exact parityCircuit_input_indices_upper_bound n
      $\cdot$ constructor
        $\cdot$ exact parityCircuit_output_has_logarithmic_depth_upper_bound n
        $\cdot$ exact parityCircuit_output_has_quadratic_upper_bound n n.pos$\rangle$
\end{lstlisting}

\section{Lower Bounds for Constant-Depth Formulas Computing PARITY}

Now that several properties of the PARITY function have been established, the
switching lemma can be used to prove circuit size lower bounds for
constant-depth formulas computing PARITY. Several transformations are required
to enable this.

\subsection{NOT-Gate Elimination}

Applying the switching lemma requires that the circuit must not contain any NOT
gates.  There is a standard transformation that pushes NOT gates toward the
inputs using De Morgan's laws, replacing a negated AND or OR by the dual gate
with negated children. Repetition leaves negation only at input literals, where
it is represented by a Boolean flag.  This algorithm is implemented in Lean as
the \texttt{elimNotGates} function in listing \ref{lst:pushNeg}.

\begin{lstlisting}[caption={NOT-Gate Elimination Algorithm for Formulas},
                   mathescape=true,
                   label={lst:pushNeg}]{}
mutual
def pushNeg (neg : Bool) (f : UnboundedFanInFormula) : UnboundedFanInFormula :=
  match f with
  | .inputGate i b  => .inputGate i (Bool.xor neg b)
  | .constant c lbl => .constant (if neg then not c else c) lbl
  | .notGate g      => pushNeg (!neg) g
  | .andGate gs     => if neg then
                         .orGate (pushNegList neg gs)
                       else
                         .andGate (pushNegList neg gs)
  | .orGate gs      => if neg then
                         .andGate (pushNegList neg gs)
                       else
                         .orGate (pushNegList neg gs)

def pushNegList (neg : Bool) (l : List UnboundedFanInFormula) :
    List UnboundedFanInFormula :=
  match l with
  | []      => []
  | g :: gs => pushNeg neg g :: pushNegList neg gs
end

def elimNotGates (f : UnboundedFanInFormula) : UnboundedFanInFormula :=
  pushNeg false f
\end{lstlisting}

The resulting formula has no NOT gates, i.e. it contains no \texttt{notGate}
constructor anywhere in its syntax tree. Negation is carried only on
\texttt{inputGate} literals. We define the \texttt{HasNoNotGates} proposition so
that we can prove this property of the transformation. We then prove that
\texttt{elimNotGates} outputs a formula for which the \texttt{HasNoNotGates}
proposition holds.

\noindent\begin{minipage}{\linewidth}
\begin{lstlisting}[caption={Absence of NOT Gates as a Proposition},
                   mathescape=true,
                   label={lst:HasNoNotGates}]{}
def HasNoNotGates : UnboundedFanInFormula $\rightarrow$ Prop
  | .inputGate _ _ => True
  | .constant _ _  => True
  | .notGate _     => False
  | .andGate gates => $\forall$ g $\in$ gates, HasNoNotGates g
  | .orGate  gates => $\forall$ g $\in$ gates, HasNoNotGates g

theorem elimNotGates_hasNoNotGates (f : UnboundedFanInFormula) :
  HasNoNotGates (elimNotGates f)
\end{lstlisting}
\end{minipage}

This transformation does not change the function computed by the formula.  We
define a proposition, \texttt{AgreesOn}, for the equivalence of two formulas at
a given input length: two formulas are equivalent on $n$-bit inputs when they
agree on every length-$n$ input list.

\noindent\begin{minipage}{\linewidth}
\begin{lstlisting}[caption={Formula Agreement Proposition},
                   mathescape=true,
                   label={lst:AgreesOn}]{}
def AgreesOn (f g : UnboundedFanInFormula) (n : Nat) : Prop :=
  $\forall$ inputs : List Bool,
    inputs.length = n
      $\rightarrow$ ufiFormulaEval f inputs = ufiFormulaEval g inputs
\end{lstlisting}
\end{minipage}

The NOT-gate elimination contract then states that the output formula has at
most the same circuit size and depth as the original formula, it has no NOT
gates, and it agrees with the original formula. This is the
\texttt{elimNotGates\_contract} theorem in Lean.

\begin{lstlisting}[caption={NOT-Gate Elimination Contract},
                   mathescape=true,
                   label={lst:elimNotGates_contract}]{}
theorem elimNotGates_contract (d : Nat) (_hd : 1 $\leq$ d) :
  $\forall$ a b : Nat, $\forall$ (n : Nat) (f : UnboundedFanInFormula),
    ufiLargestInput f $<$ n $\rightarrow$
    ufiFormulaDepth f $\leq$ d $\rightarrow$
    ufiFormulaCircuitSize f $\leq$ a * n ^ b $\rightarrow$
    $\exists$ g : UnboundedFanInFormula,
      HasNoNotGates g
      $\land$ ufiLargestInput g $<$ n
      $\land$ ufiFormulaDepth g $\leq$ d
      $\land$ ufiFormulaCircuitSize g $\leq$ a * n ^ b
      $\land$ AgreesOn f g n
\end{lstlisting}

\subsection{Formula Leveling}

The second main transformation is leveling. Each AND or OR gate is assigned a
natural-number level, and each level contains only one gate type, with AND and
OR layers alternating. Associativity permits a flattening operation to combine
an AND of AND gates into a single AND gate, and similarly for OR gates.  We
define the alternation proposition in listing
\ref{lst:IsAlternatingAndLeveledAt} for these requirements (we assume that NOT
gates have all been eliminated from the formula).

\noindent\begin{minipage}{\linewidth}
\begin{lstlisting}[caption={Proposition for Alternating AND/OR Gates},
                   mathescape=true,
                   label={lst:HasAlternatingAndOrGates}]{}
def HasAlternatingAndOrGates : UnboundedFanInFormula $\rightarrow$ Prop
  | UnboundedFanInFormula.inputGate _ _ => True
  | UnboundedFanInFormula.constant  _ _ => True
  | UnboundedFanInFormula.notGate   _   => False
  | UnboundedFanInFormula.andGate   gates =>
      ($\forall$ g $\in$ gates, HasAlternatingAndOrGates g) $\land$
      ($\forall$ g $\in$ gates, $\forall$ inner, g $\neq$ UnboundedFanInFormula.andGate inner)
  | UnboundedFanInFormula.orGate    gates =>
      ($\forall$ g $\in$ gates, HasAlternatingAndOrGates g) $\land$
      ($\forall$ g $\in$ gates, $\forall$ inner, g $\neq$ UnboundedFanInFormula.orGate inner)
\end{lstlisting}
\end{minipage}

Next, we define the \texttt{flatten} function to recursively contract nested
same-type gates.  It is mutual with \texttt{flattenList} so that children are
contracted first. Each \texttt{andGate} then absorbs any immediate
\texttt{andGate} children of its already-contracted children via
\texttt{List.flatMap}. \texttt{orGate}s are handled in the same way.

\noindent\begin{minipage}{\linewidth}
\begin{lstlisting}[caption={AND/OR Gate Flattening},
                   mathescape=true,
                   label={lst:flatten}]{}
def asAndChildren : UnboundedFanInFormula $\rightarrow$ List UnboundedFanInFormula
  | andGate gs => gs
  | g          => [g]

def asOrChildren : UnboundedFanInFormula $\rightarrow$ List UnboundedFanInFormula
  | orGate gs => gs
  | g         => [g]

mutual
def flatten : UnboundedFanInFormula $\rightarrow$ UnboundedFanInFormula
  | inputGate i n  => inputGate i n
  | constant v lbl => constant v lbl
  | notGate g      => notGate (flatten g)
  | andGate gs     => andGate ((flattenList gs).flatMap asAndChildren)
  | orGate gs      => orGate ((flattenList gs).flatMap asOrChildren)

def flattenList :
    List UnboundedFanInFormula $\rightarrow$ List UnboundedFanInFormula
  | []      => []
  | g :: gs => flatten g :: flattenList gs
end
\end{lstlisting}
\end{minipage}

We prove that the flatten operation produces a formula with alternating AND
gates and OR gates, preserves the size and depth upper bounds of the original
formula, and does not introduce NOT gates, i.e. it preserves the
\texttt{HasNoNotGates} proposition.

\begin{lstlisting}[caption={Flattening Produces Alternating Circuit},
                   mathescape=true,
                   label={lst:flatten_hasAlternatingAndOrGates}]{}
theorem flatten_hasAlternatingAndOrGates (g : UnboundedFanInFormula)
                               (h_nn : HasNoNotGates g) :
  (flatten g).HasAlternatingAndOrGates

theorem flatten_depth_le (g : UnboundedFanInFormula) :
  ufiFormulaDepth (flatten g) $\leq$ ufiFormulaDepth g

theorem flatten_size_le (g : UnboundedFanInFormula) :
  ufiFormulaCircuitSize (flatten g) $\leq$ ufiFormulaCircuitSize g
\end{lstlisting}

Once an alternating AND/OR formula has been created, the leveling requirement
still remains. The \texttt{IsAlternatingAndLeveledAt} proposition covers both of
these requirements.

\begin{lstlisting}[caption={Alternation and Leveling Proposition},
                   mathescape=true,
                   label={lst:IsAlternatingAndLeveledAt}]{}
def IsAlternatingAndLeveledAt :
    UnboundedFanInFormula $\rightarrow$ Nat $\rightarrow$ Prop
  | UnboundedFanInFormula.inputGate _ _, _ => True
  | UnboundedFanInFormula.constant  _ _, _ => True
  | UnboundedFanInFormula.notGate   _,   _ => False
  | UnboundedFanInFormula.andGate   gates, n =>
      ($\forall$ g $\in$ gates, $\forall$ inner, g $\neq$ UnboundedFanInFormula.andGate inner) $\land$
      ($\forall$ g $\in$ gates, IsAndOr g $\rightarrow$ 1 $\leq$ n) $\land$
      ($\forall$ g $\in$ gates, IsAlternatingAndLeveledAt g (n - 1))
  | UnboundedFanInFormula.orGate   gates, n =>
      ($\forall$ g $\in$ gates, $\forall$ inner, g $\neq$ UnboundedFanInFormula.orGate inner) $\land$
      ($\forall$ g $\in$ gates, IsAndOr g $\rightarrow$ 1 $\leq$ n) $\land$
      ($\forall$ g $\in$ gates, IsAlternatingAndLeveledAt g (n - 1))
\end{lstlisting}

We show that the \texttt{IsAlternatingAndLeveledAt} proposition holds for any
formula that has alternating AND gates and OR gates and is bounded by depth $d$.

\begin{lstlisting}[caption={Alternation and Depth Bound Imply the Leveled Proposition},
                   mathescape=true,
                   label={lst:hasAlternatingAndOrGates_depth_imp_strict}]{}
theorem hasAlternatingAndOrGates_depth_imp_strict
    (g : UnboundedFanInFormula)
    (d : Nat)
    (h_leveled : g.HasAlternatingAndOrGates)
    (hdepth : ufiFormulaDepth g $\leq$ d) :
  IsAlternatingAndLeveledAt g d
\end{lstlisting}

Note that a formula for which \texttt{IsAlternatingAndLeveledAt} holds must not
have NOT gates. This fact is proved as the theorem in listing
\ref{lst:isAlternatingAndLeveledAt_hasNoNotGates}.

\begin{lstlisting}[caption={Alternation and Depth Bound Imply the Leveled Proposition},
                   mathescape=true,
                   label={lst:isAlternatingAndLeveledAt_hasNoNotGates}]{}
theorem isAlternatingAndLeveledAt_hasNoNotGates
    (f : UnboundedFanInFormula)
    (lvl : Nat)
    (h : IsAlternatingAndLeveledAt f lvl) :
  HasNoNotGates f
\end{lstlisting}

All these properties combine into the formula leveling contract:

\begin{lstlisting}[caption={Alternation and Depth Bound Imply the Leveled Proposition},
                   mathescape=true,
                   label={lst:levelize_contract}]{}
theorem levelize_contract (d n : Nat)
                          (f : UnboundedFanInFormula)
                          (h_nn : HasNoNotGates f)
                          (hlarge : ufiLargestInput f < n)
                          (hdepth : ufiFormulaDepth f $\leq$ d) :
  $\exists$ g : UnboundedFanInFormula,
    IsAlternatingAndLeveledAt g d
    $\land$ ufiLargestInput g < n
    $\land$ ufiFormulaDepth g $\leq$ d
    $\land$ ufiFormulaCircuitSize g $\leq$ ufiFormulaCircuitSize f
    $\land$ AgreesOn f g n
\end{lstlisting}

\subsection{Generating Proper Formula Bottoms}

The final normalization transformation ensures that the bottoms of the formula
are proper. Recall that a proper CNF/DNF has no empty clauses and no duplicate
literals in any clause.

\subsubsection{Deduplication of Gate Children}
\texttt{dedupChildren} recursively removes duplicate children without collapsing
gate constructors, which ensures that \texttt{IsAlternatingAndLeveledAt} is
preserved.  It uses the custom \texttt{formulaBEq} comparator for deciding
formula equivalence (see section \ref{sec:formulaBEq} of the appendix for
details).

\begin{lstlisting}[caption={Deduplicating Gate Children},
                   mathescape=true,
                   label={lst:dedupChildren}]{}
def dedupFormulas : List UnboundedFanInFormula $\rightarrow$ List UnboundedFanInFormula
  | []      => []
  | g :: gs => let rest := dedupFormulas gs
               if rest.any (fun x => formulaBEq x g) then
                 rest
               else
                 g :: rest

mutual
def dedupChildren : UnboundedFanInFormula $\rightarrow$ UnboundedFanInFormula
  | .inputGate i b => .inputGate i b
  | .constant b m  => .constant b m
  | .notGate g.    => .notGate (dedupChildren g)
  | .andGate gs    => .andGate (dedupFormulas (dedupChildrenList gs))
  | .orGate gs     => .orGate  (dedupFormulas (dedupChildrenList gs))

def dedupChildrenList : List UnboundedFanInFormula $\rightarrow$ List UnboundedFanInFormula
  | [] => []
  | g :: gs => dedupChildren g :: dedupChildrenList gs
end
\end{lstlisting}

Gate deduplication must preserve formula size and depth bounds, input width,
evaluation, AND/OR alternation, and leveling. The theorems in
Listing~\ref{lst:dedupChildrenProperties} establish these properties.

\begin{lstlisting}[caption={Properties of Gate Deduplication},
                   mathescape=true,
                   label={lst:dedupChildrenProperties}]{}
theorem dedupChildren_alternating : $\forall$ f lvl,
  IsAlternatingAndLeveledAt f lvl $\rightarrow$
    IsAlternatingAndLeveledAt (dedupChildren f) lvl

theorem dedupChildren_depth_le (f : UnboundedFanInFormula) :
  ufiFormulaDepth (dedupChildren f) $\leq$ ufiFormulaDepth f

theorem dedupChildren_circuit_size_le (f : UnboundedFanInFormula) :
  ufiFormulaCircuitSize (dedupChildren f) $\leq$ ufiFormulaCircuitSize f

theorem dedupChildren_largest_input_lt
  {f : UnboundedFanInFormula} {n : Nat} (h : ufiLargestInput f < n) :
    ufiLargestInput (dedupChildren f) < n

theorem dedupChildren_eval (f : UnboundedFanInFormula) (xs : List Bool) :
  ufiFormulaEval (dedupChildren f) xs = ufiFormulaEval f xs
\end{lstlisting}

\subsubsection{Generating Formulas with Proper Bottoms}

We define a proposition to represent the non-empty clauses and unique literals
requirements for the bottom gates.

\begin{lstlisting}[caption={HasProperBottomsAt},
                   mathescape=true,
                   label={lst:HasProperBottomsAt}]{}
def HasProperBottomsAt : UnboundedFanInFormula $\rightarrow$ Nat $\rightarrow$ Prop
  | .inputGate _ _, _   => True
  | .constant _ _, _    => True
  | .notGate _, _       => False
  | .andGate gates, lvl =>
      if lvl $\leq$ 2 then
        Circuits.CnfDnf.IsProperCNF (.andGate gates)
      else
        $\forall$ g $\in$ gates, HasProperBottomsAt g (lvl - 1)
  | .orGate gates, lvl  =>
      if lvl $\leq$ 2 then
        Circuits.CnfDnf.IsProperDNF (.orGate gates)
      else
        $\forall$ g $\in$ gates, HasProperBottomsAt g (lvl - 1)
\end{lstlisting}

To transform a formula into this form, we start by performing constant folding
using the \texttt{andClauses} and \texttt{orClauses} functions of listings
\ref{lst:andClauses} and \ref{lst:orClauses}.  The \texttt{orChildLiterals}
function creates a clause without constants from a list of input gates and
constant gates (all other gates are dropped). Input gates to this function are
preserved, any \texttt{constant false} gates are dropped (since \texttt{false}
is the identity for the OR function), and any \texttt{constant true} gates force
the function to return \texttt{none} indicating that no OR gate is required for
this clause (because it is identically true).

\noindent\begin{minipage}{\linewidth}
\begin{lstlisting}[caption={constant Folding in AND Gates},
                   mathescape=true,
                   label={lst:andClauses}]{}
def orChildLiterals : List UnboundedFanInFormula
                    $\rightarrow$ Option (List (Nat $\times$ Bool))
  | []                          => some []
  | (.inputGate i b) :: rest    => (orChildLiterals rest).map
                                     (fun clause => (i, b) :: clause)
  | (.constant true _) :: _     => none
  | (.constant false _) :: rest => orChildLiterals rest
  | _ :: rest                   => orChildLiterals rest

def andClauses : List UnboundedFanInFormula
               $\rightarrow$ Option (List (List (Nat $\times$ Bool)))
  | []        => some []
  | g :: rest => match andClauses rest with
                 | none    => none
                 | some cs => match g with
                              | .inputGate i b    => some ([(i, b)] :: cs)
                              | .constant true  _ => some cs
                              | .constant false _ => none
                              | .orGate literals  =>
                                  match orChildLiterals literals with
                                  | none    => some cs
                                  | some [] => none
                                  | some c  => some (c :: cs)
                              | _ => some cs
\end{lstlisting}
\end{minipage}

A similar pattern is used to implement constant folding of OR gates in listing
\ref{lst:orClauses}.

\begin{lstlisting}[caption={constant Folding in OR Gates},
                   mathescape=true,
                   label={lst:orClauses}]{}
def andChildLiterals : List UnboundedFanInFormula
                     $\rightarrow$ Option (List (Nat $\times$ Bool))
  | []                         => some []
  | (.inputGate i b) :: rest   => (andChildLiterals rest).map
                                    (fun c => (i, b) :: c)
  | (.constant false _) :: _   => none
  | (.constant true _) :: rest => andChildLiterals rest
  | _ :: rest                  => andChildLiterals rest

def orClauses : List UnboundedFanInFormula
              $\rightarrow$ Option (List (List (Nat $\times$ Bool)))
  | []        => some []
  | g :: rest => match orClauses rest with
                 | none    => none
                 | some cs => match g with
                              | .inputGate i b    => some ([(i, b)] :: cs)
                              | .constant false _ => some cs
                              | .constant true _  => none
                              | .andGate literals =>
                                  match andChildLiterals literals with
                                  | none => some cs
                                  | some [] => none
                                  | some c => some (c :: cs)
                              | _ => some cs
\end{lstlisting}

These functions are then be used to make a bottom CNF or DNF proper.  A bottom
CNF in the formula is replaced with the constant \texttt{false} value if
\texttt{andClauses} returns \texttt{none} as shown in listing
\ref{lst:properizeBottomAnd}. Similarly, a bottom DNF in the formula is replaced
with the constant \texttt{true} value if \texttt{orClauses} returns
\texttt{none} (listing \ref{lst:properizeBottomOr}).  The
\texttt{cnfFromClauses} and \texttt{dnfFromClauses} helper functions are defined
in the appendix \ref{sec:cnfDnfFromClauses}.

\begin{lstlisting}[caption={Properizing a Bottom CNF},
                   mathescape=true,
                   label={lst:properizeBottomAnd}]{}
def pairLevelDedup : List (Nat $\times$ Bool) $\rightarrow$ List (Nat $\times$ Bool)
  | [] => []
  | x :: xs => if x $\in$ pairLevelDedup xs then
                 pairLevelDedup xs
               else
                 x :: pairLevelDedup xs

def properizeClauses (clauses : List (List (Nat $\times$ Bool)))
  : List (List (Nat $\times$ Bool)) :=
  (clauses.map pairLevelDedup).filter
    (fun c => decide ((c.map Prod.fst).Nodup))

def properizeCNF (g : UnboundedFanInFormula) : UnboundedFanInFormula :=
  let cs := properizeClauses (cnfClauses g)
  if [] $\in$ cs then cnfFromClauses [[(0, true)], [(0, false)]]
  else cnfFromClauses cs

def properizeBottomAnd (gs : List UnboundedFanInFormula) : UnboundedFanInFormula :=
  match andClauses gs with
  | none    => constant false 0
  | some cs => properizeCNF (cnfFromClauses cs)
\end{lstlisting}

\begin{lstlisting}[caption={Properizing a Bottom DNF},
                   mathescape=true,
                   label={lst:properizeBottomOr}]{}
def properizeDNF (g : UnboundedFanInFormula) : UnboundedFanInFormula :=
  let cs := properizeClauses (dnfClauses g)
  if [] $\in$ cs then
    dnfFromClauses [[(0, true)], [(0, false)]]
  else
    dnfFromClauses cs

def properizeBottomOr (gs : List UnboundedFanInFormula) : UnboundedFanInFormula :=
  match orClauses gs with
  | none    => constant true 0
  | some cs => properizeDNF (dnfFromClauses cs)
\end{lstlisting}

The \texttt{properizeTree} function in listing \ref{lst:properizeTree} shows how
to perform these operations recursively on an entire formula.

\noindent\begin{minipage}{\linewidth}
\begin{lstlisting}[caption={Properizing a Formula/Tree},
                   mathescape=true,
                   label={lst:properizeTree}]{}
mutual
def properizeTree : Nat $\rightarrow$ UnboundedFanInFormula $\rightarrow$ UnboundedFanInFormula
  | _, .inputGate i b  => .inputGate i b
  | _, .constant c l   => .constant c l
  | _, .notGate g      => .notGate g
  | lvl, .andGate gs   =>
      if lvl $\leq$ 2 then
        properizeBottomAnd gs
      else
        .andGate (properizeTreeList (lvl - 1) gs)
  | lvl, .orGate gs   =>
      if lvl $\leq$ 2 then
        properizeBottomOr gs
      else
        .orGate (properizeTreeList (lvl - 1) gs)

def properizeTreeList : Nat $\rightarrow$ List UnboundedFanInFormula $\rightarrow$ List UnboundedFanInFormula
  | _, []        => []
  | lvl, g :: gs => properizeTree lvl g :: properizeTreeList lvl gs
end
\end{lstlisting}
\end{minipage}

Converting the gates into proper CNFs and DNFs recursively yields a formula with
these properties:

\begin{enumerate}
  \item{
    Alternation of AND gates and OR gates and their levels is preserved.
  }
  \item{
    The AND gates and OR gates at every level are proper.
  }
  \item{
    The formula depth upper bound is preserved.
  }
  \item{
    The input width upper bound is preserved.
  }
  \item{
    Formula evaluation on the original formula and the output formula is identical.
  }
\end{enumerate}

These properties are proved as the theorems in listing
\ref{lst:strict_properizeTree}.

\noindent\begin{minipage}{\linewidth}
\begin{lstlisting}[caption={Properization Properties},
                   mathescape=true,
                   label={lst:strict_properizeTree}]{}
theorem strict_properizeTree (lvl : Nat)
                             (f : UnboundedFanInFormula)
                             (hstrict : IsAlternatingAndLeveledAt f lvl)
                             (hlvl : 2 $\leq$ lvl)
                             (hdepth : ufiFormulaDepth f $\leq$ lvl) :
  IsAlternatingAndLeveledAt (properizeTree lvl f) lvl

theorem proper_properizeTree (lvl : Nat)
                             (f : UnboundedFanInFormula)
                             (hnn : Circuits.Leveling.HasNoNotGates f) :
  HasProperBottomsAt (properizeTree lvl f) lvl

theorem depth_properizeTree_le (lvl : Nat)
                               (f : UnboundedFanInFormula)
                               (hlvl : 2 $\leq$ lvl)
                               (hdepth : ufiFormulaDepth f $\leq$ lvl) :
  ufiFormulaDepth (properizeTree lvl f) $\leq$ lvl

theorem vars_properizeTree (lvl : Nat)
                           (f : UnboundedFanInFormula)
                           (n : Nat)
                           (hn : 0 < n)
                           (hbound : $\forall$ i $\in$ ufiCollectInputIndices f,
                                        i < n) :
  $\forall$ i $\in$ ufiCollectInputIndices (properizeTree lvl f), i < n

theorem eval_properizeTree (inputs : List Bool)
                           (lvl : Nat)
                           (f : UnboundedFanInFormula)
                           (hnn : Circuits.Leveling.HasNoNotGates f)
                           (hstrict : IsAlternatingAndLeveledAt f lvl)
                           (hlvl : 2 $\leq$ lvl)
                           (hdepth : ufiFormulaDepth f $\leq$ lvl)
                           (hpos : 0 < inputs.length)
                           (hbound : $\forall$ i $\in$ ufiCollectInputIndices f,
                                        i < inputs.length) :
  ufiFormulaEval (properizeTree lvl f) inputs =
  ufiFormulaEval f inputs
\end{lstlisting}
\end{minipage}

Given an unbounded fan-in formula of size at most $an^b$ and depth at most $d$
and alternating AND gates and OR gates, the
\texttt{proper\_bottoms\_contract\_explicit} theorem of listing
\ref{lst:proper_bottoms_contract_explicit} provides proof of all the properties
from listing \ref{lst:strict_properizeTree}.

\noindent\begin{minipage}{\linewidth}
\begin{lstlisting}[caption={The Formula Proper Bottoms Contract},
                   mathescape=true,
                   label={lst:proper_bottoms_contract_explicit}]{}
theorem proper_bottoms_contract_explicit (d : Nat)
                                         (hd : 2 $\leq$ d) :
  $\forall$ a b : Nat,
    $\exists$ a' b' : Nat,
      $\forall$ (n : Nat) (f : UnboundedFanInFormula),
        IsAlternatingAndLeveledAt f d $\rightarrow$
        ufiLargestInput f < n $\rightarrow$
        ufiFormulaDepth f $\leq$ d $\rightarrow$
        ufiFormulaCircuitSize f $\leq$ a * n ^ b $\rightarrow$
        $\exists$ g : UnboundedFanInFormula,
          IsAlternatingAndLeveledAt g d
          $\land$ HasProperBottomsAt g d
          $\land$ ufiLargestInput g < n
          $\land$ ufiFormulaDepth g $\leq$ d
          $\land$ ufiFormulaCircuitSize g $\leq$ a' * n ^ b'
          $\land$ AgreesOn f g n
\end{lstlisting}
\end{minipage}

We define the proposition \texttt{IsProperlyLeveled}
(listing~\ref{lst:IsProperlyLeveled}) to state that a formula meets all the
requirements introduced above. We then prove that combining the transformations
discussed so far produces a formula that \texttt{IsProperlyLeveled}.

\begin{lstlisting}[caption={Defining Formula Leveling},
                   mathescape=true,
                   label={lst:IsProperlyLeveled}]{}
def IsProperlyLeveled : UnboundedFanInFormula $\rightarrow$ Nat $\rightarrow$ Prop
  | .inputGate _ _,   _ => True
  | .constant _ _, _    => True
  | .notGate _,    _    => False
  | .andGate gates, lvl =>
      ($\forall$ g $\in$ gates, $\forall$ inner, g $\neq$ .andGate inner) $\land$
      ($\forall$ g $\in$ gates, UnboundedFanInFormula.IsAndOr g $\rightarrow$ 1 $\leq$ lvl) $\land$
      (if lvl $\leq$ 2 then
        Circuits.CnfDnf.IsProperCNF (.andGate gates)
      else
        $\forall$ g $\in$ gates, IsProperlyLeveled g (lvl - 1))
  | .orGate gates, lvl  =>
      ($\forall$ g $\in$ gates, $\forall$ inner, g $\neq$ .orGate inner) $\land$
      ($\forall$ g $\in$ gates, UnboundedFanInFormula.IsAndOr g $\rightarrow$ 1 $\leq$ lvl) $\land$
      (if lvl $\leq$ 2 then
        Circuits.CnfDnf.IsProperDNF (.orGate gates)
      else
        $\forall$ g $\in$ gates, IsProperlyLeveled g (lvl - 1))

theorem isProperlyLeveled_of_strict_proper :
  $\forall$ (f : UnboundedFanInFormula) (lvl : Nat),
    IsAlternatingAndLeveledAt f lvl $\rightarrow$
    HasProperBottomsAt f lvl $\rightarrow$ IsProperlyLeveled f lvl

theorem isProperlyLeveled_imp_proper :
  $\forall$ (f : UnboundedFanInFormula) (n : Nat),
    IsProperlyLeveled f n $\rightarrow$ HasProperBottomsAt f n
\end{lstlisting}

\subsubsection{Constant Simplification}

The final transformation performed on the formula is the replacement of
\texttt{constant} leaves by canonical empty gates. This involves replacing
AND/OR gates whose immediate children include canonical empty gates with a
single canonical empty gate where possible as shown in listing
\ref{lst:simplifyConstants}.

\begin{lstlisting}[caption={Syntactic Tests for the Canonical True and False Constants in a Formula},
                   mathescape=true,
                   label={lst:isCanonicalTrueFalse}]{}
def isCanonicalTrue : UnboundedFanInFormula $\rightarrow$ Bool
  | .andGate [] => true
  | _           => false

def isCanonicalFalse : UnboundedFanInFormula $\rightarrow$ Bool
  | .orGate [] => true
  | _          => false
\end{lstlisting}

\begin{lstlisting}[caption={Constant Simplification in a Formula},
                   mathescape=true,
                   label={lst:simplifyConstants}]{}
mutual
def simplifyConstants : UnboundedFanInFormula $\rightarrow$ UnboundedFanInFormula
  | .inputGate x b => .inputGate x b
  | .constant b _  => match b with
                      | true  => .andGate []
                      | false => .orGate []
  | .notGate g     => .notGate (simplifyConstants g)
  | .andGate gs    => let cs := simplifyConstantsList gs
                      if cs.any isCanonicalFalse then
                        .orGate []
                      else
                        .andGate (cs.filter (fun g => !isCanonicalTrue g))
  | .orGate gs     => let cs := simplifyConstantsList gs
                      if cs.any isCanonicalTrue then
                        .andGate []
                      else
                        .orGate (cs.filter (fun g => !isCanonicalFalse g))

def simplifyConstantsList : List UnboundedFanInFormula $\rightarrow$
  List UnboundedFanInFormula
  | [] => []
  | g :: gs => simplifyConstants g :: simplifyConstantsList gs
\end{lstlisting}

\begin{lstlisting}[caption={Expressing Constant-Freeness},
                   mathescape=true,
                   label={lst:IsConstantFree}]{}
def IsConstantFree : UnboundedFanInFormula $\rightarrow$ Prop
  | .inputGate _ _ => True
  | .constant _ _  => False
  | .notGate g     => IsConstantFree g
  | .andGate gates => $\forall$ g $\in$ gates, IsConstantFree g
  | .orGate gates  => $\forall$ g $\in$ gates, IsConstantFree g
\end{lstlisting}

\subsection{Existence of Properly Leveled Formulas}

The four transformations compose to give a properly leveled formula without
increasing depth. The quantitative lower bound uses the explicit size estimate
in Listing~\ref{lst:exists_strictly_leveled_proper_form_core}.  An input formula
of size at most $cn^k$ produces a formula of size at most
$80(c+1)^2n^{2(k+1)}$. For an individual formula of size $S$, taking $c=S$ and
$k=0$ bounds the normalization cost by $80(S+1)^2n^2$.

\noindent\begin{minipage}{\linewidth}
\begin{lstlisting}[caption={Final Leveled Core},
                   mathescape=true,
                   label={lst:exists_strictly_leveled_proper_form_core}]{}
theorem exists_strictly_leveled_proper_form_core_explicit
    (c k d : Nat) (hd : 2 $\leq$ d) (n : Nat)
    (circuit : UnboundedFanInFormula)
    (hlarge : ufiLargestInput circuit < n)
    (hdepth : ufiFormulaDepth circuit $\leq$ d)
    (hbound : ufiFormulaCircuitSize circuit $\leq$ c * n ^ k) :
  $\exists$ g : UnboundedFanInFormula,
    ufiFormulaDepth g $\leq$ d
    $\land$ ufiLargestInput g < n
    $\land$ IsProperlyLeveled g d
    $\land$ ufiFormulaCircuitSize g $\leq$
        (80 * (c + 1) * (c + 1)) * n ^ (2 * (k + 1))
    $\land$ AgreesOn circuit g n
    $\land$ IsConstantFree g
\end{lstlisting}
\end{minipage}

We then define \texttt{LeveledUFIFormulaOfSizePolyNAndDepthD}, a subtype of
formulas that have no NOT gates, whose AND/OR layers strictly alternate with the
root at level $d$, and whose layers form proper CNF/DNF clauses.  For
polynomial-size formula families of depth at most $d\geq2$, the explicit
estimate gives uniform parameters $c'=80(c+1)^2$ and $k'=2(k+1)$. The
family-level statement below packages the resulting formulas as members of
\texttt{LeveledUFIFormulaOfSizePolyNAndDepthD}.

\begin{lstlisting}[caption={Leveled Constant Depth Formulas},
                   mathescape=true,
                   label={lst:exists_leveled_form}]{}
def LeveledUFIFormulaOfSizePolyNAndDepthD
  (n c k d : Nat) :=
  {
    circuit : UnboundedFanInFormula //
    ufiLargestInput circuit < n
    $\land$ ufiFormulaDepth circuit $\leq$ d
    $\land$ ufiFormulaCircuitSize circuit $\leq$ c * n ^ k
    $\land$ d > 0
    $\land$ IsProperlyLeveled circuit d
  }

lemma exists_leveled_form (c k d : Nat) (hd : 2 $\leq$ d) :
  $\exists$ c' k' : Nat,
    $\forall$ (n : Nat)
      (circuit : UFIFormulaOfSizeAtMostPolyNAndDepthAtMostD n c k d),
      $\exists$ (lc : LeveledUFIFormulaOfSizePolyNAndDepthD n c' k' d),
          ufiLargestInput lc.val < n
          $\land$ AgreesOn circuit.val lc.val n
\end{lstlisting}

\subsection{Switching Lemma Parameter Selection}

After normalization, the switching lemma applies simultaneously to the bottom
CNFs and DNFs. For a DNF of width at most $w$, the fraction of restrictions
whose canonical decision tree has depth greater than $t$ is at most $(10\sigma
w)^t$. A restriction good for every bottom formula lets us replace those
formulas by CNFs or DNFs that merge with their parent gates, reducing the
formula depth by one.

Let $S$ bound the number of bottom formulas to which the switching lemma is
applied. The union bound gives a common good restriction whenever
\[
  S(10\sigma w)^t<1.
\]
Choose an integer $t\geq2$ with $S<2^t$. Once the bottom width satisfies
$w\leq t$, any density $\sigma\leq1/(20t)$ gives
\[
  S(10\sigma w)^t\leq S\left(\frac12\right)^t<1.
\]
On a domain of $m$ live variables, the formal proof uses the exact-cardinality
density
\[
  \sigma=\frac{\lfloor m/(20t)\rfloor}{m}.
\]
The local hypothesis $20t(t+1)\leq m$ ensures that this density is positive and
retains more than $t$ variables. It therefore supplies both the counting bound
and the live-variable requirement for the next stage.

During iteration, $S$ bounds the maintained switching-gate budget, which
controls how many bottom formulas are extracted. The budget stays below $2^t$
even though replacing bottom formulas can increase the total formula size. The
calibration applies to arbitrary finite size bounds. In the quantitative PARITY
argument, an integer $a\geq1$ with $S<2^a$ supplies the common cutoff $t=2a$ for
round 0 and all subsequent nonterminal rounds.

\subsubsection{Parameters for Killing Wide Clauses}

Depth reduction maintains bottom width at most $t$. To establish this invariant
initially, round 0 kills every clause whose width exceeds $t$.  A properly
leveled formula can initially have clauses of width $n$, so applying the
switching restriction immediately could leave only a constant number of
variables live. Instead, round 0 retains exactly $\lfloor n/3\rfloor$ variables
and uses the cutoff $t=2a$ to kill all wide clauses at once.  Note that other
fractions like $p = \frac{1}{5}$ can also be used. The choice of fraction
affects the constants in the exponent in the final formula size lower bound.

To show that all wide clauses are killed, we bound the probability that a single
wide clause survives.  Every variable on the domain $x_{[n]}$ is live with
probability $p$. Dead variables have a 1/2 probability of surviving since
exactly 1 of the bit assignments kills the clause. Let $q$ be the probability
that a single variable keeps its clause alive.

\begin{align*}
  q
    &= \text{Pr}[\text{Initial Restriction Keeps } x_i \text{ alive}] + \\
    &~~~~ \text{Pr}[\text{Initial Restriction Kills } x_i \text{ and the killing bit keeps the whole clause alive}] \\
    &= p + (1-p) \cdot \frac{1}{2} \\
    &= \frac{p + 1}{2} \\
\end{align*}

For $p = \frac{1}{3}$, $q = \frac{2}{3}$.  A finite counting argument shows that
a clause of length $L$ survives with probability at most $q^L$.  Since every
wide clause has length at least $t+1$, its survival probability is at most
$q^{t+1}$.  For the round 0 restriction, we consider a restriction good for a
wide clause $c$ if it kills $c$.  Let $W$ be the number of wide clauses. Note
that $W$ is at most the formula size, $S$. To show that there exists a single
restriction that kills all wide clauses, we need this union bound inequality
(which uses the highest survival probability of a single clause):

$$ W \cdot q^{t+1} < 1 $$

Since $W\leq S$, it suffices to prove that $S q^{t+1}<1$.  Choose an integer
$a$ such that $S<2^a$ and set $t=2a$.  For $q=2/3$,
\begin{align*}
 S\left(\frac{2}{3}\right)^{2a+1}
 &< 2^a\left(\frac{4}{9}\right)^a\frac{2}{3} \\
 &= \left(\frac{8}{9}\right)^a\frac{2}{3} < 1.
\end{align*}
The integral choice $t=2a$ therefore satisfies both the round-zero counting
bound and the later switching bound. Taking the least integer $a\geq1$ with
$S<2^a$ gives a cutoff of order $\log(S+1)$.  The parameters are grouped into
the \texttt{RoundZeroCalibration} structure in
Listing~\ref{lst:RoundZeroCalibration}.  The
\texttt{exists\_restriction\_width\_le\_card} theorem in listing
\ref{lst:Round0RestrictionWidthLECardOfBound} then uses it to prove the
existence of a good restriction killing all wide clauses in a formula of
polynomial size.

\begin{lstlisting}[caption={Round 0 Calibration Data},
                   mathescape=true,
                   label={lst:RoundZeroCalibration}]{}
structure RoundZeroCalibration (q : $\mathbb{Q}$) (n s : Nat) : Prop where
  q_pos    : 0 < q
  q_le_one : q $\leq$ 1
  n_pos    : 0 < n
  live_le  : s $\leq$ n
  sigma_exists : $\exists$ $\sigma$ : OpenUnitIntervalQ,
                    Nat.ceil ($\sigma$.val * (n : $\mathbb{Q}$)) = s $\land$
                    (1 + (s : $\mathbb{Q}$) / n) / 2 $\leq$ q
\end{lstlisting}

\begin{lstlisting}[caption={Round 0 Fan-in Reduction Properties},
                   mathescape=true,
                   label={lst:Round0RestrictionWidthLECardOfBound}]{}
theorem RoundZeroCalibration.exists_restriction_width_le_card
    {q      : $\mathbb{Q}$}
    {n s    : Nat}
    (cal    : RoundZeroCalibration q n s)
    (t      : Nat)
    (dnfs   : List UnboundedFanInFormula)
    (hdnf   : $\forall$ d $\in$ dnfs, isDNF d = true)
    (hnd    : $\forall$ d $\in$ dnfs, $\forall$ c $\in$ dnfClauses d, (c.map Prod.fst).Nodup)
    (hvar   : $\forall$ d $\in$ dnfs, $\forall$ c $\in$ dnfClauses d, $\forall$ p $\in$ c, p.1 < n)
    (hcount : ((wideClauses t dnfs).length : $\mathbb{Q}$) * q ^ (t + 1) < 1) :
  $\exists$ ($\sigma$ : OpenUnitIntervalQ) ($\rho$ : AssignedRandomRestriction $\sigma$ n),
    $\rho$.starAssignment.val.val.card = s $\land$
    $\forall$ d $\in$ dnfs,
      dnfWidth (simpleRestrictDNF (randomRestrictionToMap n $\sigma$ $\rho$) d) $\leq$ t
\end{lstlisting}

The example in Listing~\ref{lst:roundZeroCalibration_one_third} supplies a
\texttt{RoundZeroCalibration} for $p=\frac{1}{3}$ and uses it to obtain a restriction that kills
every wide clause.

\noindent\begin{minipage}{\linewidth}
\begin{lstlisting}[caption={Sample Round 0 Calibration Data for 1/3 Liveness Ratio},
                   mathescape=true,
                   label={lst:roundZeroCalibration_one_third}]{}
theorem roundZeroCalibration_one_third (n : Nat) (hn : 3 $\leq$ n) :
    RoundZeroCalibration (2 / 3) n (n / 3)

-- Get a restriction $\sigma$ killing all wide clauses
obtain $\langle\sigma$, $\rho$, hcard, hwidth$\rangle$ :=
          (roundZeroCalibration_one_third n hn).exists_restriction_width_le_card
                t dnfs hdnf hnd hvar hcount
\end{lstlisting}
\end{minipage}

\subsubsection{Ensuring Live Variable Sufficiency}
\label{subsec:LiveVariableSufficiency}

Fix a formula depth $d\geq2$ and an integer bottom-width cutoff $t\geq2$.  Round
0 leaves exactly $\lfloor n/3\rfloor$ variables live. Each of the next $d-2$
switching rounds reduces the depth by one and retains at least $\lfloor
m/(20t)\rfloor$ of the current $m$ live variables. Such a round requires
\[
  20t(t+1)\leq m,
\]
which ensures that at least $t+1$ variables survive while the new bottom width
remains at most $t$. This quadratic requirement applies to the nonterminal
rounds, where the width invariant must be carried to the next stage.

At depth two, the whole formula is a single DNF or CNF of width at most $t$. The
final restriction can use a decision-tree depth bound of one and retain exactly
two variables. For a DNF, choose $\sigma=2/m$. The switching lemma then bounds
the fraction of bad restrictions by
\[
  (10\sigma t)^1=\frac{20t}{m}<1
\]
whenever $20(t+1)<m$. Hence a good restriction exists, and its decision tree
converts to a DNF of width at most one on two live variables. For a CNF, apply
the same argument to its DNF dual and negate the resulting tree.  The final
width is strictly smaller than the live-variable count, as required by
\texttt{narrow\_dnf\_misclassifies\_parity}
(Listing~\ref{lst:narrow_dnf_misclassifies_parity}). This terminal step needs
only a number of live variables linear in $t$.  We therefore choose the
\emph{base-case reserve}, denoted by $B(t)$, to be the number of live variables
sufficient for the final depth-two collapse:
\[
  B(t)=40(t+1).
\]
The factor $40$ is a convenient choice that lets the induction use the
non-strict hypothesis $B(t)\leq m$ while guaranteeing the strict inequality
$20(t+1)<m$ required by the depth-two argument above.  With $j\geq2$ layers
remaining, define
\[
  R_j(t)=(20t)^{j-2}B(t).
\]
At depth two, $R_2(t)=B(t)$ supplies the terminal reserve. At a nonterminal
stage $j\geq3$, having $m\geq R_j(t)$ live variables guarantees the condition
$20t(t+1)\leq m$ needed for the next switching round:
\[
  m\geq R_j(t)\geq(20t)B(t)\geq20t(t+1).
\]
This condition is quadratic in the width cutoff $t$. It ensures that the round
retains at least $\lfloor m/(20t)\rfloor\geq t+1$ variables, so the surviving
variable count strictly exceeds the new bottom-width bound $t$.  For $j\geq3$,
we also have $R_j(t)=(20t)R_{j-1}(t)$. Thus $m\geq R_j(t)$ implies
\[
  \left\lfloor\frac{m}{20t}\right\rfloor\geq R_{j-1}(t),
\]
so the next stage has its full reserve even after integer rounding.  Immediately
after round 0, the formula has depth at most $d$ and exactly $m=\lfloor
n/3\rfloor$ live variables. Applying the reserve condition with $j=d$ therefore
requires $R_d(t)\leq\lfloor n/3\rfloor$. Expanding the definitions of $R_d(t)$
and $B(t)$ gives the sufficient condition
\[
  R_d(t)=(20t)^{d-2}B(t)
  =(20t)^{d-2}\,40(t+1)\leq\left\lfloor\frac n3\right\rfloor.
\]
This is the threshold in \texttt{exists\_iterated\_switching\_depth\_collapse\_sharp}
(Listing~\ref{lst:exists_iterated_switching_depth_collapse}).

To combine this condition with the size estimate, take an integer $a\geq1$ such
that $S<2^a$ and set $t=2a$, as in the round-zero argument. Since $2a+1\leq3a$,
\[
  3R_d(2a)
  =3(40a)^{d-2}\,40(2a+1)
  \leq360\cdot40^{d-2}a^{d-1}.
\]
Therefore $360\cdot40^{d-2}a^{d-1}\leq n$ suffices to meet the live-variable
requirements throughout the restriction process. The linear base-case reserve
makes this sufficient threshold have degree $d-1$ in $a$.

\subsection{Initial Restriction Killing Wide Bottom Gates}

We now examine the process used to kill wide gates at the bottom of a leveled
formula.

\subsubsection{Bottom Layer Extraction}

To apply a restriction to all the gates at the bottom of a formula, we first use
the \texttt{extractBottomLayer} function in listing
\ref{lst:BottomLayerExtraction} to extract the subformulas at levels 1 and 2. We
then replace them with leaves.  This function does not simplify gates or
evaluate the formula. It maps a formula to a triple comprising:

\begin{enumerate}
  \item a left-to-right list of bottom subformulas
  \item the remaining top skeleton with each bottom replaced by a fresh positive input
  \item the next unused placeholder index.
\end{enumerate}

\begin{figure}[H]
  \centering
  \begin{forest}
    decision tree,
    for tree={s sep=6mm},
    formula left edge/.style={edge path'={(!u.south west) -- (.north)}},
    formula right edge/.style={edge path'={(!u.south east) -- (.north)}},
    before drawing tree={
      tikz+={
        \coordinate (formula-level-left) at (formula-xzero.west);
        \coordinate (formula-level-right) at (formula-xfive.east);
        \coordinate (formula-root-left) at
          (formula-level-left |- formula-root.center);
        \coordinate (formula-root-right) at
          (formula-level-right |- formula-root.center);
        \coordinate (formula-or-left) at
          (formula-level-left |- formula-or-left-node.center);
        \coordinate (formula-or-right) at
          (formula-level-right |- formula-or-left-node.center);
        \coordinate (formula-and-left) at
          (formula-level-left |- formula-and-left-node.center);
        \coordinate (formula-and-right) at
          (formula-level-right |- formula-and-left-node.center);
        \begin{pgfonlayer}{background}
          \node[fill=paleblue, rounded corners=7pt,
                fit=(formula-root-left)(formula-root-right)(formula-root),
                inner xsep=16mm, inner ysep=3mm]
                (formula-level-four-box) {};
          \node[fill=paleteal, rounded corners=7pt,
                fit=(formula-or-left)(formula-or-right)
                    (formula-or-left-node)(formula-or-right-node),
                inner xsep=16mm, inner ysep=3mm]
                (formula-level-three-box) {};
          \node[fill=paleorange, rounded corners=7pt,
                fit=(formula-and-left)(formula-and-right)
                    (formula-and-left-node)(formula-and-right-node),
                inner xsep=16mm, inner ysep=3mm]
                (formula-level-two-box) {};
          \node[fill=lightgray, rounded corners=7pt,
                fit=(formula-xzero)(formula-xfive),
                inner xsep=16mm, inner ysep=3mm]
                (formula-level-one-box) {};
          \node[anchor=north west, font=\scriptsize\bfseries, text=muted]
            at ($(formula-level-four-box.north west)+(2mm,-1.5mm)$) {Level 4};
          \node[anchor=north west, font=\scriptsize\bfseries, text=muted]
            at ($(formula-level-three-box.north west)+(2mm,-1.5mm)$) {Level 3};
          \node[anchor=north west, font=\scriptsize\bfseries, text=muted]
            at ($(formula-level-two-box.north west)+(2mm,-1.5mm)$) {Level 2};
          \node[anchor=north west, font=\scriptsize\bfseries, text=muted]
            at ($(formula-level-one-box.north west)+(2mm,-1.5mm)$) {Level 1};
        \end{pgfonlayer}
      }
    }
    [$\land$, variable, name=formula-root
      [$\lor$, variable, formula left edge, name=formula-or-left-node
        [$\land$, variable, formula left edge, name=formula-and-left-node
          [$x_0$, leaf, formula left edge, name=formula-xzero]
          [$\overline{x_1}$, leaf, formula right edge]
        ]
        [$\land$, variable, formula right edge
          [$x_2$, leaf]
        ]
      ]
      [$\lor$, variable, formula right edge, name=formula-or-right-node
        [$\land$, variable, formula left edge
          [$x_3$, leaf]
        ]
        [$\land$, variable, formula right edge, name=formula-and-right-node
          [$\overline{x_4}$, leaf, formula left edge]
          [$x_5$, leaf, formula right edge, name=formula-xfive]
        ]
      ]
    ]
  \end{forest}
  \caption{Example leveled formula for bottom-layer extraction.}
  \label{fig:bottom-layer-extraction-example}
\end{figure}
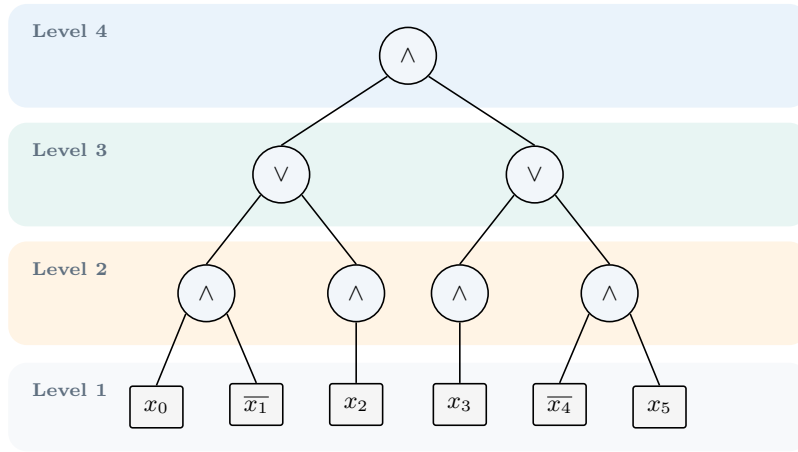

\begin{callout}
Consider the example formula in figure
\ref{fig:bottom-layer-extraction-example}. The function call
\texttt{extractBottomLayer 4 10 formula} specifies a starting level for the
formula and a starting index for the new inputs to the formula (since each
subformula at level 2 will be replaced with single input).  The function
recurses down to level 2 then extracts the subformulas.
\end{callout}

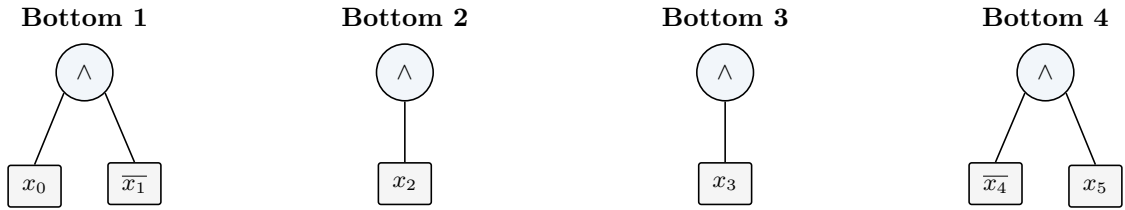
\begin{figure}[H]
  \centering
  \begin{minipage}[t]{0.23\textwidth}
    \centering
    \textbf{Bottom 1}

    \vspace{2mm}
    \begin{forest}
      decision tree,
      for tree={s sep=6mm},
      formula left edge/.style={edge path'={(!u.south west) -- (.north)}},
      formula right edge/.style={edge path'={(!u.south east) -- (.north)}}
      [$\land$, variable
        [$x_0$, leaf, formula left edge]
        [$\overline{x_1}$, leaf, formula right edge]
      ]
    \end{forest}
  \end{minipage}
  \hfill
  \begin{minipage}[t]{0.23\textwidth}
    \centering
    \textbf{Bottom 2}

    \vspace{2mm}
    \begin{forest}
      decision tree
      [$\land$, variable
        [$x_2$, leaf]
      ]
    \end{forest}
  \end{minipage}
  \hfill
  \begin{minipage}[t]{0.23\textwidth}
    \centering
    \textbf{Bottom 3}

    \vspace{2mm}
    \begin{forest}
      decision tree
      [$\land$, variable
        [$x_3$, leaf]
      ]
    \end{forest}
  \end{minipage}
  \hfill
  \begin{minipage}[t]{0.23\textwidth}
    \centering
    \textbf{Bottom 4}

    \vspace{2mm}
    \begin{forest}
      decision tree,
      for tree={s sep=6mm},
      formula left edge/.style={edge path'={(!u.south west) -- (.north)}},
      formula right edge/.style={edge path'={(!u.south east) -- (.north)}}
      [$\land$, variable
        [$\overline{x_4}$, leaf, formula left edge]
        [$x_5$, leaf, formula right edge]
      ]
    \end{forest}
  \end{minipage}
  \caption{The four Level-2 bottom subformulas drawn individually.}
  \label{fig:bottom-layer-subformulas}
\end{figure}

\begin{figure}[H]
  \centering
  \begin{forest}
    decision tree,
    for tree={s sep=8mm},
    formula left edge/.style={edge path'={(!u.south west) -- (.north)}},
    formula right edge/.style={edge path'={(!u.south east) -- (.north)}},
    before drawing tree={
      tikz+={
        \coordinate (skeleton-level-left) at (skeleton-xten.west);
        \coordinate (skeleton-level-right) at (skeleton-xthirteen.east);
        \coordinate (skeleton-root-left) at
          (skeleton-level-left |- skeleton-root.center);
        \coordinate (skeleton-root-right) at
          (skeleton-level-right |- skeleton-root.center);
        \coordinate (skeleton-or-left) at
          (skeleton-level-left |- skeleton-or-left-node.center);
        \coordinate (skeleton-or-right) at
          (skeleton-level-right |- skeleton-or-left-node.center);
        \begin{pgfonlayer}{background}
          \node[fill=paleblue, rounded corners=7pt,
                fit=(skeleton-root-left)(skeleton-root-right)(skeleton-root),
                inner xsep=16mm, inner ysep=3mm]
                (skeleton-level-four-box) {};
          \node[fill=paleteal, rounded corners=7pt,
                fit=(skeleton-or-left)(skeleton-or-right)
                    (skeleton-or-left-node)(skeleton-or-right-node),
                inner xsep=16mm, inner ysep=3mm]
                (skeleton-level-three-box) {};
          \node[fill=paleorange, rounded corners=7pt,
                fit=(skeleton-xten)(skeleton-xthirteen),
                inner xsep=16mm, inner ysep=3mm]
                (skeleton-level-two-box) {};
          \node[anchor=north west, font=\scriptsize\bfseries, text=muted]
            at ($(skeleton-level-four-box.north west)+(2mm,-1.5mm)$) {};
          \node[anchor=north west, font=\scriptsize\bfseries, text=muted]
            at ($(skeleton-level-three-box.north west)+(2mm,-1.5mm)$) {};
          \node[anchor=north west, font=\scriptsize\bfseries, text=muted]
            at ($(skeleton-level-two-box.north west)+(2mm,-1.5mm)$) {};
        \end{pgfonlayer}
      }
    }
    [$\land$, variable, name=skeleton-root
      [$\lor$, variable, formula left edge, name=skeleton-or-left-node
        [$x_{10}$, leaf, formula left edge, name=skeleton-xten]
        [$x_{11}$, leaf, formula right edge]
      ]
      [$\lor$, variable, formula right edge, name=skeleton-or-right-node
        [$x_{12}$, leaf, formula left edge]
        [$x_{13}$, leaf, formula right edge, name=skeleton-xthirteen]
      ]
    ]
  \end{forest}
  \caption{The top formula after replacing the four Level-2 bottoms with fresh inputs.}
  \label{fig:bottom-layer-top-skeleton}
\end{figure}
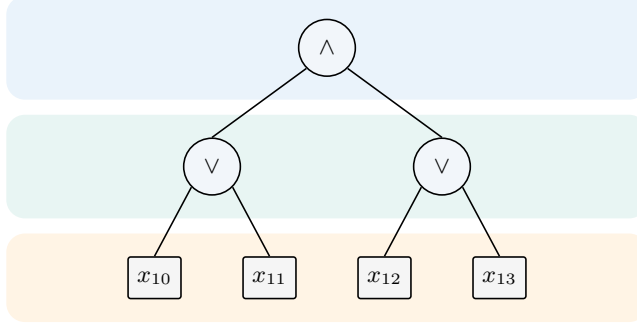

\noindent\begin{minipage}{\linewidth}
\begin{lstlisting}[caption={Bottom Layer Extraction},
                   mathescape=true,
                   label={lst:BottomLayerExtraction}]{}
mutual
def extractBottomLayer : Nat $\rightarrow$ Nat $\rightarrow$ UnboundedFanInFormula $\rightarrow$
    List UnboundedFanInFormula $\times$ UnboundedFanInFormula $\times$ Nat
  | lvl, start, .andGate gates =>
      if lvl $\leq$ 2 then
        ([.andGate gates], .inputGate start false, start + 1)
      else
        let r := extractBottomLayerList (lvl - 1) start gates
        (r.1, .andGate r.2.1, r.2.2)
  | lvl, start, .orGate gates =>
      if lvl $\leq$ 2 then
        ([.orGate gates], .inputGate start false, start + 1)
      else
        let r := extractBottomLayerList (lvl - 1) start gates
        (r.1, .orGate r.2.1, r.2.2)
  | _, start, other =>
      ([other], .inputGate start false, start + 1)

def extractBottomLayerList : Nat $\rightarrow$ Nat $\rightarrow$ List UnboundedFanInFormula $\rightarrow$
    List UnboundedFanInFormula $\times$ List UnboundedFanInFormula $\times$ Nat
  | _,   start, []      => ([], [], start)
  | lvl, start, g :: gs =>
      let r$_1$ := extractBottomLayer lvl start g
      let r$_2$ := extractBottomLayerList lvl r$_1$.2.2 gs
      (r$_1$.1 ++ r$_2$.1, r$_1$.2.1 :: r$_2$.2.1, r$_2$.2.2)
end
\end{lstlisting}
\end{minipage}

The next step processes the list of bottom gates returned by
\texttt{extractBottomLayer}.

\subsubsection{Conversion of AND Gates to DNFs}

The switching lemma as stated earlier is a statement about DNFs. Therefore, we
convert CNFs extracted from the bottom level of a formula into DNFs using the
\texttt{toBottomDNF} function in listing \ref{lst:toBottomDNF}.

\begin{lstlisting}[caption={Conversion of AND Gates to DNFs},
                   mathescape=true,
                   label={lst:toBottomDNF}]{}
def negLit : UnboundedFanInFormula $\rightarrow$ UnboundedFanInFormula
  | .inputGate i b => .inputGate i (!b)
  | g => g

def dualClause : UnboundedFanInFormula $\rightarrow$ UnboundedFanInFormula
  | .orGate lits => .andGate (lits.map negLit)
  | g => g

def cnfDual : UnboundedFanInFormula $\rightarrow$ UnboundedFanInFormula
  | .andGate clauses => .orGate (clauses.map dualClause)
  | g => g

def toBottomDNF : UnboundedFanInFormula $\rightarrow$ UnboundedFanInFormula
  | orGate terms => orGate terms
  | andGate clauses => cnfDual (andGate clauses)
  | _ => orGate []
\end{lstlisting}

\subsubsection{Killing Wide Clauses}

Next, we apply theorem \texttt{exists\_restriction\_width\_le\_card} (of listing
\ref{lst:Round0RestrictionWidthLECardOfBound}) to obtain a restriction
($\sigma$) that kills the wide clauses in the list of DNFs obtained from the
bottom level of the formula. This restriction can be decomposed into a set of
live variables, \texttt{live$\_0$} and a list of bits for the dead variables,
\texttt{dead$\_0$}.

\subsubsection{Reindexing Surviving Variables}

Since applying the wide-clause-killing restriction $\sigma$ kills variables, the
surviving variables need to be reindexed to start from 0 so that the resulting
formula is in canonical form for subsequent operations. We define a rank
function $\phi : \mathbb{N} \rightarrow \mathbb{N}$ that relabels live
variables.  We prove that $\phi$ is injective then use it to reindex the
surviving variables in the formula using the \texttt{ufiRestrictRekey} function
in listing \ref{lst:ufiRestrictRekey}. Every \texttt{inputGate i b} leaf in the
formula is replaced with a \texttt{constant} leaf if it refers to a variable
killed by $\sigma$ or an input referring to the variable $\phi(i)$,
i.e. \texttt{inputGate ($\phi$ i) b}.

\begin{lstlisting}[caption={Reindexing Surviving Variables},
                   mathescape=true,
                   label={lst:ufiRestrictRekey}]{}
set $\phi$ : Nat $\rightarrow$ Nat :=
  fun i =>
    (live$_0$.findIdx? ($\cdot$ = i)).getD i with h$\phi$def

def ufiRestrictRekey (asgn : Nat $\rightarrow$ Option Bool)
                     ($\phi$ : Nat $\rightarrow$ Nat) :
    UnboundedFanInFormula $\rightarrow$ UnboundedFanInFormula
  | UnboundedFanInFormula.inputGate i b =>
      match asgn i with
      | none     => UnboundedFanInFormula.inputGate ($\phi$ i) b
      | some bit => UnboundedFanInFormula.constant
                    (if b then
                      Bool.not bit
                    else
                      bit) 0
  | UnboundedFanInFormula.constant b lbl =>
      UnboundedFanInFormula.constant b lbl
  | UnboundedFanInFormula.notGate g  =>
      UnboundedFanInFormula.notGate (ufiRestrictRekey asgn $\phi$ g)
  | UnboundedFanInFormula.andGate gs =>
      UnboundedFanInFormula.andGate (gs.map (ufiRestrictRekey asgn $\phi$))
  | UnboundedFanInFormula.orGate gs  =>
      UnboundedFanInFormula.orGate (gs.map (ufiRestrictRekey asgn $\phi$))
\end{lstlisting}

Applying the restriction $\sigma$ introduces constants in place of dead
variables. A further invocation of \texttt{simplifyConstants} restores the
constant-free invariant required by subsequent operations.

\subsection{Depth Reduction Invariants}

Recall that we defined the switching lemma parameter $t$ such that the formula
size $cn^k < 2 ^ t$.  We define a function \texttt{switchingGateBudget} to
compute the number of gates involved in applying the switching lemma for depth
reduction at a given level. This will enable us to enforce the $cn^k < 2 ^ t$
requirement.

\noindent\begin{minipage}{\linewidth}
\begin{lstlisting}[caption={Switching Budget Definition},
                   mathescape=true,
                   label={lst:switchingGateBudget}]{}
def switchingGateBudget : Nat $\rightarrow$ UnboundedFanInFormula $\rightarrow$ Nat
  | _,   .inputGate _ _ => 0
  | _,   .constant _ _  => 0
  | lvl, .notGate g     => switchingGateBudget lvl g + 1
  | _,   .andGate []    => 0
  | _,   .orGate []     => 0
  | lvl, .andGate gates =>
      if lvl $\leq$ 2 then
        1
      else
        (gates.map (switchingGateBudget (lvl - 1))).sum + 1
  | lvl, .orGate gates =>
      if lvl $\leq$ 2 then
        1
      else
        (gates.map (switchingGateBudget (lvl - 1))).sum + 1
\end{lstlisting}
\end{minipage}

Once all the formula preprocessing steps described in the previous sections are
complete, we can prove that the following properties hold:

\begin{enumerate}
  \item{ The formula must be constant-free, i.e. it contains no
    \texttt{constant} leaf.  This ensures that every gate extracted from the
    bottom is either a proper CNF or DNF or an input literal (the switching
    lemma is stated for proper DNFs, and the final parity argument compares DNF
    width with the number of live variables so constants inside clauses would
    violate those interfaces).  }
  \item{ The formula must either be the canonical true or canonical false gate
    (an empty AND/OR gate) or it must have no empty AND/OR gates.  }
  \item{ Every gate at the bottom level of the formula must have fan-in of at
    most $t$, which is why we applied the initial restriction to kill wide
    clauses.  }
\end{enumerate}

Note that every application of the switching lemma during depth reduction
requires and maintains these properties as invariants.  The formal definition of
these properties is shown in listing \ref{lst:Round0FanInReductionProperties}.

\begin{lstlisting}[caption={Round 0 Fan-in Reduction Properties},
                   mathescape=true,
                   label={lst:Round0FanInReductionProperties}]{}
def HasNoEmptyAndOrGate : UnboundedFanInFormula $\rightarrow$ Prop
  | inputGate _ _ => True
  | constant _ _  => True
  | notGate g     => HasNoEmptyAndOrGate g
  | andGate gs    => gs $\neq$ [] $\land$ $\forall$ g $\in$ gs, HasNoEmptyAndOrGate g
  | orGate gs     => gs $\neq$ [] $\land$ $\forall$ g $\in$ gs, HasNoEmptyAndOrGate g

def IsCleanFormula (g : UnboundedFanInFormula) : Prop :=
  (g = andGate [] $\lor$ g = orGate []) $\lor$ HasNoEmptyAndOrGate g

def ufiBottomFanIn : UnboundedFanInFormula $\rightarrow$ Nat
  | .andGate gates => max 1 (cnfWidth (.andGate gates))
  | .orGate gates  => max 1 (dnfWidth (.orGate gates))
  | _ => 1

def HasBottomFanInLE (lvl : Nat)
                     (circuit : UnboundedFanInFormula)
                     (t : Nat) : Prop :=
  $\forall$ g $\in$ (extractBottomLayer lvl 0 circuit).1, ufiBottomFanIn g $\leq$ t
\end{lstlisting}

These properties are packaged into a single structure,
\texttt{SwitchingRoundState} for consumption by the iterated depth reduction
implementation.

\begin{lstlisting}[caption={Packaged Depth Reduction Invariant},
                   mathescape=true,
                   label={lst:SwitchingRoundState}]{}
structure SwitchingRoundState (n c k d t : Nat) where
  circuit       : LeveledUFIFormulaOfSizePolyNAndDepthD n c k d
  constant_free : IsConstantFree circuit.val
  clean         : IsCleanFormula circuit.val
  bottom_fan_in : HasBottomFanInLE d circuit.val t
  bottom_budget : switchingGateBudget d circuit.val < 2 ^ t
\end{lstlisting}

The lemma in listing \ref{lst:ExistsRound0FanInReducedOfParametersCore} proves
that applying the initial restriction to kill wide clauses is sufficient to set
up the initial \texttt{SwitchingRoundState} invariant.

\begin{lstlisting}[caption={Core Round 0 Width Reduction Lemma},
                   mathescape=true,
                   label={lst:ExistsRound0FanInReducedOfParametersCore}]{}
lemma exists_round_zero_fanIn_reduced_of_parameters_core
      {c k d n t s : Nat}
      (q           : $\mathbb{Q}$)
      (hd          : 1 $\leq$ d)
      (h_two_le_t  : 2 $\leq$ t)
      (h_count_m   : (((c * n ^ k : $\mathbb{N}$) : $\mathbb{Q}$) * q ^ (t + 1) < 1))
      (h_bot_m     : c * n ^ k < 2 ^ t)
      (h_thresh_m  : (20 * t) ^ (d - 2) * (40 * (t + 1)) $\leq$ s)
      (cal         : FaninReduction.RoundZeroCalibration q n s)
      (circuit     : LeveledUFIFormulaOfSizePolyNAndDepthD n c k d) :
  $\exists$ (live$_0$ : List Nat)
    (_h_live$_0$_lt : $\forall$ v $\in$ live$_0$, v < n)
    (_h_live$_0$_nodup : live$_0$.Nodup)
    (dead$_0$ : List Bool)
    (c' k' : Nat)
    (state : SwitchingRoundState live$_0$.length c' k' d t)
    (_h_thresh : (20 * t) ^ (d - 2) * (40 * (t + 1)) $\leq$ live$_0$.length),
    $\forall$ (liveBits : List Bool),
      liveBits.length = live$_0$.length $\rightarrow$
        ufiFormulaEval circuit.val (assembleInput n live$_0$ liveBits dead$_0$) =
        ufiFormulaEval state.circuit.val liveBits
\end{lstlisting}

\subsection{Repeated Application of Switching Restrictions and Depth Reduction}

Given a \texttt{SwitchingRoundState} and a valid live variable threshold, the
switching lemma is used to repeatedly reduce the depth of formula. The result of
this process is a DNF with at least 2 live variables whose width is less than
its live variable count while still computing the same function of the live
variables as the original formula with the same killing bits applied to reduce
the size of its domain.  This is proved by the
\texttt{exists\_iterated\_switching\_depth\_collapse\_sharp} lemma in listing
\ref{lst:exists_iterated_switching_depth_collapse}.  Note that the live variable
threshold (\texttt{h\_thresh}) is excluded from the fields of
\texttt{SwitchingRoundState} because it decreases according to how many rounds
are still to be performed (unlike the other fields of
\texttt{SwitchingRoundState}).

\noindent\begin{minipage}{\linewidth}
\begin{lstlisting}[caption={Iterated Switching Depth Collapse Lemma},
                   mathescape=true,
                   label={lst:exists_iterated_switching_depth_collapse}]{}
lemma exists_iterated_switching_depth_collapse_sharp
    {c k d n t  : Nat}
    (hd         : 1 $\leq$ d)
    (state      : SwitchingRoundState n c k d t)
    (h          : 2 $\leq$ t)
    (h_thresh   : (20 * t) ^ (d - 2) * (40 * (t + 1)) $\leq$ n) :
  $\exists$ (live           : List Nat)
    (_h_live_lt    : $\forall$ v $\in$ live, v < n)
    (_h_live_nodup : live.Nodup)
    (_h_live_big   : 2 $\leq$ live.length)
    (deadBits      : List Bool)
    (w             : Nat)
    (_hw           : w < live.length)
    (g             : UnboundedFanInDNF live.length),
      dnfWidth g.val $\leq$ w
        $\land$ $\forall$ (liveBits : List Bool),
            liveBits.length = live.length $\rightarrow$
            ufiFormulaEval state.circuit.val
                          (assembleInput n live liveBits deadBits) =
            ufiFormulaEval g.val liveBits
\end{lstlisting}
\end{minipage}

The round 0 restriction from the last subsection sets up the
\texttt{SwitchingRoundState} invariant required by this lemma. The proof is by
induction on the formula depth $d$.

\subsubsection{Depth 1 Case}

When $d=1$, the structural lemma
\texttt{exists\_depth\_one\_collapse\_of\_three\_le} requires only $n\geq3$. It
produces a narrow DNF on at least two live variables by case analysis on the
formula. The iterated reserve implies this input-count hypothesis, and this case
spends no switching round.

\begin{enumerate}
  \item If the formula is an input, then there exists a 1-literal DNF of width 1
    on the set of 2 or more live variables as required by the lemma.
  \item The formula cannot be a NOT of another formula because the switching
    invariant ensures there are no NOT gates in any of the formulas.
  \item \texttt{constant}s are represented by one of the canonical width-zero
    DNFs (\texttt{orGate []} for false and \texttt{orGate [andGate []]} for
    true).
  \item If the formula is an AND gate, $d = 1$ implies that its children are
    either inputs or constants. If any child of an AND gate is a
    \texttt{constant false \_} leaf, then the canonical width-zero DNF
    \texttt{orGate []} suffices for the proof. Otherwise, we kill one of the
    live variables (leaving $n - 1 \geq 2$ live variables) with a value that
    kills the AND gate in the original formula.
  \item If the formula is an OR gate, $d = 1$ also implies that its children are
    either inputs or constants as well. If any child of an OR gate is a
    \texttt{constant true \_} leaf, then the canonical width-zero DNF
    \texttt{orGate [andGate []]} suffices for the proof. Otherwise, we kill one
    of the live variables (leaving $n - 1 \geq 2$ live variables) with a value
    that satisfies the OR gate in the original formula.
\end{enumerate}

This case analysis enables us to obtain the set of live variables and dead bits
required by the depth 1 collapse lemma of listing \ref{lst:ExistsDepth1Collapse}
along with the properties required by the lemma.

\noindent\begin{minipage}{\linewidth}
\begin{lstlisting}[caption={Depth 1 Depth Collapse Lemma},
                   mathescape=true,
                   label={lst:ExistsDepth1Collapse}]{}
lemma exists_depth_one_collapse_of_three_le
      {c k n : Nat}
      (circuit : LeveledUFIFormulaOfSizePolyNAndDepthD n c k 1)
      (h_three_le_n : 3 $\leq$ n) :
  $\exists$ (live           : List Nat)
    (_h_live_lt    : $\forall$ v $\in$ live, v < n)
    (_h_live_nodup : live.Nodup)
    (_h_live_big   : 2 $\leq$ live.length)
    (deadBits      : List Bool)
    (w             : Nat)
    (_hw           : w < live.length)
    (g             : UnboundedFanInDNF live.length),
      dnfWidth g.val $\leq$ w
        $\land$ $\forall$ (liveBits : List Bool),
            liveBits.length = live.length $\rightarrow$
                ufiFormulaEval circuit.val
                              (assembleInput n live liveBits deadBits) =
                ufiFormulaEval g.val liveBits
\end{lstlisting}
\end{minipage}

\subsubsection{Induction Base Case: Depth 2}

The goal is to find a live-set size $L\geq2$ and a DNF of width less than $L$
such that the DNF agrees with the restricted depth-at-most-two formula. We take
$L=2$, so it suffices to produce a restricted DNF of width at most one.

For a proper DNF of width at most $w$, apply the switching lemma with
$\sigma=2/n$ and decision-tree cutoff one. The bad-restriction bound is
\[
  (10\sigma w)^1=20w/n<1.
\]
The theorem \texttt{exists\_switching\_lemma\_pigeonhole\_exact}
(Listing~\ref{lst:exists_switching_lemma_pigeonhole_exact}) then provides a good
restriction leaving exactly two variables live and a canonical decision tree of
depth at most one. Here the parameter \texttt{d} in the switching lemma is
instantiated with one, independently of the source formula's depth.  A proper
CNF is handled through its DNF dual.

The terminal hypothesis $20(w+1)<n$ supplies this strict probability bound and
makes $\sigma=2/n$ an admissible density. When $w=t$, the iterated reserve
$40(t+1)\leq n$ implies this hypothesis directly, as explained in
Section~\ref{subsec:LiveVariableSufficiency}.

\begin{lstlisting}[caption={Switching Lemma Bound Implying Existence of a Good Restriction},
                   mathescape=true,
                   label={lst:exists_switching_lemma_pigeonhole_exact}]{}
lemma exists_switching_lemma_pigeonhole_exact
    {n         : Nat}
    (w d       : Nat)
    (f         : UnboundedFanInProperDNF n)
    (hwidth    : dnfWidth f.val $\leq$ w)
    ($\sigma$         : OpenUnitIntervalQ)
    (h$\sigma$        : $\sigma$.val $\leq$ 1 / 5)
    (h_s_exact : (Nat.ceil ($\sigma$.val * (n : $\mathbb{Q}$)) : $\mathbb{Q}$) = $\sigma$.val * (n : $\mathbb{Q}$))
    (hbound    : (10 * $\sigma$.val * (w : $\mathbb{Q}$)) ^ d < 1) :
  $\exists$ $\rho$ : AssignedRandomRestriction $\sigma$ n,
    $\lnot$ isBadRestriction d n $\sigma$ f $\rho$
\end{lstlisting}

The depth-at-most-2 case is then expressed as the
\texttt{exists\_depth\_two\_collapse} lemma in listing
\ref{lst:ExistsDepth2Collapse}.

\noindent\begin{minipage}{\linewidth}
\begin{lstlisting}[caption={Depth 2 Depth Collapse Lemma},
                   mathescape=true,
                   label={lst:ExistsDepth2Collapse}]{}
lemma exists_depth_two_collapse
      {c k n          : Nat}
      (circuit        : LeveledUFIFormulaOfSizePolyNAndDepthD n c k 2)
      (h_inputs_bound : ufiLargestInput circuit.val < n)
      (w              : Nat)
      (h_node_w       : ufiBottomFanIn circuit.val $\leq$ w)
      (h_thresh       : 20 * (w + 1) < n) :
  $\exists$ (live           : List Nat)
    (_h_live_lt    : $\forall$ v $\in$ live, v < n)
    (_h_live_nodup : live.Nodup)
    (_h_live_big   : 2 $\leq$ live.length)
    (deadBits      : List Bool)
    (w             : Nat)
    (_hw           : w < live.length)
    (g             : UnboundedFanInDNF live.length),
      dnfWidth g.val $\leq$ w
        $\land$ $\forall$ (liveBits : List Bool),
            liveBits.length = live.length $\rightarrow$
              ufiFormulaEval circuit.val
                            (assembleInput n live liveBits deadBits) =
              ufiFormulaEval g.val liveBits
\end{lstlisting}
\end{minipage}

The formula in the lemma is properly leveled, which implies that it
\texttt{HasProperBottoms} at level 2. The liveness threshold of $20(w + 1) < n$
also implies that $n > 20$. Using these facts, we prove this lemma by a case
analysis on the formula once again.

\begin{enumerate}
  \item For an input literal, retain its coordinate and one distinct
    coordinate. Rekeying gives a DNF of width one on two live variables.
  \item A NOT gate is excluded by the proper-leveling invariant.
  \item For a constant, retain any two coordinates and use the corresponding
    width-zero DNF, \texttt{orGate []} or \texttt{orGate [andGate []]}.
  \item For an OR gate, properness gives a DNF whose width is at most $w$ by
    \texttt{h\_node\_w}. The switching restriction with $\sigma=2/n$ and cutoff
    one gives a decision tree of depth at most one.  Converting it to a DNF
    yields width at most one, strictly less than the two remaining live
    variables.
  \item For an AND gate, take the DNF dual computing the negation and apply the
    same restriction argument. Flip the decision tree's leaves to recover the
    original function, preserving its depth. The resulting DNF again has width
    at most one on two live variables.
\end{enumerate}

\subsubsection{Inductive Hypothesis Case}

When $d > 2$, we use the switching lemma to combine the bottom two levels of the
formula. The first step is to extract the bottom AND/OR gates of the formula
using the \texttt{extractBottomLayer} function. Since other gate types like
inputs are ignored by this step, we define the inductive type
\texttt{BottomFormula} to indicate this restriction and simplify the tracking of
the polarity of the extracted gates. The list of extracted gates is converted
into a list of \texttt{BottomFormula}s.

\noindent\begin{minipage}{\linewidth}
\begin{lstlisting}[caption={Representing the Bottom Formulas},
                   mathescape=true,
                   label={lst:BottomFormula}]{}
inductive BottomFormula (n : Nat) : Type
  | dnf (f : UnboundedFanInProperDNF n) : BottomFormula n
  | cnf (f : UnboundedFanInProperCNF n) : BottomFormula n

def BottomFormula.polarity {n : Nat} : BottomFormula n $\rightarrow$ Bool
  | .dnf _ => true
  | .cnf _ => false

def BottomFormula.toUFI {n : Nat} : BottomFormula n $\rightarrow$ UnboundedFanInFormula
  | .dnf f => f.val
  | .cnf f => f.val

def BottomFormula.width {n : Nat} : BottomFormula n $\rightarrow$ Nat
  | .dnf f => dnfWidth f.val
  | .cnf f => cnfWidth f.val
\end{lstlisting}
\end{minipage}

The list of \texttt{BottomFormula}s is then converted to a list of DNFs.  The
\texttt{exists\_bottomFormula\_dnf\_view} lemma in listing
\ref{lst:exists_bottomFormula_dnf_view} is used in the conversion.  It takes
advantage of the \texttt{exists\_properDNF\_dual\_of\_properCNF} lemma, which
proves that every CNF has a DNF dual of the same width.

\begin{lstlisting}[caption={Converting Bottom Formulas to DNFs},
                   mathescape=true,
                   label={lst:exists_bottomFormula_dnf_view}]{}
lemma exists_properDNF_dual_of_properCNF
    {n : Nat} (f : UnboundedFanInProperCNF n) :
  $\exists$ (fd : UnboundedFanInProperDNF n),
    dnfWidth fd.val = cnfWidth f.val
    $\land$ $\forall$ xs,
        xs.length = n $\rightarrow$
        ufiFormulaEval fd.val xs = not (ufiFormulaEval f.val xs)

lemma exists_bottomFormula_dnf_view {n : Nat} (f : BottomFormula n) :
  $\exists$ (fv : UnboundedFanInProperDNF n),
    dnfWidth fv.val $\leq$ f.width
    $\land$ $\forall$ xs,
        xs.length = n $\rightarrow$
          ufiFormulaEval fv.val xs =
          (if f.polarity then
            ufiFormulaEval f.toUFI xs
          else
            not (ufiFormulaEval f.toUFI xs))
\end{lstlisting}

For the list of bottom DNFs, use the density $\sigma=\lfloor
n/(20t)\rfloor/n\leq1/(20t)$.  The lemma in
Listing~\ref{lst:exists_switching_round_restriction} finds a common good
restriction, with decision-tree depth at most $t$ for every DNF in the list. Its
local hypothesis $20t(t+1)\leq n$ ensures that more than $t$ variables
survive. At every stage of depth greater than two, the iterated reserve implies
this quadratic local hypothesis.

\noindent\begin{minipage}{\linewidth}
\begin{lstlisting}[caption={Existence of a Good Restriction for Recursive Depth Reduction},
                   mathescape=true,
                   label={lst:exists_switching_round_restriction}]{}
lemma exists_switching_round_restriction
      (c k t  : Nat)
      {n      : Nat}
      (fs     : List (UnboundedFanInProperDNF n))
      (ht     : 2 $\leq$ t)
      (hwidth : $\forall$ f $\in$ fs, dnfWidth f.val $\leq$ t)
      (hcount : fs.length $\leq$ c * n ^ k)
      (h_t_s  : c * n ^ k < 2 ^ t)
      (hn     : 20 * t * (t + 1) $\leq$ n) :
  $\exists$ ($\sigma$ : OpenUnitIntervalQ)
    ($\rho$ : AssignedRandomRestriction $\sigma$ n),
      ($\forall$ f $\in$ fs, $\lnot$ isBadRestriction t n $\sigma$ f $\rho$)
      $\land$ t < Nat.ceil ($\sigma$.val * (n : $\mathbb{Q}$))
      $\land$ $\sigma$.val = ((n / (20 * t) : Nat) : $\mathbb{Q}$) / (n : $\mathbb{Q}$)
\end{lstlisting}
\end{minipage}

Similar to the base case, each bounded depth decision tree created from the
bottom layer can be converted into a CNF/DNF or a negated CNF/DNF of width
bounded by the decision tree depth. This is proved by the lemma in listing
\ref{lst:decisionTreeToDNFWidthLETreeDepth}. See
\ref{subsec:DecisionTreeAppendix} for the function converting a decision tree to
a DNF. The resulting decision-tree depth is at most $t$. This cutoff is
maintained through the nonterminal rounds. The depth-two endpoint uses cutoff
one.

\begin{lstlisting}[caption={A DNF from a Decision Tree Has Width Bounded by Decision Tree Height},
                   mathescape=true,
                   label={lst:decisionTreeToDNFWidthLETreeDepth}]{}
lemma decisionTreeToDNF_dnfWidth_le_decisionTreeDepth
    (tree : DecisionTree) :
  dnfWidth (decisionTreeToDNF tree)
           $\leq$ decisionTreeDepth tree
\end{lstlisting}

We also need to show that every variable in the canonical decision tree is live
as shown by the lemma in listing \ref{lst:CanonicalDecisionTreeVarsAllLive}.

\begin{lstlisting}[caption={Liveness of Canonical Decision Tree Variables},
                   mathescape=true,
                   label={lst:CanonicalDecisionTreeVarsAllLive}]{}
lemma mem_live_of_mem_dtCollectInputIndices_canonicalDecisionTree
    {n : Nat} {$\sigma$ : OpenUnitIntervalQ}
    (f : UnboundedFanInProperDNF n)
    ($\rho$ : AssignedRandomRestriction $\sigma$ n)
    (live : List Nat)
    (_h_live_eq : (live : Multiset Nat) = $\rho$.starAssignment.val.val.val) :
  $\forall$ i $\in$ DecisionTrees.dtCollectInputIndices
      (canonicalDecisionTree f.val
        (mkAssignmentList $\rho$.starAssignment.val.val $\rho$.varAssignments n)),
    i $\in$ live
\end{lstlisting}

These two properties underpin the argument that CNFs/DNFs with the desired width
upper bound of $t$ can be derived from the canonical decision trees created from
the bottom layer.  Just like in the round 0 step (which killed wide clauses),
the variable indices in the formula need to be adjusted after applying the
restriction. This ensures that all variables in the generated formula have
indices in the range \texttt{[0, live.length)}, the smaller domain on which the
  output formula is defined.  Finally, the bounded width CNFs/DNFs are
  substituted into the skeleton that was returned by
  \texttt{extractBottomLayer}.

The \texttt{exists\_switching\_depth\_reduction} lemma below encapsulates the
composition of all these operations to perform a depth reduction step in the
induction.

\noindent\begin{minipage}{\linewidth}
\begin{lstlisting}[caption={Recursive Reduction of Switching Depth},
                   mathescape=true,
                   label={lst:exists_switching_depth_reduction}]{}
lemma exists_switching_depth_reduction {c k d n t    : Nat}
      (hd           : 2 $\leq$ d)
      (state        : SwitchingRoundState n c k (d + 1) t)
      (ht           : 2 $\leq$ t)
      (h_and_or     : UnboundedFanInFormula.IsAndOr state.circuit.val)
      (h_ne_circuit : HasNoEmptyAndOrGate state.circuit.val)
      (h_thresh     : 20 * t * (t + 1) $\leq$ n) :
  $\exists$ (live           : List Nat)
    (_h_live_lt    : $\forall$ v $\in$ live, v < n)
    (_h_live_nodup : live.Nodup)
    (deadBits      : List Bool)
    (c' k'         : Nat)
    (next          : SwitchingRoundState live.length c' k' d t)
    (_h_live'      : n / (20 * t) $\leq$ live.length),
    $\forall$ (liveBits : List Bool),
      liveBits.length = live.length $\rightarrow$
        ufiFormulaEval state.circuit.val
                       (assembleInput n live liveBits deadBits) =
        ufiFormulaEval next.circuit.val liveBits
\end{lstlisting}
\end{minipage}

\subsection{Obtaining a Narrow Proper DNF from Depth Reduction}

Once depth reduction is complete, the restriction that was used to kill wide
clauses in round 0 and the one used by the inductive depth reduction are
combined to give a single restriction that is valid for the entire process. From
this single restriction, we derive the set of live variables and the DNF (of
narrow width) on these live variables. The lemma in listing
\ref{lst:exists_good_restriction_reduces_ac0_to_narrow_dnf} shows that the good
restriction formed from the aforementioned combination yields a DNF of width
narrower than the live variable count.  For our PARITY lower bound, we invoke it
with the parameters $q = 2/3$ and $s = n/3$.

\begin{lstlisting}[caption={A Good Restriction Implies a Reduction to a Narrow \texorpdfstring{$\AC^0$}{AC0} Circuit},
                   mathescape=true,
                   label={lst:exists_good_restriction_reduces_ac0_to_narrow_dnf}]{}
lemma exists_good_restriction_reduces_ac0_to_narrow_dnf_of_parameters_core
      {c k d n t s : Nat}
      (q          : $\mathbb{Q}$)
      (hd         : 1 $\leq$ d)
      (ht         : 2 $\leq$ t)
      (h_count_m  : (((c * n ^ k : $\mathbb{N}$) : $\mathbb{Q}$) * q ^ (t + 1) < 1))
      (h_bot_m    : c * n ^ k < 2 ^ t)
      (h_thresh_m : (20 * t) ^ (d - 2) * (40 * (t + 1)) $\leq$ s)
      (cal        : FaninReduction.RoundZeroCalibration q n s)
      (formula    : LeveledUFIFormulaOfSizePolyNAndDepthD n c k d) :
  $\exists$ (live           : List Nat)
    (deadBits      : List Bool)
    (_h_live_lt    : $\forall$ v $\in$ live, v < n)
    (_h_live_nodup : live.Nodup)
    (_h_card       : deadBits.length + live.length = n)
    (g             : UnboundedFanInProperDNF live.length),
      dnfWidth g.val < live.length
        $\land$ $\forall$ (liveBits : List Bool), liveBits.length = live.length $\rightarrow$
              ufiFormulaEval formula.val
                            (assembleInput n live liveBits deadBits) =
              ufiFormulaEval g.val liveBits
\end{lstlisting}

\subsection{Formula Size Lower Bounds for PARITY}

The reduction of the original formula to a DNF of narrow width relative to the
number of remaining live variables is a key ingredient in the PARITY lower
bounds proof. The depth reduction process yielded a restriction that killed
enough variables to give us this property. To obtain the PARITY lower bounds:

\begin{enumerate}
  \item Apply the \texttt{exists\_offset\_odd\_countP\_assembleInput\_iff} lemma
    of Section~\ref{subsec:ParityUnderRestrictions} to determine the Boolean
    offset required for the formula to compute the parity of the remaining bits.
  \item Use the \texttt{narrow\_dnf\_misclassifies\_parity} theorem in listing
    \ref{lst:narrow_dnf_misclassifies_parity} to find a live assignment on which
    the narrow DNF disagrees with the PARITY function.
  \item Assemble the full inputs on which the narrow DNF and the PARITY function
    disagree.
\end{enumerate}

The depth-reduction iteration uses the reserve
\[
  R_d(t)=(20t)^{d-2}\,40(t+1).
\]
At depth two the terminal collapse needs only $20(t+1)<m$ live variables.  Each
preceding switching round retains at least $\lfloor m/(20t)\rfloor$
variables. At every nonterminal stage, $R_d(t)$ also implies the local quadratic
reserve $20t(t+1)\leq m$. The theorem
\texttt{exists\_iterated\_switching\_depth\_collapse\_sharp} therefore replaces
the earlier quadratic terminal reserve by a linear one.

The lemma below uses these steps to prove the existence of inputs where this DNF
disagrees with PARITY.

\begin{lstlisting}[caption={constant Depth Formulas Misclassify PARITY},
                   mathescape=true,
                   label={lst:ConstantDepthFormulasMisclassifyPARITY}]{}
lemma parity_misclassified_of_explicit_leveled_bounds_sharp
      {c k d n t    : Nat}
      (hd           : 1 $\leq$ d)
      (h_three_le_n : 3 $\leq$ n)
      (ht           : 2 $\leq$ t)
      (h_count_m    : (((c * n ^ k : $\mathbb{N}$) : $\mathbb{Q}$) * (2 / 3) ^ (t + 1) < 1))
      (h_bot_m      : c * n ^ k < 2 ^ t)
      (h_thresh_m   : (20 * t) ^ (d - 2) * (40 * (t + 1)) $\leq$ n / 3)
      (formula      : LeveledUFIFormulaOfSizePolyNAndDepthD n c k d) :
  $\exists$ (inputs : List Bool),
    inputs.length = n $\land$
      ((ufiFormulaEval formula.val inputs == false
        $\land$ Odd (inputs.countP ($\cdot$ == true)))
      $\lor$
        (ufiFormulaEval formula.val inputs == true
        $\land$ $\lnot$ Odd (inputs.countP ($\cdot$ == true))))
\end{lstlisting}

The maximum bottom fan-in $t$ remains a free parameter in depth reduction.  The
next theorem shows that a formula computing PARITY (or its complement) has size
at least $2^r$ whenever $r\leq t$ and the stated calibration inequalities hold.
Its proof uses the lemma in listing
\ref{lst:ConstantDepthFormulasMisclassifyPARITY} above to find an input on which
the constant-depth formula and PARITY disagree whenever
$\texttt{ufiFormulaCircuitSize formula} < 2^r$.

\noindent\begin{minipage}{\linewidth}
\begin{lstlisting}[caption={parityBit Lower Bound for Formulas},
                   mathescape=true,
                   label={lst:FormulaParitySizeLowerBoundOfOneThird}]{}
theorem leveled_formula_parity_size_lower_bound_binary_of_parameters_sharp
    (n d r t   : Nat)
    (ht        : 2 $\leq$ t)
    (h_decay   : (2 : $\mathbb{Q}$) ^ r * (2 / 3 : $\mathbb{Q}$) ^ (t + 1) < 1)
    (h_thresh  : (20 * t) ^ (d - 2) * (40 * (t + 1)) $\leq$ n / 3)
    (formula   : UnboundedFanInFormula)
    (h_inputs  : ufiLargestInput formula < n)
    (h_depth   : ufiFormulaDepth formula $\leq$ d)
    (h_leveled : Circuits.Leveling.IsProperlyLeveled formula d)
    (h_parity  : FormulaComputesParity n formula) :
  (2 : $\mathbb{Q}$) ^ r $\leq$ (ufiFormulaCircuitSize formula : $\mathbb{Q}$)
\end{lstlisting}
\end{minipage}

For the final PARITY formula-size lower bound, set
$r=\operatorname{nthRoot}_{d-1}\!\left(\left\lfloor
n/(360\cdot40^{d-2})\right\rfloor\right)$ for $d\geq2$ and $r\geq1$.  With
$t=2r$, the decay condition holds because $2^r(2/3)^{2r+1}=(2/3)(8/9)^r<1$, and
the reserve fits because $3R_d(2r)\leq360\cdot40^{d-2}r^{d-1}\leq n$.  The
resulting lower bound is $2^r$, as stated in
Listing~\ref{lst:FormulaParitySizeLeveledLowerBoundRoot}.  The linear base-case
reserve makes the sufficient live-variable threshold have degree $d-1$ in $r$,
as calculated in Section~\ref{subsec:LiveVariableSufficiency}. Consequently, for
fixed $d$, $r$ grows as $n^{1/(d-1)}$, giving the PARITY size lower bound
$\exp(\Omega_d(n^{1/(d-1)}))$.

\begin{lstlisting}[caption={parityBit Lower Bound for Leveled Formulas},
                   mathescape=true,
                   label={lst:FormulaParitySizeLeveledLowerBoundRoot}]{}
theorem leveled_formula_parity_size_lower_bound_root_sharp
    (n d : Nat) (hd : 2 $\leq$ d)
    (hn        : 360 * 40 ^ (d - 2) $\leq$ n)
    (formula   : UnboundedFanInFormula)
    (h_inputs  : ufiLargestInput formula < n)
    (h_depth   : ufiFormulaDepth formula $\leq$ d)
    (h_leveled : Circuits.Leveling.IsProperlyLeveled formula d)
    (h_parity  : FormulaComputesParity n formula) :
  (2 : $\mathbb{Q}$) ^ Nat.nthRoot (d - 1) (n / (360 * 40 ^ (d - 2))) $\leq$
    (ufiFormulaCircuitSize formula : $\mathbb{Q}$)
\end{lstlisting}

For a general formula of size $S$, apply the explicit normalization estimate
with $c=S$ and $k=0$. The leveled lower bound gives
\[
  2^r\leq80(S+1)^2n^2.
\]
This is the conclusion of
\texttt{formula\_parity\_size\_lower\_bound\_root\_of\_large\_sharp}.  For each
fixed $d\geq2$, the root exponent $r$ eventually dominates $\log n$. Absorbing
the normalization overhead therefore gives $2^{\lfloor r/4\rfloor}\leq S$, as
stated below.

\begin{lstlisting}[caption={parityBit Lower Bound for Formulas},
                   mathescape=true,
                   label={lst:FormulaParitySizeLowerBoundRoot}]{}
theorem formula_parity_size_lower_bound_root_sharp
    (d : Nat) (hd : 2 $\leq$ d) :
  $\exists$ N, $\forall$ n, N $\leq$ n $\rightarrow$
    $\forall$ (formula : UnboundedFanInFormula),
      ufiLargestInput formula < n $\rightarrow$
      ufiFormulaDepth formula $\leq$ d $\rightarrow$
      FormulaComputesParity n formula $\rightarrow$
      2 ^ (Nat.nthRoot (d - 1)
          (n / (360 * 40 ^ (d - 2))) / 4) $\leq$
        ufiFormulaCircuitSize formula
\end{lstlisting}

From this result, it follows that for all sufficiently large $n$, constant-depth
polynomial-size $\AC^0$ formulas do not compute PARITY. This is proved as the
\texttt{hastad\_parity\_lower\_bound\_general\_direct} theorem.

\noindent\begin{minipage}{\linewidth}
\begin{lstlisting}[caption={parityBit Lower Bound for Formulas: Formulation 1},
                   mathescape=true,
                   label={lst:ParityLowerBoundFormulas2}]{}
lemma hastad_parity_lower_bound_general_direct (c k d : Nat) :
  $\exists$ n$_0$, $\forall$ n,
    n$_0$ < n $\rightarrow$
      $\forall$ (circuit : UFIFormulaOfSizeAtMostPolyNAndDepthAtMostD n c k d),
        $\exists$ (inputs : List Bool),
          inputs.length = n $\land$
          ((ufiFormulaEval circuit.val inputs == false $\land$
              Odd (inputs.countP ($\cdot$ == true)))
            $\lor$
            (ufiFormulaEval circuit.val inputs == true $\land$
              $\lnot$ Odd (inputs.countP ($\cdot$ == true))))
\end{lstlisting}
\end{minipage}

Using this result, it follows that any family of polynomial size $\AC^0$
formulas cannot compute the PARITY function. This is proved as the
\texttt{parity\_does\_not\_have\_ac0\_formulas} theorem.

\begin{lstlisting}[caption={parityBit Lower Bound for Formulas: Formulation 2},
                   mathescape=true,
                   label={lst:ParityDoesNotHaveAC0Formulas}]{}
theorem parity_does_not_have_ac0_formulas :
  $\forall$ (c k d : Nat), $\exists$ N,
    $\forall$ n (ac0Family : AC0FormulaFamily c k d),
      n.val > N $\rightarrow$
        $\lnot$ FormulaComputesParity n.val (ac0Family n).val

\end{lstlisting}

\section{Circuit Lower Bounds from Formula Lower Bounds}

This section proves the quantitative circuit lower bound stated in
Listing~\ref{lst:MainQuantitativeParityCircuitLowerBound}. The argument uses the
formula lower bound established in the preceding section, where the tree
structure supports recursive normalization and restriction transformations.
Unfolding a circuit into an equivalent formula duplicates shared gates.
Controlling this increase in size, together with the resulting formula depth,
transfers the lower bound to circuits. The exclusion of polynomial-size,
constant-depth PARITY circuits and circuit families then follows as a corollary.

The circuit model consists of finite, topologically ordered directed acyclic
graphs with unbounded-fan-in AND and OR gates, unary NOT gates, and separate
output vertices. Positive and negated inputs have canonical identifiers, and
shared internal gates are permitted. Circuit size counts each non-input gate
once, including NOT gates and output vertices. Circuit depth is the maximum
number of edges on a path ending at an output vertex. These are the conventions
used in both the quantitative theorem and its asymptotic corollaries.

A circuit consists of a finite list of gates and a finite list of directed
edges. Each gate has a natural-number identifier and a type. An edge from
\lstinline|src| to \lstinline|dst| represents a wire from the source gate's
output to one input of the destination gate.

\noindent\begin{minipage}{\linewidth}
\begin{lstlisting}[caption={Defining DAG Circuits},
                   mathescape=true,
                   label={lst:DAGCircuitDefs}]{}
inductive GateType where
  | input (negated : Bool) : GateType
  | output : GateType
  | andGate : GateType
  | orGate : GateType
  | notGate : GateType
  deriving Repr, BEq, DecidableEq

structure Gate where
  id : Nat
  type : GateType
  deriving Repr, BEq, DecidableEq

structure Edge where
  src : Nat
  dst : Nat
  deriving Repr, BEq, DecidableEq

structure Circuit where
  gates : List Gate
  edges : List Edge
  deriving Repr
\end{lstlisting}
\end{minipage}

\subsection{Well-Formed Circuits}

The gate and edge lists are constrained by a well-formedness proposition.  A
circuit $c$ is well formed if the following properties hold.

\begin{enumerate}
  \item It has no duplicate gate identifiers. This is expressed as \lstinline|c.gateIds.Nodup|.
  \item{ It has closed edges (wires). The source and destination gates of any
    wire must both be gates with IDs in the circuit.  }
  \item{
    It has no duplicate edges.
  }
  \item{
    The circuit has no cycles.
  }
  \item{
    All input gates have a fan-in of 0.
  }
  \item{
    Every NOT gate has a fan-in of 1.
  }
  \item{
    Every output gate has a fan-in of 1.
  }
  \item{
    Every output gate has fan-out zero.
  }
  \item{
    The gates all have consecutive ids.
  }
  \item{
    The graph is topologically sorted.
  }
  \item{ There is a path from every non-input gate in the circuit to an output
    gate. This requirement ensures there are no dangling gates.  }
  \item{ The input ids are ordered canonically with the first $n$ ids starting
    from 0 assigned to the non-negated inputs and the next $n$ ids starting from
    $n$ assigned to the negated inputs.  }
  \item{ There is at least one path from an input gate to an output gate. This
    excludes empty, disconnected, and constant-only circuits.  }
\end{enumerate}

\begin{lstlisting}[caption={Well Formed DAG Circuits},
                   mathescape=true,
                   label={lst:WellFormed}]{}
structure WellFormed (c : Circuit) : Prop where
  unique_ids   : c.gateIds.Nodup
  edges_closed : $\forall$ e $\in$ c.edges, e.src $\in$ c.gateIds $\land$ e.dst $\in$ c.gateIds
  edges_nodup  : c.edges.Nodup
  acyclic      : c.IsAcyclic

  fanin_input  : $\forall$ g $\in$ c.gates, g.type.isInput = true $\rightarrow$ c.fanIn g.id = 0
  fanin_not    : $\forall$ g $\in$ c.gates, g.type = GateType.notGate $\rightarrow$ c.fanIn g.id = 1
  fanin_output : $\forall$ g $\in$ c.gates, g.type = GateType.output $\rightarrow$
                                c.fanIn g.id = 1
  fanout_output : $\forall$ g $\in$ c.gates, g.type = GateType.output $\rightarrow$
                                c.fanOut g.id = 0

  cons_ids     : $\forall$ k (hk : k < c.gates.length), (c.gates[k]'hk).id = k
  topo         : $\forall$ e $\in$ c.edges, e.src < e.dst

  non_input_reaches_output :
    $\forall$ g $\in$ c.gates,
      g.type.isInput = false $\rightarrow$ g.type = GateType.output
      $\lor$ $\exists$ o $\in$ c.gates, o.type = GateType.output $\land$ c.Reachable g.id o.id

  has_canonical_input_ids :
    let posIds := (c.gates.filter
                    (fun g => g.type == GateType.input false)).map Gate.id
    let negIds := (c.gates.filter
                    (fun g => g.type == GateType.input true)).map Gate.id
    let n := max posIds.length negIds.length
    posIds = List.range n
      $\land$
    negIds = (List.range n).map (fun i => n + i)

  input_reaches_output :
    $\exists$ i $\in$ c.gates, i.type.isInput = true $\land$
      $\exists$ o $\in$ c.gates, o.type = GateType.output $\land$ c.Reachable i.id o.id
\end{lstlisting}

\subsection{Families of General Circuits}

For fixed natural numbers $c$, $k$, and $d>0$, the subtype below contains
well-formed $n$-input circuits with size at most $c n^k$ and depth at most $d$.
A circuit family chooses one such circuit for each positive input length.  No
uniformity condition is imposed on this choice.

\noindent\begin{minipage}{\linewidth}
\begin{lstlisting}[caption={AC$^0$ Circuit Family},
                   mathescape=true,
                   label={lst:AC0Circuit}]{}
def UFICircuitOfSizeAtMostPolyNAndDepthAtMostD
  (n c k d : Nat) :=
  {
    circuit : Circuit //
    circuit.WellFormed
    $\land$ circuit.inputWidth = n
    $\land$ circuit.depth $\leq$ d
    $\land$ circuit.circuitSize $\leq$ c * n ^ k
    $\land$ d > 0
  }

def AC0CircuitFamily (c k d : Nat) :=
  (n : PNat) ->
    UFICircuitOfSizeAtMostPolyNAndDepthAtMostD n c k d
\end{lstlisting}
\end{minipage}

\subsection{Unfolding Circuits into Formulas}

For a gate identifier $k$, \lstinline|Circuit.toUFIByPos|
(Listing~\ref{lst:toUFIByPos} in the appendix) recursively replaces the gate by
a formula for the subcircuit feeding it. Input gates become formula inputs, NOT
gates recurse on their unique predecessor, and AND and OR gates recurse over
their incoming edges.  Output gates recurse on their unique input.  The
recursion carries a fuel argument. Topological ordering ensures that each
recursive call moves to a smaller gate identifier. We prove that, for a valid
gate identifier $k$, any two fuel values greater than $k$ produce the same
formula.

The conversion preserves evaluation and uses only indices below the circuit's
input width. To preserve computation depth, we additionally apply
\lstinline|normalizeNullary|, which replaces empty AND and OR gates by constant
leaves. This preserves node count, input occurrences, and evaluation, and gives
formula depth at most the corresponding gate depth. Removing the unique output
wire then gives formula depth at most $d$ when \lstinline|circuit.depth| $\leq
d+1$.

If the circuit has $G$ total vertices, including inputs, this formula has at
most $(G+1)^{d+3}$ nodes. This bound accounts for duplicated subformulas created
when shared circuit gates are unfolded. Applying the formula lower bound at
depth $d$ and absorbing this polynomial overhead gives
$\exp(\Omega_d(n^{1/(d-1)}))$ for each fixed computation depth $d\geq2$.

\subsection{Constant-Depth Circuit Lower Bound for PARITY}

A well-formed DAG circuit computes PARITY when its output list is the singleton
containing the parity bit on every input of the declared width.

\begin{lstlisting}[caption={What it Means for a DAG Circuit to Compute parityBit},
                   mathescape=true,
                   label={lst:CircuitComputesParity}]{}
def CircuitComputesParity (n : Nat) (c : Circuit) : Prop :=
  c.WellFormed $\land$
  $\forall$ inputs : List Bool, inputs.length = n $\rightarrow$
    c.evalCanonical inputs = [parityBit inputs]
\end{lstlisting}

The main theorem gives an explicit size lower bound for every sufficiently large
input length. Let $r=\operatorname{nthRoot}_{d-1}\!\left(\left\lfloor
n/(360\cdot40^{d-2})\right\rfloor\right)$, where $d\geq2$.  After accounting for
unfolding, a circuit with at most $d$ computation layers computing PARITY has
size at least $2^{\lfloor r/(8(d+3))\rfloor}$, as stated in
Listing~\ref{lst:DAGParityCircuitLowerBound}. For fixed $d$, this is
$\exp(\Omega_d(n^{1/(d-1)}))$, the quantitative bound announced in the
introduction.

\noindent\begin{minipage}{\linewidth}
\begin{lstlisting}[caption={Main Theorem: Quantitative PARITY Circuit Lower Bound},
                   mathescape=true,
                   label={lst:DAGParityCircuitLowerBound}]{}
theorem circuit_parity_size_lower_bound_root_sharp
    (d : Nat) (hd : 2 $\leq$ d) :
    $\exists$ N, $\forall$ n, N $\leq$ n $\rightarrow$
      $\forall$ (circuit : Circuit),
        circuit.inputWidth = n $\rightarrow$
        circuit.depth $\leq$ d + 1 $\rightarrow$
        CircuitComputesParity n circuit $\rightarrow$
        2 ^ (Nat.nthRoot (d - 1) (n / (360 * 40 ^ (d - 2))) /
            (8 * (d + 3))) $\leq$ circuit.circuitSize
\end{lstlisting}
\end{minipage}

For every fixed $d$, the quantitative bound eventually exceeds any polynomial $c
n^k$. Thus, for each choice of $c$, $k$, and $d$, there is a threshold $N$ such
that no circuit with these size and depth bounds computes PARITY on any input
length $n>N$. Listing~\ref{lst:DAGParityCircuitLowerBound1} states this
corollary.

\begin{lstlisting}[caption={Corollary: No Polynomial-Size Constant-Depth PARITY Circuits},
                   mathescape=true,
                   label={lst:DAGParityCircuitLowerBound1}]{}
theorem hastad_parity_lower_bound_general (c k d : Nat) :
    $\exists$ N, $\forall$ n, N < n $\rightarrow$
      $\forall$ (circuit : UFICircuitOfSizeAtMostPolyNAndDepthAtMostD n c k d),
        $\lnot$ CircuitComputesParity n circuit.val
\end{lstlisting}

Applying this corollary to each member of a circuit family shows that PARITY is
not in nonuniform $\AC^0$ for the formalized model. The threshold depends only
on $c$, $k$, and $d$, so it works for every family with those size and depth
bounds, as recorded in Listing~\ref{lst:ParityLowerBoundCircuits}.

\begin{lstlisting}[caption={Corollary: PARITY Has No AC$^0$ Circuit Family},
                   mathescape=true,
                   label={lst:ParityLowerBoundCircuits}]{}
theorem parity_does_not_have_ac0_circuits :
  $\forall$ (c k d : Nat),
    $\exists$ N,
      $\forall$ (n : PNat)
        (ac0Circuit : AC0CircuitFamily c k d),
          n.val > N $\rightarrow$
            $\lnot$ CircuitComputesParity n.val (ac0Circuit n).val
\end{lstlisting}

\section{Conclusion}

We have formalized a quantitative circuit lower bound for PARITY in Lean.  For
every fixed computation depth $d\geq2$ and all sufficiently large $n$, the main
theorem requires size at least $\exp(\Omega_d(n^{1/(d-1)}))$ for well-formed
$n$-input circuits computing PARITY. The exponent $1/(d-1)$ is optimal. The
stored circuit depth includes the output wire, so the formal depth hypothesis is
$\texttt{circuit.depth}\leq d+1$.  The exclusion of polynomial-size,
constant-depth circuit families follows as a corollary.

The proof combines canonical decision trees and an injective encoding of bad
restrictions with formula normalization, fan-in reduction, and iterated depth
reduction. These transformations establish the formula lower bound. Unfolding
DAG circuits into evaluation-equivalent formulas, with explicit control of depth
and the size increase from duplicated shared gates, transfers that bound to
circuits.

A polynomial-size, logarithmic-depth bounded-fan-in formula family for PARITY
provides the upper bound witnessing $\NC^1 \nsubseteq \AC^0$ for the formalized
models. The development also supplies reusable circuit and formula definitions,
restriction lemmas, and checked transformations for further formalizations of
circuit lower bounds.

\bibliographystyle{alpha}
\bibliography{new}

\clearpage
\appendix
\section{Supplementary Definitions and the Switching-Lemma Proof}\label{sec:appendix}

This appendix collects supporting definitions and listings deferred from the
main development.  Beginning with Section~\ref{sec:CoreSwitchingLemmaIdea}, it
gives the canonical decision-tree construction and the encoder--decoder counting
argument underlying Theorem~\ref{SwitchingLemmaThm}.

\subsection{Formula}

\subsubsection{Formula Evaluation}

The following definition illustrates evaluation for unbounded-fan-in formulas.
Bounded-fan-in evaluation is analogous and is omitted.

\begin{lstlisting}[caption={Unbounded Formula Evaluation},
                   mathescape=true,
                   label={lst:ufiFormulaEval}]{}
def ufiFormulaEval (f : UnboundedFanInFormula) (inputs : List Bool) : Bool :=
  match f with
  | inputGate index negated => match inputs[index]? with
                           | none => false
                           | some input => match negated with
                                           | true => not input
                                           | false => input
  | constant bit _ => bit
  | notGate gate   => not (ufiFormulaEval gate inputs)
  | andGate gates  => match gates with
                      | .nil                  => true
                      | .cons gate otherGates =>
                          match (ufiFormulaEval gate inputs) with
                          | false => false
                          | true => ufiFormulaEval (andGate otherGates) inputs
  | orGate gates   => match gates with
                      | .nil                  => false
                      | .cons gate otherGates =>
                          match (ufiFormulaEval gate inputs) with
                          | false => ufiFormulaEval (orGate otherGates) inputs
                          | true => true
\end{lstlisting}

\begin{lstlisting}[caption={Bounded Formula Evaluation},
                   mathescape=true,
                   label={lst:bfiFormulaEval}]{}
def bfiFormulaEval (formula : BoundedFanInFormula) (inputs : List Bool) : Bool :=
  match formula with
  | .inputGate index negated => match inputs[index]? with
                                | none => false
                                | some input => if negated then
                                                  !input
                                                else
                                                  input
  | .constant bit _     => bit
  | .notGate gate       => !(bfiFormulaEval gate inputs)
  | .andGate left right => if bfiFormulaEval left inputs then
                             bfiFormulaEval right inputs
                           else
                             false
  | .orGate left right  => if bfiFormulaEval left inputs then
                             true
                           else
                             bfiFormulaEval right inputs
\end{lstlisting}

\subsubsection{Obtaining a CNF/DNF From a List of Clauses}\label{sec:cnfDnfFromClauses}

\begin{lstlisting}[caption={Obtaining a CNF/DNF From a List of Clauses},
                   mathescape=true,
                   label={lst:cnfDnfFromClauses}]{}
def litToInput : Nat $\times$ Bool $\rightarrow$ UnboundedFanInFormula :=
  fun p => inputGate p.1 p.2

def clauseToOr (clause : List (Nat $\times$ Bool)) : UnboundedFanInFormula :=
    orGate (clause.map litToInput)

def cnfFromClauses (clauses : List (List (Nat $\times$ Bool)))
  : UnboundedFanInFormula :=
    andGate (clauses.map clauseToOr)

def clauseToAnd (clause : List (Nat $\times$ Bool)) : UnboundedFanInFormula :=
  andGate (clause.map litToInput)

def dnfFromClauses (clauses : List (List (Nat $\times$ Bool)))
  : UnboundedFanInFormula :=
    orGate (clause.map clauseToAnd)
\end{lstlisting}

\subsubsection{A Custom Formula Equality Comparator}\label{sec:formulaBEq}

Lean can derive \texttt{BEq} for the \texttt{UnboundedFanInFormula}
nested-recursive type, but it cannot derive \texttt{DecidableEq},
\texttt{ReflBEq}, or \texttt{LawfulBEq} through the \texttt{List
  UnboundedFanInFormula} fields.  The gate deduplication implementation needs
explicit reflexivity and soundness proofs, so those properties are proved
directly as shown below.

\begin{lstlisting}[caption={Formula Equality Comparator},
                   mathescape=true,
                   label={lst:dedupFormulas}]{}
mutual
def formulaBEq : UnboundedFanInFormula $\rightarrow$ UnboundedFanInFormula $\rightarrow$ Bool
  | inputGate i n,  inputGate i' n'  => decide (i = i') && decide (n = n')
  | constant v lbl, constant v' lbl' => (v == v') && decide (lbl = lbl')
  | notGate g,      notGate g'       => formulaBEq g g'
  | andGate gs,     andGate gs'      => formulaListBEq gs gs'
  | orGate gs,      orGate gs'       => formulaListBEq gs gs'
  | _,              _                => false

def formulaListBEq :
    List UnboundedFanInFormula $\rightarrow$ List UnboundedFanInFormula $\rightarrow$ Bool
  | [],     []        => true
  | g :: gs, g' :: gs' => formulaBEq g g' && formulaListBEq gs gs'
  | _,      _         => false
end

theorem formulaBEq_refl : $\forall$ g, formulaBEq g g = true

theorem formulaBEq_sound :
    $\forall$ g g', formulaBEq g g' = true $\rightarrow$ g = g'
\end{lstlisting}

\subsection{Circuit-to-Formula Conversion}

The following definition unfolds the subcircuit feeding gate $k$ into a
formula. Input indices are positions in the positive or negated input-gate
lists, so the formula reads the same input bits as the circuit.

\begin{lstlisting}[caption={Unfolding a Circuit Gate into a Formula},
                   mathescape=true,
                   label={lst:toUFIByPos}]{}
def Circuit.toUFIByPos (c : Circuit)
    (fuel : Nat) (k : Nat) : UnboundedFanInFormula :=
  match fuel with
  | 0 => inputGate 0 false
  | fuel' + 1 =>
    if hk : k < c.gates.length then
      match (c.gates[k]'hk).type with
      | GateType.input neg =>
          let posIds := (c.gates.filter
            (fun g => g.type == GateType.input false)).map Gate.id
          let negIds := (c.gates.filter
            (fun g => g.type == GateType.input true)).map Gate.id
          if neg then inputGate (negIds.findIdx ($\cdot$ == k)) true
          else inputGate (posIds.findIdx ($\cdot$ == k)) false
      | GateType.output =>
          match (c.inEdges k).head? with
          | some e => c.toUFIByPos fuel' e.src
          | none   => constant false 0
      | GateType.notGate =>
          match (c.inEdges k).head? with
          | some e => notGate (c.toUFIByPos fuel' e.src)
          | none   => constant false 0
      | GateType.andGate =>
          andGate ((c.inEdges k).map fun e => c.toUFIByPos fuel' e.src)
      | GateType.orGate =>
          orGate ((c.inEdges k).map fun e => c.toUFIByPos fuel' e.src)
    else inputGate 0 false
\end{lstlisting}

\subsection{Restrictions}

\subsubsection{Generating All Restrictions}\label{sec:GeneratingAllRestrictions}

Generation of all restrictions requires generation of all possible dead bit
assignments.  This is done by the \lstinline|allBitLists| function shown in
listing \ref{lst:allBitLists}, along with a lemma stating that for every input
$k$, every list in its output list has the required length $k$.  The proof is by
induction on $k$ but the implementation tactics are omitted in the listing.

\begin{lstlisting}[caption={Generating all bit lists of length k},
                   mathescape=true,
                   label={lst:allBitLists}]{}
def allBitLists : (k : Nat) $\rightarrow$ List (List Bool)
  | 0     => [[]]
  | k + 1 => ((allBitLists k).map
              fun l =>
                [false :: l, true :: l]).flatten

lemma allBitLists_mem_length (k : Nat) :
    $\forall$ l $\in$ allBitLists k, l.length = k
\end{lstlisting}

The \lstinline|generateAllRestrictions| helper function (listing
\ref{lst:generateAllRestrictions}) uses the Mathlib \lstinline|Finset.range|
function to get the set of natural numbers less than $n$ and
\lstinline|powersetCard| to get the set of subsets of $[n]$ of cardinality $s$.
These subsets represent the set of ways to choose $s$ live variables out of $n$.
\lstinline|subsetsWithMembership| is a subtype whose values are the subsets
(from \lstinline|powersetCard|) and whose properties are proofs that they are
members of \lstinline|subsets|.  The proofs in these subtypes are used by
\lstinline|Finset.mem_powersetCard| to construct \lstinline|h_s|, a proof that
each such set is an element of \lstinline|subsets| and has cardinality equal to
$n$.  The left and right conjuncts are denoted \lstinline|h_s.1| and
\lstinline|h_s.2| respectively.  \lstinline|h_s| is required to show that all
non-starred variables in the restrictions are fully assigned.  The
\ref{lst:generateAllRestrictions_card} lemma shows that
$|generateAllRestrictions ~ n ~ \sigma| = |\mathcal{R}_s| = \binom{n}{s} 2^{n -
  s}$ thus verifying the correctness of \lstinline|generateAllRestrictions|.

\noindent\begin{minipage}{\linewidth}
\begin{lstlisting}[caption={Cardinality of generateAllRestrictions},
                   mathescape=true,
                   label={lst:generateAllRestrictions_card}]{}
lemma generateAllRestrictions_card (n : Nat) ($\sigma$ : OpenUnitIntervalQ) :
    (generateAllRestrictions n $\sigma$).card =
      Nat.choose n (Nat.ceil ($\sigma$.val * n)) *
        2 ^ (n - Nat.ceil ($\sigma$.val * n))

lemma generateAllRestrictions_card_eq_totalRestrictionCount
    (n : Nat)
    ($\sigma$ : OpenUnitIntervalQ) :
  (generateAllRestrictions n $\sigma$).card = totalRestrictionCount n $\sigma$
\end{lstlisting}
\end{minipage}

\begin{lstlisting}[caption={Generating All Restrictions},
                   mathescape=true,
                   label={lst:generateAllRestrictions}]{}
lemma ceil_sigma_n_le ($\sigma$ : OpenUnitIntervalQ) (n : Nat) :
    Nat.ceil ($\sigma$.val * (n : $\mathbb{Q}$)) $\leq$ n

def generateAllRestrictions (n : Nat) ($\sigma$ : OpenUnitIntervalQ) :
    Multiset (AssignedRandomRestriction $\sigma$ n) :=
  let s := Nat.ceil ($\sigma$.val * n)
  let subsets := (Finset.range n).powersetCard s
  let subsetsWithMembership := subsets.attach

  (subsetsWithMembership.val.bind
    fun $\langle$s_live, h_s_mem$\rangle$
      let h_s := Finset.mem_powersetCard.mp h_s_mem
      let bitLists := allBitLists (n - s)
      let bitListsWithMembership := bitLists.attach

      (bitListsWithMembership.map fun bitsWithMembership =>
        let bits := bitsWithMembership.val
        {
          starAssignment := $\langle\langle$s_live, h_s.1$\rangle$, h_s.2$\rangle$
          varAssignments := bits
          non_starred_vars_fully_assigned := by
            have h : bits.length = n - s :=
              by
                apply allBitLists_mem_length
                apply bitsWithMembership.property
            rw [h, h_s.2]
            have hsn : s $\leq$ n := ceil_sigma_n_le $\sigma$ n
            apply Nat.add_sub_of_le hsn
        } : List (AssignedRandomRestriction $\sigma$ n)))
\end{lstlisting}

\begin{lstlisting}[caption={Representing Random Restrictions as Lists},
                   mathescape=true,
                   label={lst:mkAssignmentList}]{}
def mkAssignmentList (live : Finset Nat) (deadBits : List Bool) (n : Nat)
    : List (Nat $\times$ Bool) :=
  (List.range n).filterMap
    (fun v =>
      if v $\in$ live then
        none
      else
        let j := (Finset.range v \ live).card
        match deadBits[j]? with
        | some b => some (v, b)
        | none => none)
\end{lstlisting}

\subsection{DNFs}

\subsubsection{DNF Clause Manipulation}

\begin{lstlisting}[caption={Extracting DNF Clauses},
                   mathescape=true,
                   label={lst:dnfClauses}]{}
def dnfClauses (dnf : UnboundedFanInFormula) : List (List (Nat $\times$ Bool)) :=
  match dnf with
  | .orGate gates =>
      gates.map fun gate =>
        match gate with
        | .andGate literals => literals.filterMap fun lit =>
            match lit with
            | .inputGate i b => some (i, b)
            | _ => none
        | _ => []
  | _ => []
\end{lstlisting}

\begin{lstlisting}[caption={Restrictions as Functions},
                   mathescape=true,
                   label={lst:restrictionAsFunction}]{}
def restrictionAsFunction (asgn : List (Nat $\times$ Bool)) : Nat $\rightarrow$ Option Bool :=
  fun v =>
    match asgn.find? (fun p => p.1 == v) with
    | some (_, b) => some b
    | none => none
\end{lstlisting}

\subsubsection{DNF Clause Restriction}
The restriction operation is defined compositionally.  The function
\texttt{simpleRestrictLiteral} retains live literals and replaces assigned
literals by constants.  The function \texttt{simpleRestrictTerm} returns
\texttt{none} when an assigned literal falsifies the term. Otherwise it returns
the conjunction of the surviving literals.  Finally, \texttt{restrictDNF} maps
this operation over all terms and drops falsified terms.  The development proves
that this transformation preserves evaluation when an assignment to the
remaining variables is combined with the fixed part of the restriction.

\noindent\begin{minipage}{\linewidth}
\begin{lstlisting}[caption={DNF Clause Restriction},
                   mathescape=true,
                   label={lst:restrictDNF}]{}
def simpleRestrictLiteral (asgn : Nat $\rightarrow$ Option Bool)
                          (lit : UnboundedFanInFormula)
                          : UnboundedFanInFormula :=
  match lit with
  | .inputGate i negated =>
    match asgn i with
    | none => .inputGate i negated
    | some b => .constant (if negated then Bool.not b else b) 0
  | c => c

def simpleRestrictTerm (asgn : Nat $\rightarrow$ Option Bool)
                       (term : UnboundedFanInFormula)
                      : Option UnboundedFanInFormula :=
  match term with
  | .andGate lits =>
    let applied := lits.map (simpleRestrictLiteral asgn)
    if applied.any
      (fun l => match l with
                | .constant false _ => true
                | _ => false) then
      none
    else
      some (.andGate (applied.filter
                        (fun l => match l with
                                  | .constant _ _ => false
                                  | _ => true)))
  | c => some c

def restrictDNF (dnf : UnboundedFanInFormula)
                ($\rho$ : AssignedRandomRestriction $\sigma$ n)
                : UnboundedFanInFormula :=
  let asgn := mkAssignmentList $\rho$.starAssignment.val.val
                               $\rho$.varAssignments
                               n
  let asgnFn := restrictionAsFunction asgn
  match dnf with
  | .orGate terms => .orGate (terms.filterMap (simpleRestrictTerm asgnFn))
  | c => c
\end{lstlisting}
\end{minipage}

\subsection{Decision Trees}\label{subsec:DecisionTreeAppendix}

\begin{lstlisting}[caption={Converting a Decision Tree to a DNF},
                   mathescape=true,
                   label={lst:decisionTreeToDNF}]{}
def decisionTreeToDNFClauses (tree : DecisionTree)
                             (path : List UnboundedFanInFormula)
                             : List UnboundedFanInFormula :=
  match tree with
  | dtLeaf true  => [andGate path]
  | dtLeaf false => []
  | dtNode i left right =>
      (decisionTreeToDNFClauses left (path ++ [(inputGate i true)])) ++
      (decisionTreeToDNFClauses right (path ++ [inputGate i false]))

def decisionTreeToDNF (tree : DecisionTree) : UnboundedFanInFormula :=
  orGate (decisionTreeToDNFClauses tree [])
\end{lstlisting}

\subsection{The Core Idea of the Switching Lemma Proof}\label{sec:CoreSwitchingLemmaIdea}
The core idea in the proof of the switching lemma is that given the DNF $f$ and
its accompanying switching lemma arguments, we can define a function from the
set $\mathcal{B}$ of bad restrictions to a set that is much smaller than
$\mathcal{R}_s$. If this function is injective, then $\mathcal{B}$ must be a
small set, which is what we are trying to prove. We formalize Razborov's
construction of such a function. Since it is injective, we refer to it as an
encoding function and its (well defined) inverse as a decoding function. The
natural way to map a restriction into a smaller domain is to fix some of its
variables.  Therefore, a natural smaller target set for the encoding function is
the set $\mathcal{R}_{s - d}$ obtained by fixing $d$ variables of a bad
restriction, $\beta$. The resulting restriction $\beta' \in \mathcal{R}_{s-d}$
has exactly $d$ fewer free variables. However, this leaves open how a decoder
can recover $\beta$ from $\beta'$: it must identify which $d$ variables the
encoder fixed. Therefore, the encoder must also output auxiliary information
sufficient for recovery: Naively encoding the $d$ variable identifiers would
require $d \lceil \lg n \rceil$ bits of auxiliary information:

$$ Enc_{naive} : \mathcal{B} \rightarrow \mathcal{R}_{s - d} \times \{0,1\}^{d \lceil \lg n \rceil} $$

\begin{align*}
  \left\| \mathcal{R}_{s - d} \times \{0,1\}^{d \lceil \lg n \rceil} \right\|
    &= \left\| \mathcal{R}_{s - d} \right\| \cdot
       \left\| \{0,1\}^{d \lceil \lg n \rceil} \right\| \\
    &= \binom{n}{s - d} 2^{n - (s - d)} \cdot 2^{d \lceil \lg n \rceil} \\
\end{align*}

From this expression, it is clear that computing the ratio of the size of this
range to the size of the set of all possible restrictions with $s$ stars (as the
switching lemma does) will result in a factor of $n^d$ in the
expression. Therefore, an alternative encoding is required for the switching
lemma.  Razborov's construction reduces the number of auxiliary bits to $d \lg w
+ 2d$, which is independent of $n$. With this injective encoder, we get the
desired ratio for the switching lemma:

$$ Enc : \mathcal{B} \rightarrow \mathcal{R}_{s - d} \times \{0,1\}^{d \lg w + 2d } $$

\begin{align*}
  \frac{\left\| \mathcal{R}_{s - d} \times \{0,1\}^{d \lg w + 2d} \right\|}
       {\left\| \mathcal{R}_{s} \right\|}
    &=\frac{\binom{n}{s - d} 2^{n - (s - d)} \cdot 2^{d \lg w + 2d}}
       {\binom{n}{s} 2^{n - s}} \\
    &= \frac{\frac{n!}{(n - s + d)! (s - d)!}
             2^{n - (s - d) - (n - s)} \cdot 2^{d \lg w + 2d}}
            {\frac{n!}{(n - s)! s!}} \\
    &= \frac{(n - s)! s!}{(n - s + d)! (s - d)!}
       \cdot 2^{d + d \lg w + 2d} \\
    &= \frac{s(s - 1)\dotsm(s - d + 1)}
            {(n - s + d)(n - s + d - 1)\dotsm(n - s + 1)}
       \cdot (8w)^d \\
    &\leq \left(\frac{s}{n - s + d}\right)^d \cdot (8w)^d \\
    &= \left(\frac{\sigma n}{n - \sigma n + d}\right)^d \cdot (8w)^d \\
    &= \left(\frac{n \sigma}{n(1 - \sigma + d/n)}\right)^d \cdot (8w)^d \\
    &\leq \left(\frac{\sigma}{1 - \sigma}\right)^d \cdot (8w)^d \\
    &\leq (10 \sigma w)^d.
\end{align*}

The first inequality holds because $s = \sigma n$ and $\sigma \leq \frac{1}{5}$
imply that $n - s + d \geq n - s \geq \frac{4}{5}n \geq s$. It follows that for
$i \in [0.. d-1]$:

\begin{align*}
  n - s + d \geq s
    &\Rightarrow i (n - s + d) \geq is \\
    &\Rightarrow -i(n - s + d) \leq -is \\
    &\Rightarrow s(n - s + d) - i(n - s + d)
                \leq s(n - s + d) - is \\
    &\Rightarrow (s - i)(n - s + d) \leq s(n - s + d - i) \\
    &\Rightarrow \frac{s - i}{n - s + d - i} \leq \frac{s}{n - s + d}
\end{align*}

The last inequality follows from these implications:

\begin{align*}
  \sigma \leq \frac{1}{5}
    &\Rightarrow -\sigma \geq -\frac{1}{5} \\
    &\Rightarrow 1 - \sigma \geq \frac{4}{5} > 0 \\
    &\Rightarrow \frac{1}{1 - \sigma} \leq \frac{1}{4/5} = \frac{5}{4} \\
    &\Rightarrow \sigma \cdot \frac{1}{1 - \sigma}
        \leq \sigma \cdot \frac{5}{4}
        \qquad (\text{since } \sigma \geq 0) \\
    &\Rightarrow \frac{\sigma}{1 - \sigma} \leq \frac{5}{4} \sigma .
\end{align*}

The proof that $\frac{\binom{n}{s - d}}{\binom{n}{s}} \leq \frac{5}{4} \sigma$
is stated in Lean as the \lstinline|choose_ratio_bound_exact| lemma:

\noindent\begin{minipage}{\linewidth}
\begin{lstlisting}[caption={Binomial Ratio Upper Bound},
                   mathescape=true,
                   label={lst:choose_ratio_bound_exact}]{}
lemma choose_ratio_bound_exact (n s d : Nat)
                            ($\sigma$ : OpenUnitIntervalQ)
                            (h$\sigma$ : $\sigma$.val $\leq$ 1 / 5)
                            (hs_exact : (s : $\mathbb{Q}$) = $\sigma$.val * n)
                            (hsd : d < s)
                            (hsn : s $\leq$ n) :
  (Nat.choose n (s - d) : $\mathbb{Q}$)
    $\leq$ ((5 / 4) * $\sigma$.val) ^ d
      * Nat.choose n s
\end{lstlisting}
\end{minipage}

All that remains to complete the switching lemma proof is the definition of an
injective encoder and decoder using $d \lg w + 2d$ auxiliary bits of
information. This is achieved using canonical decision trees.

\subsection{Canonical Decision Trees}
The domain of the encoder is the set of bad restrictions.  To determine whether
a restriction is bad, \lstinline|isBadRestriction| (listing
\ref{lst:isBadRestriction}) builds a decision tree for the function obtained by
fixing the variables in the given assigned random restriction then compares its
depth to the given constant $d$.  The switching lemma is a statement about the
decision tree complexity of the function, but does not place any conditions on
the shape of the decision tree.  Therefore, we use a canonical decision tree,
which enables injective encoding and decoding with the desired maximum number of
auxiliary bits.

\begin{lstlisting}[caption={Detecting Bad Restrictions},
                   mathescape=true,
                   label={lst:isBadRestriction}]{}
def isBadRestriction (d n : Nat)
                     ($\sigma$ : OpenUnitIntervalQ)
                     (f : UnboundedFanInProperDNF n)
                     ($\rho$ : AssignedRandomRestriction $\sigma$ n) : Bool :=
  decisionTreeDepth (properDNFCanonicalDecisionTree f $\rho$) > d
\end{lstlisting}

Consider the DNF $f : \{0, 1\}^6 \rightarrow \{0, 1\}$ defined below and the
restriction $\beta$ mapping $x_0 \rightarrow 0$ and $x_5 \rightarrow 1$ and
leaving $x_1$, $x_2$, $x_3$, and $x_4$ live. Note that we are using 0 and 1 in
place of the Lean \texttt{false} and \texttt{true} values for succinctness. The
switching lemma is a statement about the depth of the decision tree
corresponding to the function obtained by applying this restriction to the DNF.
This is done by the \lstinline|restrictDNF| Lean function (see listing
\ref{lst:restrictDNF} in the appendix \ref{sec:appendix} for details).  The
clause by clause restriction is shown in figure~\ref{fig:restriction-step}.

\newcommand{\clauseCone}{
  x_1 \land x_2 \land \overline{x_3}
}

$$ f(x) = (x_0 \land x_2) \lor (\clauseCone) \lor (x_4 \land \overline{x_2}) \lor (x_3 \land x_5) \lor \overline{x_4} $$

\begin{figure}[H]
  \centering
  \begin{align*}
    x_0 \land x_2
      &\longrightarrow 0 ~(\text{since}~ x_0=0) \\
    \clauseCone
      &\longrightarrow \clauseCone \\
    x_4 \land \overline{x_2}
      &\longrightarrow x_4 \land \overline{x_2} \\
    x_3 \land x_5
      &\longrightarrow x_3 ~(\text{since}~ x_5=1) \\
    \overline{x_4}
      &\longrightarrow \overline{x_4}
  \end{align*}
  \caption{Applying the non-trivial restriction to the original DNF.}
  \label{fig:restriction-step}
\end{figure}

\newcommand{\restrictedDNF}{
  (\clauseCone)
  \lor (x_4 \land \overline{x_2}) \lor x_3 \lor \overline{x_4}}

Thus the restricted DNF is $f'(x) = \restrictedDNF$.  Section
\ref{sec:DecisionTrees} presented an example of a decision tree for this DNF.
We treat the clauses in a DNF as ordered so we will refer to these clauses as
$C_1$ through $C_4$ from left to right. These restricted clauses are the inputs
to the canonical tree construction algorithm as shown in listing
\ref{lst:properDNFCanonicalDecisionTree}. We first present the algorithm for
canonical tree construction then show how canonical decision trees meet the
injective encoding/decoding requirements.

\begin{lstlisting}[caption={Creating a Canonical Decision Tree},
                   mathescape=true,
                   label={lst:properDNFCanonicalDecisionTree}]{}
def properDNFCanonicalDecisionTree (dnf : UnboundedFanInProperDNF n)
                          ($\rho$ : AssignedRandomRestriction $\sigma$ n)
  : DecisionTree :=
    let restrictedDNF := restrictDNF dnf.val $\rho$
    let restrictedClauses := dnfClauses restrictedDNF
    canonicalDecisionTreeAuxPreciseFull restrictedClauses.length restrictedClauses
\end{lstlisting}

\subsubsection{Step 1: Full Query Path Tree Construction}

In a canonical decision tree for a DNF, a full binary tree is constructed for
each clause. Each leaf of the tree is labeled with the value the clause
evaluates to if its variables are assigned the values indicated by the path from
the root of the clause tree to that leaf. Since every variable is queried along
every path, we refer to this tree as the full-query path tree.
Figure~\ref{fig:c1-path-tree} shows the full-query path tree for $C_1 =
\clauseCone$. Its unique 1-leaf is reached by the assignment
$(x_1,x_2,x_3)=(1,1,0)$.

\begin{figure}[H]
  \centering
  \begin{forest}
    decision tree
    [$x_1$, variable
      [$x_2$, variable, zero edge
        [$x_3$, variable, zero edge
          [$0$, leaf, zero edge]
          [$0$, leaf, one edge]
        ]
        [$x_3$, variable, one edge
          [$0$, leaf, zero edge]
          [$0$, leaf, one edge]
        ]
      ]
      [$x_2$, variable, one edge
        [$x_3$, variable, zero edge
          [$0$, leaf, zero edge]
          [$0$, leaf, one edge]
        ]
        [$x_3$, variable, one edge
          [$1$, leaf, zero edge]
          [$0$, leaf, one edge]
        ]
      ]
    ]
  \end{forest}
  \caption{The full-query path tree for $C_1 = \clauseCone$.}
  \label{fig:c1-path-tree}
\end{figure}

Recall that a literal is represented as a pair comprising a natural number
identifying the variable and a Boolean flag indicated whether the literal is
negated. A clause is therefore a list of such pairs. The full-query path tree is
generated by recursively creating a node with a full dead tree on one side (one
in which all paths lead to a 0-leaf) and a full-query path tree of the remaining
literals on the side that satisfies the literal in the node.  The unique
satisfying path reaches 1-leaf and every other path reaches a 0-leaf after
querying all remaining variables.  Recall that the canonical decision tree is a
valid decision tree for the DNF $f'$. Therefore, the unique path leading to a
1-leaf ensures that any input that satisfies that clause will necessarily force
evaluation of the canonical decision tree to a 1-leaf.  On the other hand, the
dead subtree construction ensures that every root-to-leaf path queries all
clause variables so that every possible way that the clause can fail to be
satisfied is accounted for by the canonical decision tree when evaluating
subsequent clauses of $f'$.  The Lean implementation of full-path tree
construction is shown in listing \ref{lst:clauseToPathTreeFull}.

\begin{lstlisting}[caption={Full Path Tree Construction from a Clause},
                   mathescape=true,
                   label={lst:clauseToPathTreeFull}]{}
def deadTree : List (Nat $\times$ Bool) $\rightarrow$ DecisionTree
  | [] => .dtLeaf .false
  | (v, _) :: rest => .dtNode v (deadTree rest) (deadTree rest)

def literalSatisfyingBit (negated : Bool) : Bool :=
  if negated then
    false
  else
    true

def clauseToPathTreeFull : List (Nat $\times$ Bool) $\rightarrow$ DecisionTree
  | [] => .dtLeaf .true
  | (v, neg) :: rest =>
      let sat := literalSatisfyingBit neg
      match sat with
      | .false => .dtNode v (clauseToPathTreeFull rest) (deadTree rest)
      | .true  => .dtNode v (deadTree rest) (clauseToPathTreeFull rest)
\end{lstlisting}

\subsubsection{Step 2: path-aware simplification and continuation trees}

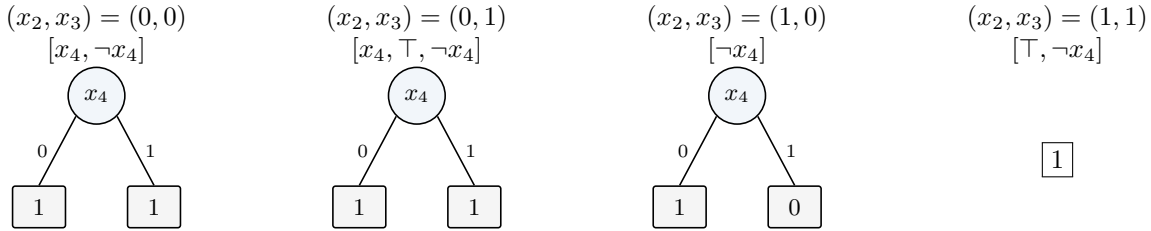
\begin{figure}[H]
  \centering
  \begin{minipage}[t]{0.23\textwidth}
    \centering
    $(x_2,x_3)=(0,0)$

    $[x_4,\neg x_4]$

    \begin{forest}
      decision tree, for tree={s sep=8mm}
      [$x_4$, variable
        [$1$, leaf, zero edge]
        [$1$, leaf, one edge]
      ]
    \end{forest}
  \end{minipage}
  \hfill
  \begin{minipage}[t]{0.23\textwidth}
    \centering
    $(x_2,x_3)=(0,1)$

    $[x_4,\top,\neg x_4]$

    \begin{forest}
      decision tree, for tree={s sep=8mm}
      [$x_4$, variable
        [$1$, leaf, zero edge]
        [$1$, leaf, one edge]
      ]
    \end{forest}
  \end{minipage}
  \hfill
  \begin{minipage}[t]{0.23\textwidth}
    \centering
    $(x_2,x_3)=(1,0)$

    $[\neg x_4]$

    \begin{forest}
      decision tree, for tree={s sep=8mm}
      [$x_4$, variable
        [$1$, leaf, zero edge]
        [$0$, leaf, one edge]
      ]
    \end{forest}
  \end{minipage}
  \hfill
  \begin{minipage}[t]{0.23\textwidth}
    \centering
    $(x_2,x_3)=(1,1)$

    $[\top,\neg x_4]$

    \vspace{10mm}
    \fbox{$1$}
  \end{minipage}
  \caption{The four path-conditioned continuation trees. The symbol $\top$
    denotes an already satisfied, hence empty, clause.}
  \label{fig:continuation-trees}
\end{figure}

From figure \ref{fig:c1-path-tree}, it is clear that any path that does not end
at the 1-leaf is an assignment of values to $C_1$'s variables such that $C_1$ is
not satisfied. Therefore, any such path corresponds to a further restriction of
the variables in the DNF $f'$ to their values from that path.  For the canonical
decision tree to compute $f'$, every 0-leaf $\gamma$ of $C_1$ needs to be
replaced with a canonical decision tree for the function resulting from the
restriction of $f'$ according to the variable assignments in $\gamma$'s path. We
refer to this as grafting canonical decision trees onto the 0-leaves.  The
grafting process works by recursively walking the tree and replacing every
0-leaf. Each recursive call simplifies the surviving clauses according to the
assignment given by the edge taken by the call.  Of the path variables in figure
\ref{fig:c1-path-tree}, $x_1$ does not occur in $C_2,C_3,C_4$, so the
continuation at every 0-leaf of the $C_1$ tree depends only on the path
variables $(x_2,x_3)$.  When the grafting process descends to the leftmost $x_3$
node in figure \ref{fig:c1-path-tree}, $x_2 = 0$ so that restriction is applied
to the remaining clauses $C_2 = x_4 \land \overline{x_2}$, $C_3 = x_3$ and $C_4
= \overline{x_4}$. The only clauses that survive this restriction are $C_2 =
x_4$ and $C_4 = \overline{x_4}$.  Figure~\ref{fig:continuation-trees} shows the
four restrictions to $x_2$ and $x_3$ (reachable along paths to 0-leaves in
$C_1$'s full-query path tree), the clause lists resulting from clause
simplification along the paths, and their corresponding canonical trees.

Listing \ref{lst:SimplifyClauses} shows the Lean implementation of clause
simplification. \lstinline|simplifyClausesLeft| uses the filter function to drop
all clauses that contain the (non-negated) variable $x_i$. This is how the $x_i
= 0$ restriction is applied because such clauses would necessarily evaluate to
false in the left branch of the node for which a left subtree is being
constructed. The negated variable $\overline{x_i}$ may still appear in the
remaining clauses but would be redundant (since they are satisfied by the left
branch). Therefore, the second filter function drops them from the surviving
clauses. Similarly, \lstinline|simplifyClausesRight| drops all clauses
containing the negated variable $x_i$ and drops the non-negated variable $x_i$
from the surviving clauses.

\noindent\begin{minipage}{\linewidth}
\begin{lstlisting}[caption={Clause Simplification},
                   mathescape=true,
                   label={lst:SimplifyClauses}]{}
def simplifyClausesLeft (clauses : List (List (Nat $\times$ Bool)))
                        (i : Nat)
                        : List (List (Nat $\times$ Bool)) :=
  (clauses.filter
    (fun c => !c.any (fun lit => lit.1 == i && !lit.2))).map
      (fun c => c.filter (fun lit => !(lit.1 == i && lit.2)))

def simplifyClausesRight (clauses : List (List (Nat $\times$ Bool)))
                         (i : Nat)
                         : List (List (Nat $\times$ Bool)) :=
  (clauses.filter
    (fun c => !c.any (fun lit => lit.1 == i && lit.2))).map
      (fun c => c.filter (fun lit => !(lit.1 == i && !lit.2)))
\end{lstlisting}
\end{minipage}

\subsubsection{Step 3: Grafting Continuations onto the Tree}

\begin{figure}[H]
  \centering
  \begin{adjustbox}{max width=0.96\textwidth,max totalheight=0.65\textheight,center}
  \begin{forest}
    decision tree, for tree={s sep=6mm}
    [$x_1$, variable
      [$x_2$, variable, zero edge
        [$x_3$, variable, zero edge
          [$x_4$, variable, zero edge
            [$1$, leaf, zero edge]
            [$1$, leaf, one edge]
          ]
          [$x_4$, variable, one edge
            [$1$, leaf, zero edge]
            [$1$, leaf, one edge]
          ]
        ]
        [$x_3$, variable, one edge
          [$x_4$, variable, zero edge
            [$1$, leaf, zero edge]
            [$0$, leaf, one edge]
          ]
          [$1$, leaf, one edge]
        ]
      ]
      [$x_2$, variable, one edge
        [$x_3$, variable, zero edge
          [$x_4$, variable, zero edge
            [$1$, leaf, zero edge]
            [$1$, leaf, one edge]
          ]
          [$x_4$, variable, one edge
            [$1$, leaf, zero edge]
            [$1$, leaf, one edge]
          ]
        ]
        [$x_3$, variable, one edge
          [$1$, leaf, zero edge]
          [$1$, leaf, one edge]
        ]
      ]
    ]
  \end{forest}
  \end{adjustbox}
  \caption{Decision tree produced by \texttt{properDNFCanonicalDecisionTree}.}
  \label{fig:final-canonical-tree}
\end{figure}
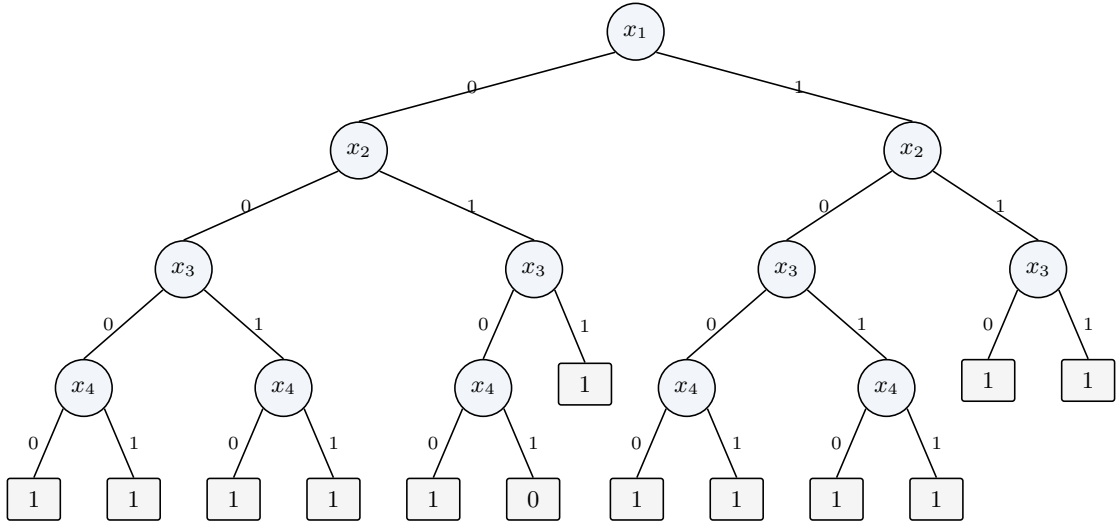

Once clause simplification is complete on every path to a 0-leaf of
Figure~\ref{fig:c1-path-tree}, the 0-leaf is replaced by the appropriate tree
from Figure~\ref{fig:continuation-trees}. The unique 1-leaf of $C_1$ is not
modified. See listing \ref{lst:canonicalDecisionTreeAuxPreciseFull} for the full
tree creation and grafting algorithm in Lean.
Figure~\ref{fig:final-canonical-tree} shows the final decision tree produced by
\texttt{properDNFCanonicalDecisionTree} for the restricted DNF $f'(x)$.

\noindent\begin{minipage}{\linewidth}
\begin{lstlisting}[caption={Creating a Canonical Decision Tree},
                   mathescape=true,
                   label={lst:canonicalDecisionTreeAuxPreciseFull}]{}
mutual
def graftOnZeroLeavesWithSimplificationFull :
    DecisionTree $\rightarrow$ List (List (Nat $\times$ Bool)) $\rightarrow$ Nat $\rightarrow$ DecisionTree
  | .dtLeaf .true,  _,       _    => .dtLeaf .true
  | .dtLeaf .false, clauses, fuel => canonicalDecisionTreeAuxPreciseFull fuel clauses
  | .dtNode v l r, clauses, fuel =>
      .dtNode v (graftOnZeroLeavesWithSimplificationFull l
                  (simplifyClausesLeft clauses v) fuel)
                (graftOnZeroLeavesWithSimplificationFull r
                  (simplifyClausesRight clauses v) fuel)

def canonicalDecisionTreeAuxPreciseFull :
    Nat $\rightarrow$ List (List (Nat $\times$ Bool)) $\rightarrow$ DecisionTree
  | 0, _                     => .dtLeaf .false
  | _, []                    => .dtLeaf .false
  | fuel + 1, clause :: rest =>
      let clauseDT := clauseToPathTreeFull clause
      graftOnZeroLeavesWithSimplificationFull clauseDT rest fuel
end
\end{lstlisting}
\end{minipage}

\subsubsection{Correctness of the Canonical Decision Tree Construction}

We need to verify that the canonical decision tree generated by
\texttt{properDNFCanonicalDecisionTree} computes the desired restriction of the
input DNF. The generic correctness theorems are stated for
\texttt{canonicalDecisionTree}, whose arguments are an unbundled formula and a
list-based assignment. The construction needs to be sound, i.e. any input that
evaluates to true in the resulting canonical decision tree should also evaluate
to true in the restricted DNF.  This soundness property is proved as the
\lstinline|canonicalDecisionTree_sound| theorem in listing
\ref{lst:canonicalDecisionTree_sound}.

\begin{lstlisting}[caption={Soundness of the Canonical Decision Tree Construction},
                   mathescape=true,
                   label={lst:canonicalDecisionTree_sound}]{}
theorem canonicalDecisionTree_sound
    (dnf : UnboundedFanInFormula)
    (asgn : List (Nat $\times$ Bool))
    (inputs : List Bool) :
  evalDecisionTree (canonicalDecisionTree dnf asgn) inputs = .true $\rightarrow$
  evalClauses inputs
    (dnfClauses (simpleRestrictDNF (restrictionAsFunction asgn) dnf)) = .true
\end{lstlisting}

The construction also needs to be complete, i.e. if the clauses of the
restricted DNF evaluate to 1 on some input, the canonical decision tree must
also evaluates to 1 on that input.  Completeness is proved as the
\lstinline|canonicalDecisionTree_complete| theorem of listing
\ref{lst:canonicalDecisionTree_complete}.  Together with soundness, this means
that the canonical decision tree computes the exact same Boolean function as the
restricted DNF.  However, we only assert this when every variable in every
clause has an associated input bit (see \texttt{ClausesVarsInBounds}) below.

\begin{lstlisting}[caption={Completeness of the Canonical Decision Tree Construction},
                   mathescape=true,
                   label={lst:canonicalDecisionTree_complete}]{}
def ClauseVarsInBounds (clause : List (Nat $\times$ Bool)) (inputs : List Bool) : Prop :=
  $\forall$ lit $\in$ clause, lit.1 < inputs.length

def ClausesVarsInBounds (clauses : List (List (Nat $\times$ Bool))) (inputs : List Bool) : Prop :=
  $\forall$ c $\in$ clauses, ClauseVarsInBounds c inputs

theorem canonicalDecisionTree_complete
    (dnf : UnboundedFanInFormula)
    (asgn : List (Nat $\times$ Bool))
    (inputs : List Bool)
    (hbounds : ClausesVarsInBounds
      (dnfClauses (simpleRestrictDNF (restrictionAsFunction asgn) dnf)) inputs) :
  evalClauses inputs
    (dnfClauses (simpleRestrictDNF (restrictionAsFunction asgn) dnf)) = .true $\rightarrow$
  evalDecisionTree (canonicalDecisionTree dnf asgn) inputs = .true
\end{lstlisting}

\subsection{Injective Encoding and Decoding of Bad Restrictions}

Recall that the switching lemma holds under the hypothesis that there exists an
injective map from the set of bad restrictions $\mathcal{B} \rightarrow
\mathcal{R}_{s - d} \times \{0,1\}^{d \lg w + 2d }$. We now show how to define
such an injective map using a canonical decision tree. This map needs to
restrict $d$ variables. To do so, we find a path of length greater than $d$ in
the canonical decision tree arising from restricting the DNF and fix the
variables according to their values along the path. The
\texttt{leftmostPathExceedingDepth} Lean function in listing
\ref{lst:leftmostPathExceedingDepth} implements this idea.  For a bad
restriction (one with a canonical decision tree of depth exceeding $d$), fixing
$d$ variables still leaves a non-trivial decision tree. The encoding will use
assignments that lead to 1-leaves in the canonical decision tree.

\noindent\begin{minipage}{\linewidth}
\begin{lstlisting}[caption={},
                   mathescape=true,
                   label={lst:leftmostPathExceedingDepth}]{}
def leftmostPath : DecisionTree $\rightarrow$ List (Nat $\times$ Bool)
  | .dtLeaf _ => []
  | .dtNode v left _ => (v, .false) :: leftmostPath left

def leftmostPathExceedingDepth :
    DecisionTree $\rightarrow$ Nat $\rightarrow$ Option (List (Nat $\times$ Bool))
  | .dtLeaf _, _                => none
  | .dtNode v left _,     0     => some ((v, .false) :: leftmostPath left)
  | .dtNode v left right, d + 1 =>
      match leftmostPathExceedingDepth left d with
      | some path => some ((v, .false) :: path)
      | none      => match leftmostPathExceedingDepth right d with
                     | some path => some ((v, .true) :: path)
                     | none => none
\end{lstlisting}
\end{minipage}

\subsubsection{The Encoder}

The \texttt{encoderRestriction} function uses \texttt{mkAssignmentList} to
convert the random restriction into a list of variable--value pairs for use by
the core recursive encoder, which generates the auxiliary encoding bits.
Consider the canonical decision tree example in
Figure~\ref{fig:final-canonical-tree} where the restriction $\beta$ maps $x_0
\rightarrow 0$ and $x_5 \rightarrow 1$.  The \texttt{deadBits} list in
\texttt{encoderRestriction} is $[(0,0),(5,1)]$ where the pair $(v,b)$ in an
assignment means $x_v=b$.  If we take $d = 3$, then there exists a path of
length greater than three in that canonical decision tree. Otherwise,
\texttt{encoderRestriction} returns the dead bits from the restriction and an
empty list of auxiliary information.

\begin{lstlisting}[caption={},
                   mathescape=true,
                   label={lst:encoderRestriction}]{}
def encoderRestriction (d : Nat)
                      (dnf : UnboundedFanInProperDNF n)
                      ($\rho$ : AssignedRandomRestriction $\sigma$ n)
    : List (Nat $\times$ Bool) $\times$ List (List (Nat $\times$ Bool))
  :=
  let deadBits := mkAssignmentList $\rho$.starAssignment.val.val
                                   $\rho$.varAssignments
                                   n
  let cdt := properDNFCanonicalDecisionTree dnf $\rho$
  let clauses := dnfClauses dnf.val

  match leftmostPathExceedingDepth cdt d with
  | none      => (deadBits, [])
  | some path =>
      let $\pi$ := path.take d
      beameEncoderAux $\pi$.length $\pi$ clauses deadBits deadBits
\end{lstlisting}

The \texttt{beameEncoderAux} function (listing \ref{lst:beameEncoderAux}) is
responsible for the actual encoding and its implementation follows definition
2.7 of \cite{online:SwitchingLemma}.  Let $\pi$ be the first $d$ variables in
the bad path (which has length $> d$).  let $T_{i_1}$ be the first term not
killed by $\beta$, and let $U_1$ be its restriction under $\beta$. $T_{i_j}$ and
$U_j$ are computed by the \texttt{firstTermNotKilledByList} and
\texttt{restrictClauseByListAssignment} functions respectively (see listing
\ref{lst:firstTermNotKilledByList}).  In this example, $i_1 = 1$ so $T_{i_1} =
\clauseCone$ and $U_1 = \clauseCone$.

\begin{lstlisting}[caption={Finding the First Term That Survives a Restriction},
                   mathescape=true,
                   label={lst:firstTermNotKilledByList}]{}
def isClauseKilledBy (clause : List (Nat $\times$ Bool))
                     (asgn : List (Nat $\times$ Bool))
                     : Bool :=
  clause.any fun (v, neg) =>
    match asgn.find? (fun p => p.1 == v) with
    | none => false
    | some (_, b) => !(b == literalSatisfyingBit neg)

def firstTermNotKilledByList (clauses : List (List (Nat $\times$ Bool)))
                         (asgn : List (Nat $\times$ Bool))
                         : List (Nat $\times$ Bool) :=
  match clauses.findIdx? (fun c => isClauseKilledBy c asgn) with
  | some i => clauses.getD i []
  | none => []

def restrictClauseByListAssignment (clause : List (Nat $\times$ Bool))
                   (asgn : List (Nat $\times$ Bool))
                   : List (Nat $\times$ Bool) :=
  if isClauseKilledBy clause asgn then
    []
  else
    clause.filter fun (v, _) => !asgn.any fun (w, _) => w == v

def gammaBitsForClause (clause : List (Nat $\times$ Bool))
                       : List (Nat $\times$ Bool) :=
  clause.map fun (v, neg) => (v, literalSatisfyingBit neg)
\end{lstlisting}

Now let $d_1$ be the number of variables in $U_1$ and $\gamma_1$ be the setting
to the variables in $U_1$ which makes it 1. The $\gamma_i$ values are computed
by the \texttt{gammaBitsForClause} function.  Let $\pi_1$ be the part of $\pi$
which sets these variables.  Assuming $\pi_1$ is not all of $\pi$, continue the
process.  Let $T_{i_2}$ be the first term not killed by $\beta \pi_1$, and let
$U_2$ be its restriction under $\beta \pi_1$. Let $d_2$ be the number of
variables in $U_2$.  Let $\gamma_2$ be the setting to the variables in $U_2$
which makes it 1.  On the other hand, let $\pi2$ be the part of $\pi$ which sets
these variables.  Keep going, until eventually $\pi_l$ finishes all of $\pi$.
At this point, truncate $\gamma_l$ to set just the variables that $\pi_l$ sets.

In our example, $d_1 = 3$, $\gamma_1 = [(1,1),(2,1),(3,0)]$, and $\pi_1 = \pi =
[(1,0),(2,0),(3,0)]$ so the recursive function terminates after the first
iteration. At this point, the dead accumulator is the initial restriction $[(0,
  0), (5, 1)]$ concatenated with $\gamma_1 = [(1,1),(2,1),(3,0)]$. The dead bits
in the final encoding are therefore $[(0,0), (5,1), (1,1), (2,1), (3,0)]$. The
auxiliary bits are generated in chunks, with one for each $T_i$ and $\pi_i$
pair. The chunk is generated by the \texttt{encoderChunk} function (listing
\ref{lst:encoderChunk}), which replaces every variable in $\pi_i$ with its
position in $T_i$ (computed by the \texttt{findPositionInClause'} function).
For $T_{i_1} = \clauseCone$ and $\pi_1 = [(1,0),(2,0),(3,0)]$, the chunk of
auxiliary bits is $[(0,0),(1,0),(2,0)]$. The final encoding of $\beta$, which
maps $x_0 \rightarrow 0$ and $x_5 \rightarrow 1$ is therefore the pair
$([(0,0),(5,1),(1,1),(2,1),(3,0)], [(0,0),(1,0),(2,0)])$.

\begin{lstlisting}[caption={Chunk encoding},
                   mathescape=true,
                   label={lst:encoderChunk}]{}
def findPositionInClause' (clause : List (Nat $\times$ Bool)) (v : Nat) : Nat :=
  match clause.findIdx? (fun lit => lit.1 == v) with
  | some i => i
  | none   => clause.length

def encoderChunk (selectedClause : List (Nat $\times$ Bool))
                (chunk : List (Nat $\times$ Bool))
                : List (Nat $\times$ Bool) :=
  chunk.map fun (v, dir) =>
    let pos := findPositionInClause' selectedClause v
    (pos, dir)
\end{lstlisting}

\noindent
\begin{minipage}{\linewidth}
\begin{lstlisting}[caption={Core Restriction Encoding},
                   mathescape=true,
                   label={lst:beameEncoderAux}]{}
def beameEncoderAux (fuel : Nat)
                         (remaining$\pi$ : List (Nat $\times$ Bool))
                         (clauses : List (List (Nat $\times$ Bool)))
                         ($\rho$ : List (Nat $\times$ Bool))
                         (deadAcc : List (Nat $\times$ Bool))
    : List (Nat $\times$ Bool) $\times$ List (List (Nat $\times$ Bool))
  :=
  match fuel with
  | 0         => (deadAcc, [])
  | fuel' + 1 =>
    if remaining$\pi$ = [] then
      (deadAcc, [])
    else
      let selectedClause := firstTermNotKilledByList clauses $\rho$
      let restrictedClause := restrictClauseByListAssignment selectedClause $\rho$
      let restrictedVars := restrictedClause.map Prod.fst

      let $\pi_i$ := remaining$\pi$.filter
        fun x => restrictedVars.contains x.1
      let $\gamma_i$ := (gammaBitsForClause restrictedClause).take $\pi_i$.length
      let deadAcc' := deadAcc ++ $\gamma_i$

      if $\pi_i$.length = 0 then
        (deadAcc', [])
      else
        let chunk := encoderChunk selectedClause $\pi_i$

        let $\rho$' := combineRestrictions $\rho$ $\pi_i$
        let remaining' := remaining$\pi$.filter
          fun (w, _) => !$\pi_i$.any fun (w', _) => w' == w

        let (finalDead, restEnc) :=
          beameEncoderAux fuel' remaining' clauses $\rho$' deadAcc'

        (finalDead, chunk :: restEnc)
\end{lstlisting}
\end{minipage}

\subsubsection{The Decoder}

The pair generated by the encoder is the input to the decoder.  Along with the
original DNF, the decoder implementation in Listing~\ref{lst:beameDecoder}
receives the dead variables from the encoder as \texttt{baseAssignment} and the
auxiliary data as \texttt{auxBits}. It computes the first term that survives the
current restriction, initially \texttt{baseAssignment}. Once this term has been
identified, the auxiliary chunk determines which variables of the term it
represents and their assigned bits. This process repeats until all chunks have
been processed. In our example, the auxiliary bits $[(0,0),(1,0),(2,0)]$ map to
$x_1$, $x_2$, and $x_3$, respectively, in $\clauseCone$. Therefore, the decoder
recovers $\pi = [(1,0),(2,0),(3,0)]$.

\noindent
\begin{minipage}{\linewidth}
\begin{lstlisting}[caption={Chunk encoding},
                   mathescape=true,
                   label={lst:beameDecoder}]{}
def beameDecoder (dnf : UnboundedFanInFormula)
                      (baseAssignment : List (Nat $\times$ Bool))
                      (auxBits : List (List (Nat $\times$ Bool)))
    : List (Nat $\times$ Bool) :=
  let clauses := dnfClauses dnf
  let (_, recovered) :=
    auxBits.foldl
      (fun (currB, accVars) chunk =>
         let selectedClause := firstTermNotKilledByList clauses currB
         chunk.foldl
          (fun (innerB, innerVars) (pos, $\pi$bit) =>
            let v := (selectedClause.getD pos (0, false)).1
            ((v, $\pi$bit) :: innerB, (v, $\pi$bit) :: innerVars))
          (currB, accVars))
      (baseAssignment, ([] : List (Nat $\times$ Bool)))
  recovered.reverse
\end{lstlisting}
\end{minipage}

At most $\lg w$ bits are required to encode a variable because the variables in
the auxiliary information are encoded as offsets within terms of the DNF.  Chunk
$i$ of auxiliary data encodes as many variables as $\pi_i$ (the variables fixed
by $\gamma_i$). The encoder runs until all the $\pi_i$ cover the $d$ variables
being restricted.  Therefore,

$$ \sum\limits_{i} \texttt{auxBits}[i].\texttt{length} = \sum\limits_{i} |\pi_i| = d$$

Since the variables must also be encoded, at most $d+d\lg w$ bits are
required. However, the encoder outputs a list of lists, whose structure
separates the chunks. Encoding those separators in a flat representation
requires at most another $d$ bits, for a total of $2d+d\lg w$ auxiliary bits,
the bound required in Section~\ref{sec:CoreSwitchingLemmaIdea}.

\subsubsection{Injectivity of the Encoding}

With the encoder and decoder defined, we prove that decoding reverses
encoding. We define the encoder's domain as \texttt{injectionTargetSet} in
Listing~\ref{lst:injectionTargetSet}.  It represents the set of all triples
$(S', \texttt{bitIndex}, \texttt{advIndex})$ such that $S'$ is an $(s -
d)$-element subset of \texttt{Finset.range n}, $\texttt{bitIndex} < 2 ^ {n - (s
  - d)}$ and $\texttt{advIndex} < (4w) ^ d$.  We also prove in the subsequent
lemma ($\texttt{injectionTargetSet\_card}$) that it has cardinality $\binom{n}{s
  - d} 2^{n - (s - d)} \cdot (4w)^d$ (as expected in section
\ref{sec:CoreSwitchingLemmaIdea}).

\begin{lstlisting}[caption={Defining the Codomain of the Encoder},
                   mathescape=true,
                   label={lst:injectionTargetSet}]{}
def injectionTargetSet (n s d w : Nat) : Finset (Finset Nat $\times$ Nat $\times$ Nat) :=
  ((Finset.range n).powersetCard (s - d))
  $\times^s$ (Finset.range (2 ^ (n - (s - d)))
  $\times^s$ Finset.range ((4 * w) ^ d))

lemma injection_target_card (n s d w : Nat) :
    (injectionTargetSet n s d w).card =
      Nat.choose n (s - d) * (4 * w) ^ d * 2 ^ (n - (s - d))
\end{lstlisting}

The \texttt{beameEncoderMap} function maps the constituents of the switching
lemma to the \texttt{injectionTargetSet}. Given a DNF on $n$ variables and a
non-negative upper bound on its width, along with a set $S$ of live variables
and a set \texttt{bits} of values for the dead variables, it computes an element
of the \texttt{injectionTargetSet} as follows:

\begin{enumerate}
\item{
  Convert $(S, bits)$ to a list-based assignment $\beta$.
}
\item{
  Build the canonical decision tree from the DNF and the restriction $\beta$.
}
\item{
  Create a finite set ($J$) from the first $d$ entries of the leftmost path of length $> d$. These entries are distinct so $|J| = d$.
}
\item{
  Run \texttt{encoderRestriction} to generate the final list of dead bits and the corresponding auxiliary bits.
}
\item{
  Compute $S' = S \setminus J$ and extracts the dead bits from the encoders output.
}
\item{
  Encode the variable positions and directions (values) and chunk separators.
}
\item{
  Output a triple $(\texttt{S'}, ~\texttt{encodeBits bits'}, ~\texttt{posIndex} \cdot 2^{2d} + \texttt{dirBoundaryIndex})$.
}
\end{enumerate}

\begin{lstlisting}[caption={Mapping Encoder Output to the injectionTargetSet},
                   mathescape=true,
                   label={lst:beameEncoderMap}]{}
def beameEncoderMap (d n : Nat)
                (f : UnboundedFanInProperDNF n)
                (w : Nat)
                (hwidth : dnfWidth f.val $\leq$ w)
                (_hw : 0 < w)
                ($\sigma$ : OpenUnitIntervalQ)
                (_hsd : d < Nat.ceil ($\sigma$.val * n))
                (s_live : Finset Nat)
                (h_s : s_live $\subseteq$ Finset.range n)
                (hcard : s_live.card = Nat.ceil ($\sigma$.val * n))
                (bits : List Bool)
                (h_bits_len : s_live.card + bits.length = n)
                (hbad : isBadRestriction d n $\sigma$ f
                  $\langle\langle\langle$s_live, h_s$\rangle$, hcard$\rangle$, bits, h_bits_len$\rangle$ = true) :
    (injectionTargetSet n (Nat.ceil ($\sigma$.val * n)) d w)
  := sorry
\end{lstlisting}

\begin{lstlisting}[caption={Mapping Encoder Output to the injectionTargetSet},
                   mathescape=true,
                   label={lst:beameDecoderMap}]{}
def beameDecoderMap (d n : Nat)
                (f : UnboundedFanInProperDNF n)
                (w : Nat)
                ($\sigma$ : OpenUnitIntervalQ)
                (s' : Finset Nat)
                (bit_idx adv_idx : Nat)
  : Finset Nat $\times$ List Bool
  := sorry
\end{lstlisting}

With \texttt{beameDecoderMap} and \texttt{beameEncoderMap} definitions
available, we can now write the roundtrip lemma
\texttt{beame\_encoder\_decoder\_injection\_roundtrip} (listing
\ref{lst:beame_encoder_decoder_injection_roundtrip}), which states that running
the decoder on the output of the encoding map recovers the original inputs to
the encoder. This proves that the bad restriction encoder is injective as
required by the core switching lemma idea in section
\ref{sec:CoreSwitchingLemmaIdea}.

\noindent
\begin{minipage}{\linewidth}
\begin{lstlisting}[caption={The Encoder and Decoder Roundtrip Correctly},
                   mathescape=true,
                   label={lst:beame_encoder_decoder_injection_roundtrip}]{}
lemma beame_encoder_decoder_injection_roundtrip
    (d n : Nat)
    (f : UnboundedFanInProperDNF n)
    (w : Nat)
    (hwidth : dnfWidth f.val $\leq$ w)
    (hw : 0 < w)
    ($\sigma$ : OpenUnitIntervalQ)
    (hsd : d < Nat.ceil ($\sigma$.val * n))
    (s_live : Finset Nat)
    (h_s : s_live $\subseteq$ Finset.range n)
    (hcard : s_live.card = Nat.ceil ($\sigma$.val * n))
    (bits : List Bool)
    (h_bits_len : s_live.card + bits.length = n)
    (hbad : isBadRestriction d n $\sigma$ f
      $\langle\langle\langle$s_live, h_s$\rangle$, hcard$\rangle$, bits, h_bits_len$\rangle$ = true)
  :
    let result := beameEncoderMap d n f w hwidth hw $\sigma$ hsd
                              s_live h_s hcard bits h_bits_len hbad

    beameDecoderMap d n f w $\sigma$ result.val.1 result.val.2.1 result.val.2.2 =
      (s_live, bits)
\end{lstlisting}
\end{minipage}

The \texttt{beame\_encoder\_decoder\_injection\_roundtrip} roundtrip lemma is a
key ingredient of the injection upper bound proved in Lean as the lemma
\texttt{injection\_bound}. This injection upper bound and the
\texttt{choose\_ratio\_bound\_exact} lemma of
Section~\ref{sec:CoreSwitchingLemmaIdea} are the core implementation of the
switching-lemma proof.

\begin{lstlisting}[caption={The Encoder and Decoder Roundtrip Correctly},
                   mathescape=true,
                   label={lst:injection_bound}]{}
lemma injection_bound (n w d : Nat)
                                        (f : UnboundedFanInProperDNF n)
                                        (hwidth : dnfWidth f.val $\leq$ w)
                                        ($\sigma$ : OpenUnitIntervalQ)
                                        (hsd : d < Nat.ceil ($\sigma$.val * n)) :
    badRestrictionCount n d f $\sigma$ $\leq$
      (Nat.choose n (Nat.ceil ($\sigma$.val * n) - d) * (4 * w) ^ d
      * 2 ^ (n - (Nat.ceil ($\sigma$.val * n) - d)))
\end{lstlisting}

This completes the appendix account of the encoding and counting argument that
supports the switching lemma stated in Section~\ref{subsec:SwitchingLemma}.

\end{document}